\documentclass[aps,pra,reprint,superscriptaddress,twocolumn,notitlepage]{revtex4-1}
\usepackage[utf8]{inputenc}

\usepackage{float}
\usepackage{graphicx}  % needed for figures
\usepackage{xcolor}
\usepackage{dcolumn}   % needed for some tables
\usepackage{bm}        % for math
\usepackage{braket}
\usepackage{amssymb}   % for math
\usepackage{epstopdf}
\usepackage{xfrac}
\usepackage{cancel}
\usepackage{soul}
\usepackage{amsmath}
\usepackage{amsthm}
\usepackage{amsfonts}
\usepackage{bbm}
\definecolor{darkblue}{rgb}{0.1,0.2,0.6}
\definecolor{darkred}{rgb}{0.8,0.1,0.2}
\definecolor{darkgreen}{rgb}{0.31,0.62,0.24}
\usepackage[colorlinks,citecolor=darkblue,linkcolor=darkblue,urlcolor=darkblue]{hyperref} 
\usepackage{tikz}
\usetikzlibrary{quantikz}

\usepackage{enumerate}
\usepackage{setspace}
\usepackage{url}  % This makes \url work
\usepackage{mathrsfs}
\usepackage{bbold}

\newcommand{\Tr}{\text{Tr}}
\newcommand{\Z}{\mathbb{Z}}

\newcommand{\rTa}{{\hat\rho^{T_2}_A} }
\newcommand{\norm}[1]{\left\lVert#1\right\rVert}

\newcommand{\bigket}[1]{ {\big|{#1}\big\rangle} }

\newcommand{\bigbra}[1]{ {\big\langle{#1}\big|} }

\numberwithin{equation}{section}
\renewcommand\theequation{\arabic{section}.\arabic{equation}}

\usepackage{tikz}
\usetikzlibrary{positioning}
\usetikzlibrary{patterns}

\usepackage{tikzit}
\tikzstyle{basic}=[fill=white, draw=black, shape=circle]
\tikzstyle{basic rect}=[fill=white, draw=black, shape=rectangle]
\tikzstyle{medium box}=[fill=white, draw=black, shape=rectangle, minimum width=0.75cm, minimum height=0.75cm]
\tikzstyle{large box}=[fill=white, draw=black, shape=rectangle, minimum height=1.5cm, minimum width=1.5cm]

\tikzstyle{dash}=[-, dashed]
\tikzstyle{dotted}=[-, densely dotted, tikzit draw=magenta, thick]

\begin{document}
\title{Symmetry protected entanglement in random mixed states}

\author{Kasra Hejazi}
\affiliation{Department of Physics, University of California Santa Barbara, CA 93106, USA}

\author{Hassan Shapourian}
\affiliation{Microsoft Station Q, Santa Barbara, CA 93106, USA}

\date{March 2021}

\begin{abstract}
Symmetry is an important property of quantum mechanical systems which may dramatically influence their behavior in and out of equilibrium. In this paper, we study the effect of symmetry on tripartite entanglement properties of typical states in symmetric sectors of Hilbert space. 
In particular, we consider Abelian symmetries and derive an explicit expression for the logarithmic entanglement negativity of systems with $\mathbb{Z}_N$ and $U(1)$ symmetry groups.
To this end, we develop a diagrammatic method to incorporate partial transpose within the random matrix theory of symmetric states and formulate a perturbation theory in the inverse of the Hilbert space dimension. We further present entanglement phase diagrams as the subsystem sizes are varied and show that there are qualitative differences between systems with and without symmetries. We also design a quantum circuit to simulate our setup.
\end{abstract}

\maketitle

\section{Introduction}

Dynamics of quantum chaotic many-body systems which ultimately leads to thermal equilibrium has been a subject of fundamental research in physics. A particular topic of great interest recently has been the dynamics of quantum entanglement in such systems. However, the study of {\it large enough} systems that exhibit interesting large scale effects has been a challenge, to some extent by definition, in quantum chaotic systems. A very useful tool, that has played an important role in our understanding of universal behavior in such systems, has been the introduction of randomness in the system; random matrix theory, as an example, has proven to be able to reproduce many universal properties in such systems \cite{BGS1984,GUHR1998,Berry_spectral_rigidity,2018PhRvX...8b1062K} and has paved the way to actually performing concrete analytical calculations.

Another paradigmatic example is the study of quantum dynamics in random unitary circuits; introduction of randomness retains the essential physics while providing analytical handles to study averaged quantities; this endeavor has been very successful in identifying universal dynamical properties and phenomena in quantum chaotic systems far from equilibrium \cite{nahum2017quantum,nahum2018operator,von2018operator,chan2018solution}. 
In many real applications, on the other hand, one expects some further structure in dynamics; in particular, a ubiquitous situation is when conservation laws are present due to e.g.~a Hamiltonian dynamics or a symmetry preserving law of motion. Such situations have also been studied widely recently in particular in the context of random unitary circuits and it has been shown that the addition of a conservation law can lead to novel phenomena and behaviors \cite{rakovszky2018diffusive,khemani2018operator}. 

\begin{figure}
\centering
\includegraphics[scale=.7]{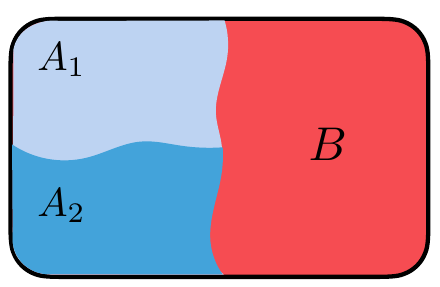}
\caption{\label{fig:geometry} Tripartite geometry of random pure states.
In this paper, we are interested in the entanglement negativity between two parties (e.g., $A_1$ and $A_2$) out of the three. There is no notion of locality in this setup.}
\end{figure}

In the present work, we study the entanglement properties of a chaotic system with a symmetry at late times starting from a simple symmetric state (or many-body eigenstates of a symmetric quantum chaotic Hamiltonian) through the lens of random matrix theory; we assume that {\it symmetric} random many-body states are capable of capturing the essential physics in such situations. Particularly, we will consider tripartite entanglement in a system having a symmetric random state.
In a previous work~\cite{Shapourian2021}, entanglement negativity, as a multipartite entanglement measure, of a random state without symmetry in a tripartite geometry (see Fig.~\ref{fig:geometry}) was studied (see also Refs.~\cite{Marko2007,Aubrun2010,Aubrun2014,Bhosale,Aubrun2012,Fukuda2013,Collins2012,Szyma_ski_2017,Collins_rev,Gray2018} for early studies on the partial transpose of random mixed states); this was, in particular, done by investigating the entanglement (encoded in $\hat\rho_A$) between $A_1$ and $A_2$. Note that subsystem $B$ can be viewed as an environment for subsystem $A$. It was shown that the entanglement between $A_1$ and $A_2$ subsystems shows different behaviors as the Hilbert space dimensions of subsystems are varied.
We briefly recapitulate the main results of this work in the following two paragraphs and next summarize our results for symmetric random states in this paper.

A large-$L$ perturbative diagrammatic approach in Ref.~\cite{Shapourian2021} was introduced and employed to calculate the entanglement negativity and the entanglement negativity spectrum in the setting outlined above (see also Refs.~\cite{Dong-Qi-2021,Dong2021,KudlerFlam2021_short,KudlerFlam2021_long,KudlerFlam2021_neg} for a similar diagrammatic approach to the entanglement negativity and relative entropy of random tensor networks and models of evaporating black holes);
it was shown that the main parameter controlling the entanglement behavior is $q = \frac{L_B}{L_A}$, where $L_{s}$ denotes the Hilbert space dimension of $s = A,B$. Looking at extreme limits of this parameter is illuminating: for $q \gg 1$ one expects the bath to be very large and thus $A$ to be almost fully entangled with $B$ resulting in a minimal entanglement between $A_1$ and $A_2$. On the other hand, for $q \ll 1$, the bath is small and not capable of thermalizing the $A$ subsystem and thus one expects a volume law entanglement between $A_1$ and $A_2$. Indeed, the results in \cite{Shapourian2021} show that this picture is correct and the reduced density matrix of $A$ shows a transition form being positive partial transpose (PPT) to being negative partial transpose (NPT) as $L_B$ is lowered from above to below the transition value $L_{B, \text{PPT}} = 4L_A$. The PPT regime by definition shows zero logarithmic negativity.

Below this transition point, where the state is NPT, it is shown that two possibilities arise (assuming without loss of generality $L_{A_1}\gg L_{A_2}$): first, if $A_1$ is not larger than half of the system (or $L_{A_1} \ll L_{A_2} L_B$), it was found that the entanglement negativity between $A_1$ and $A_2$ becomes independent of the relative sizes of these two subsystems; in this {\it entanglement plateau} regime, the logarithmic negativity  for a qubit system shows a behavior $\mathcal{E}_{A_1:A_2} \sim \frac12 (N_A-N_B)$ (independent of the ratio $N_{A_1}/N_{A_2}$). On the other hand, in the opposite limit of $L_{A_1} \gg L_{A_2} L_B$, it was shown that there is maximal entanglement between $A_1$ and $A_2$ with a logarithmic negativity of $\mathcal{E}_{A_1:A_2} \sim N_{A_2}$. The two regimes are separated by a critical point given by the relation $L_{A_1} = L_{A_2} L_B$, where the negativity spectrum exhibits a divergence at the origin.

Here, we build on this previous work and study the case where the random state of the system is symmetric. This symmetry at the level of the state could be present due to a global symmetry of the dynamics of the system. If the initial state of the system has a definite  conserved charge under a global symmetry and the dynamics preserves the symmetry, we expect the final state, as complicated as it could be, to have also the same quantum number.
Note that one example of such a conserved quantity in a Hamiltonian system could be the energy. We consider two separate classes of symmetries in this work: $\mathbb{Z}_R$ symmetry in a system consisting of qudits with $R$ onsite degrees of freedom (including $R=2$, i.e.~qubits) and a $U(1)$ symmetry in a qubit system. 
The latter can e.g.~represent systems of spin-$\frac{1}{2}$'s with rotational symmetry around $z$-axis or fermionic systems with conserved particle number.

Given that the state of the whole system has a definite symmetry charge, and as will be discussed further later, the density matrix of subsystem $A$ can be written as:
\begin{align*}
    \hat \rho_{A}= \bigoplus_{q_A} p_{q_A} \hat  \rho_A^{(q_A)}.
\end{align*}
Noting this, we focus in this work on the symmetry resolved entanglement negativity~\footnote{We should clarify that the term ``symmetry resolved entanglement negativity'' has also been used for a different quantity recently~\cite{Cornfeld2018, Murciano,Neven2021}, which has to do with the block diagonal form of $\rTa$, i.e.~after taking the partial transpose. This property is based on the observation that $[\rTa, \Delta q]=0$ where $\Delta q =q_2 -q_1$ denotes the charge imbalance between subsystems $A_1$ and $A_2$. While this is a mathematical property of $\rTa$ it is not clear what the physical meaning of these blocks is. }, i.e.~entanglement negativity of individual blocks of the density matrix that we denote by $\mathcal{E}(\hat\rho_{A}^{(q_A)})$. These quantities represent a more refined measure of mixed state entanglement than the negativity of the whole density matrix, as different symmetry sectors are considered separately. Furthermore, another quantity, i.e.~the symmetry averaged entanglement negativity can be calculated in terms of the above symmetry resolved values:
\begin{align}
\label{eq:logneg-symmetry}
    \overline{\cal E}_{A_1:A_2} := \sum_{q_A} p_{q_A}\, {\cal E}(\hat\rho_A^{(q_A)}),  
\end{align}
which is the analog of the first term in the bipartite entanglement entropy
\begin{align}
    S_{A} = \sum_{{q_A}} p_{q_A}\, S(\hat\rho_A^{({q_A})})  -\sum_{q_A} p_{q_A} \log p_{q_A}.
\end{align}
As we will see later, the symmetry averaged entanglement negativity can also be motivated as a way to identify symmetric separable states which are realized by {\it symmetric} local operations and classical communications (LOCC).

Another interesting possibility that we are considering, which can in principle access the entanglement properties of individual blocks $\hat\rho_A^{(q_A)}$ is the situation in which, only the charge of subsystem $B$ is measured in a symmetric system; if this is done properly, the state of subsystem $A$ will not completely collapse into a pure state and interesting tripartite correlations will be retained. 
It has been known that such measurement in symmetry protected topological (SPT) phases can have nontrivial outcomes; in particular, it was shown \cite{Marvian2017} that in a SPT phase, two spatially separated regions--which should show no correlations due to the SPT state being a short-range correlated state--can in fact become entangled if the symmetry charge of their complements is measured.
Similarly here in the context of symmetric random states, we will be interested in the entanglement between $A_1$ and $A_2$ when the symmetry charge of $B$ is measured. Interestingly, one can come up with quantum circuits that measure only the $B$ subsystem charges in systems with either $\mathbb{Z}_R$ or $U(1)$ symmetries. 
Details of such charge measuring circuits, which can be useful in possible quantum simulation of the setting discussed above on near term quantum devices, will also be presented in this paper.

\begin{figure*}
    \centering
    \includegraphics[scale=1.2]{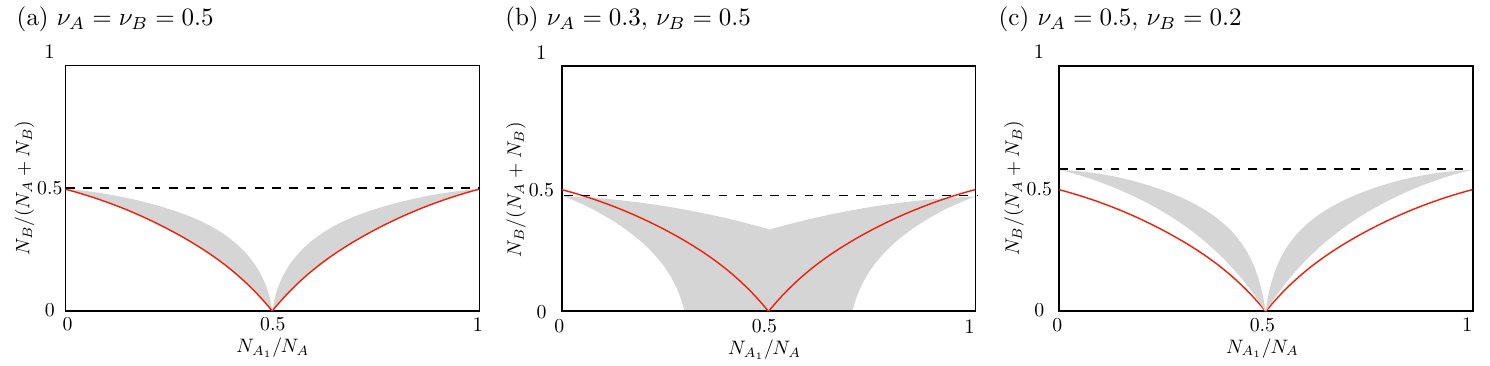}
    \caption{Entanglement phase diagram of symmetry projected random mixed states for $U(1)$ symmetric systems. The mixed state is obtained from a random pure state via partial tracing. Different panels correspond to different subsystem filling factors as indicated. Dashed lines (given by $N_A f(\nu_A)=N_B f(\nu_B)$) separate the upper part of the phase diagram, where the volume law term in the negativity is suppressed (but not fully PPT), from the lower part where the negativity obeys a volume law form. The lower region, $N_A f(\nu_A)>N_B f(\nu_B)$, consists of two phases: First, the two corners are called maximal entanglement regimes where $\braket{{\cal E}_{A_1:A_2}}= \min(N_{A_1},N_{A_2})f(\nu_A)$; second, the middle region, where the dominant diagrams break the replica symmetry leading to $\braket{{\cal E}_{A_1:A_2}}= N_A f(\nu_A)-N_B f(\nu_B)$ (to the leading order). The shaded regions represent the critical phase between the replica symmetry breaking and maximal entanglement phases where the entanglement negativity spectrum diverges at zero. Red curves are shown as a reference for critical line of non-symmetric states.}
    \label{fig:phasediag}
\end{figure*}

A summary of our approach and results follow; in order to study the mixed state entanglement of $A_1$ and $A_2$ subsystems, we calculate the ensemble averaged spectrum of the partially transposed symmetry resolved blocks using the diagrammatic approach; the calculation is carried out in different regimes, from which the entanglement negativity is also calculated. We show that entanglement regimes similar to the nonsymmetric case can be observed in the symmetric case as well. However, in the symmetric case, the fact that we need to take into account the charges of different subsystems adds to the richness and complexity of the entanglement behavior. In particular, major differences that appear between the symmetric case and the nonsymmetric one are as follows: first, the NPT to PPT transition loses its sharp definition and becomes a crossover rather than a transition. Furthermore, since the Hilbert space dimensions of subsystems depend in general on quantum numbers of those subsystems (e.g., for the $U(1)$ symmetry), we see that in general, the transition to maximal entanglement in the NPT region happens in a critical {\it region} as opposed to a critical line. This can be seen in Fig.~\ref{fig:phasediag} in the case of $U(1)$ symmetry, where several situations are considered and and the critical regions are shown as shaded, while in a nonsymmetric state we expect criticality only on the red curve.

\bigskip

The rest of our paper is organized as follows:
In Sec.~\ref{sec:Preliminary},
we review some background materials about the partial transpose, the symmetry charge, modified separability problem, and the problem setup. There, we argue that why one should use the symmetry resolved logarithmic negativity to address the separability problem in systems subject to symmetric LOCCs.
Section~\ref{sec:moments} is devoted to reviewing the diagrammatic approach to random density matrices and calculating the moments of the partially transposed density matrix; this section serves as a warm-up for the subsequent sections.
In Sec.~\ref{sec:spectraldensity}, we present the central results of this paper, where we use the resolvent function method diagrammatically to derive the spectral density of  the partially transposed density matrix in various regimes.
We also provide several numerical benchmarks.
In Sec.~\ref{sec:experiment}, we propose a quantum circuit to perform a local symmetry charge projection on a subsystem, which would ultimately be useful to simulate the symmetry resolved entanglement negativity on a quantum computer. 
Finally, we finish our paper by several closing remarks and future directions in Sec.~\ref{sec:discussion}. In several appendices, we provide more details of our calculations.

%%%%%%%%%%%%%%%%%%%%%%%%%%%%%%%%%%%%%%%%%%%%%%%%%%%%
\section{Preliminary Remarks}
\label{sec:Preliminary}

In this section, we first review the partial transpose and the definition of symmetric states and symmetry charges, and then we discuss the separability criterion for symmetric states and introduce the setup of the problem and some notations.

\subsection{Partial transpose and entanglement negativity}

In this part, we briefly review some basics about the partial transpose and the logarithimic negativity (LN) which may be skipped.
LN has proven to be useful in the study various aspects of many-body physics \cite{PhysRevA.66.042327,PhysRevLett.100.080502,PhysRevA.78.012335,Anders2008,PhysRevA.77.062102,PhysRevA.80.012325,Eisler_Neq,Sherman2016,dct-16,PhysRevA.80.010304,PhysRevLett.105.187204,PhysRevB.81.064429,PhysRevLett.109.066403,Ruggiero_1,PhysRevA.81.032311,PhysRevA.84.062307,Mbeng,Gray2019,Grover2018,java-2018,Calabrese2012,Calabrese2013,Calabrese_Ft2015,Ruggiero_2,Alba_spectrum,kpp-14,fournier-2015,bc-16,Wald_2020,PhysRevB.101.064207,lu2019structure,angelramelli2020logarithmic,Juh_sz_2018,Schreiber,rosz2020entanglement,Shapourian_FS,wu2019entanglement,10.21468/SciPostPhys.8.4.063,Wen2016_1,Wen2016_2,PhysRevA.88.042319,PhysRevA.88.042318,hc-18,YeJePark,PhysRevB.101.085136,Clabrese_network2013,Alba2013,PhysRevB.90.064401,Nobili2015,Rangamani2014,2014JHEP...09..010K,PhysRevD.99.106014,PhysRevLett.123.131603,ctc-14,ez-14,hb-15,ac-18b,wen-2015,gh-18,alba2020spreading,Gruber_2020,PhysRevB.101.245130,Kudler-Flam2019,2020JHEP...04..074K,2020arXiv200811266K,Pollmann_Turner2012,Shiozaki_Ryu2017,Shap_unoriented,Shiozaki_antiunitary,Kobayashi,Grover2020}.

Let $\hat\rho_A$ be the density matrix of subsystem $A$ which consists of subsystems $A_1$ and $A_2$ with orthonormal bases
$\bigket{e_1^{(k)}}$ and $\bigket{e_2^{(j)}} $ respectively:
\begin{align}
\hat\rho_A=\sum_{ijkl} \rho_{ijkl} \bigket{e_1^{(i)}, e_2^{(j)}}   
 \bigbra{e_1^{(k)}, e_2^{(l)}}.
\end{align}
Partial transpose of the above matrix with respect to $A_2$, which we denote as $\hat\rho_A^{T_2}$, is defined by exchanging the indices of subsystem $A_2$ in the following way:
\begin{align} \label{eq:rTb}
\hat\rho_A^{T_2}=\sum_{ijkl} \rho_{ijkl} \Ket{e_1^{(i)}, e_2^{(l)}}  \Bra{e_1^{(k)}, e_2^{(j)}}.
\end{align}
A density matrix after partial transpose is left Hermitian and its trace is preserved to be equal to 1, however it can have negative eigenvalues. If the eigenvalues of the partial transpose of a state are kept positive it is called a positive partial transpose (PPT) state and otherwise it is called negative partial transpose (NPT). The NPT property indicates that the mixed state contains quantum entanglement~\cite{Peres1996, Horodecki1996}.

Given the above properties of a partially transposed density matrix, one can define the following two measures for mixed state entanglement~\cite{PlenioEisert1999,Vidal2002,Plenio2005}:
\begin{itemize}
\item entanglement negativity:
 \begin{align} 
 \label{eq:neg_def}
{\cal N}(\hat\rho_A) &= \frac{\norm{\rTa }_1 -1}{2}, 
\end{align}
\item and logarithmic negativity:
 \begin{align} 
 \label{eq:logneg_def}
{\cal E} (\hat\rho_A) &= \log \norm{\rTa }_1.
\end{align}
\end{itemize}
Here, $\norm{O}_1= \Tr \sqrt{OO^\dag}$ is the trace norm, which for a Hermitian matrix is the sum of the absolute values of eigenvalues. %Since $\rTa$ is Hermitian, the trace norm is simply the sum of the absolute value of the eigenvalues.
Logarithmic negativity and the entanglement negativity satisfy ${\cal E}=\log (2{\cal N}+1)$.
A more informative measure, on the other hand, is the entanglement negativity spectrum that shows the probability distribution of the eigenvalues of $\rTa$:
\begin{align} \label{eq:PTdist}
P (\xi)=\sum_{i=1}^{L_A} \delta(\xi-\xi_i),
\end{align}
Having the negativity spectrum, one can also calculate the negativity:
\begin{align}
\label{eq:neg_dist}
 {\cal N}(\hat\rho_A)   =  -\int_{\xi<0} d\xi\ \xi\, P(\xi).
\end{align}

To see the relation with entanglement in mixed states, we first consider mixed states that have zero entanglement: separable states whose density matrices can be written as:
\begin{align}
\label{eq:separable}
\hat\rho_{\text{sep}} = \sum_{i,j} p_{ij} \hat\rho_{A_1,i}\otimes \hat\rho_{A_2,j},
\qquad p_{ij}>0,
\end{align}
where $\hat\rho_{A_1,i}$ and $\hat\rho_{A_2,i}$ are density matrices for subsystems $A_1$ and $A_2$. Separable states are believed to harbor no entanglement as then can be prepared through the use of local operations and classical communications (LOCC) on manifestly unentangled states, i.e.~product states, shared between $A_1$ and $A_2$. We now turn to the entanglement negativity of these states; it is straightforward to check that partially transposed separable states do not have negative eigenvalues, and as a result are PPT states and have zero negativity. This means that NPT states cannot be separable. In this paper we will employ the entanglement negativity as a measure of mixed state entanglement and as a result only distinguish NPT and PPT states as opposed to distinguishing separable states from nonseparable ones.

%%%%%%%%%%%%%%%%%%%%%%%%%%%%%%%%%%%%%%%%%%%%%%%%%%%
\subsection{Symmetry quantum number (charge) for Abelian symmetry groups}

In this part, we discuss the notion of symmetry charge in systems invariant under a symmetry group. Consider a system of $R$-dimensional qudits. Suppose the system is described by a Hamiltonian which is invariant under an Abelian symmetry group where the symmetry operation acts on-site, i.e.~the unitary operators representing the symmetry group elements take a tensor product form over the qudits of each subsystem $s$ as $\hat U^{(j)}_{s} = \bigotimes_{l \in s} \hat u^{(j)}_l$, where $\hat u^{(j)}_l$'s are single-site unitary operators and $j$ denotes symmetry group element. We use an orthonormal basis for each site $\left\{ \left| r \right\rangle \right\}$ with $r=0,\ldots,R-1$, in which the unitaries $\hat u^{(j)}$ are diagonal. Each basis element furnishes a one-dimensional representation for $\mathcal{G}$. Note that since $\mathcal{G}$ is Abelian, all its irreducible representations are one-dimensional.

For simplicity, we define symmetry charges only for the specific groups that we will be studying in this work, a general construction is similar:
\begin{itemize}
    \item 
\emph{Example 1: $\mathbb{Z}_R$ symmetry$-$}
The generator of the cyclic group $\mathbb{Z}_R$ satisfies $ \hat h^R = 1$. 
We form the qudit basis by taking each state $\ket{r}$ to satisfy $ \hat h \ket{r} = e^{i\frac{2\pi }{R} r} \ket{r}$.

A many-body basis state of $N$ qudits is denoted by
\begin{align}
\ket{r_1,\cdots, r_N} \equiv \ket{r_1}\otimes\cdots\otimes \ket{r_N},
\end{align}
where the subscripts denote the site number. Such a basis state has a definite charge, as
\begin{align}
\left[\bigotimes\hat{h}\right] \ket{r_1, \cdots, r_N} = e^{i\frac{2\pi h}{R}\sum_{i=1}^N r_i }\, \ket{r_1, \cdots, r_N},
\end{align}
we note that we can assign the total charge of $\sum_{i=1}^N r_i $ to the above state, where the sum is defined modulo $R$.

\item
\emph{Example 2: $U(1)$ symmetry$-$}
In this case, the symmetry group is continuous. Here, we consider a system of qubits, i.e.~$R=2$.
Every transformation $\hat{u}^{(\theta)}$ is specified by a continuous real parameter $\theta$ in the range $[0,2\pi)$; we define the states $\left|0 \right\rangle , \left|1 \right\rangle$ such that $\hat{u}^{(\theta)} \left| r \right\rangle = e^{i\theta r} \left| r \right\rangle$.
Furthermore, the total charge of a many-body state $\ket{r_1, \cdots, r_L}$ is defined by the integer $\sum_{i=1}^N r_i$. 

\end{itemize}

Noting the above definitions for symmetry charges, we define the projector onto the subset of all many-body states in one of the subsystems $s$, which have a given quantum number $q$ as $\hat{\Pi}^{(q)}_s$.
It is straightforward to check that with the above definitions, the total charge of a system is the sum of the charges of its constituent subsystems; in other words, a projector with a given charge $Q$ for the whole system which we denote as $AB$ subsystem can be written in terms of the projectors of its constituents $A$ and $B$ subsystems in the following way:
\begin{align}\label{eq:additivity}
\hat \Pi^{(Q)}_{AB}= \sum_{
                  q_A+q_B=Q } \hat \Pi^{(q_A)}_{A}\otimes \hat \Pi^{(q_B)}_{B}\ . 
\end{align}
Note that the above summation $q_A+q_B=Q$ should be performed mod $R$ for the $\mathbb{Z}_R$ symmetry.

In this paper, we consider symmetric random pure states $\ket{\Psi^{(Q)}}$, i.e., states with a determined symmetry charge which belong to the subspace associated with $\hat \Pi^{(Q)}$. In other words, $\hat \Pi^{(P)}\ket{\Psi^{(Q)}}= \delta_{P,Q} \ket{\Psi^{(Q)}}$.
An immediate implication of the aforementioned additivity of the symmetry charge is that any reduced density matrix $\hat\rho_A$ obtained from a symmetric pure state $\ket{\Psi^{(Q)}}$ via partial tracing $\hat\rho_A = \Tr_{\bar A} \ket{\Psi^{(Q)}}\bra{\Psi^{(Q)}}$ is also symmetric,
i.e., they commute with all symmetry opertors
$[\hat\rho_{A}, \hat U^{(j)}_A(h) ]=0$. This is because the pure state density matrix of the total system $\ket{\Psi^{(Q)}}\bra{\Psi^{(Q)}}$ is a projector as in Eq.~(\ref{eq:additivity}).
Put it differently, the reduced density matrix takes a  block diagonal form
\begin{align}
\label{eq:rho-block}
\hat \rho_{A}= \bigoplus_{q_A} p_{q_A} \hat  \rho_A^{(q_A)},
\end{align}
where $\hat  \rho_A^{(q_A)}$ is a block matrix of quantum number $q$ and $p_q = \Tr (\hat \rho_A \hat \Pi_A^{(q)})$.

\subsection{Separability problem in symmetric systems}

As mentioned, a separable state (\ref{eq:separable}) is a completely classical state which can be generated by LOCCs.
By definition, $\hat \rho_{\text{sep}}$ remains positive semi-definite even after PT, and hence it is a PPT state.
For symmetric systems, LOCCs are constrained to be locally symmetry preserving, i.e., they commute with local symmetry charge operators (or projection operators). As a result, the definition (\ref{eq:separable}) has an extra constraint on local density matrices that
\begin{align}\label{eq:symmetry_condition_symm_LOCC}
[\hat \rho_{s,j}, \hat \Pi_s^{(q_s)}]=0,
\end{align}
 where 
 $\hat \Pi_s^{(q_s)}$ denotes local projectors into the subspace with local symmetry charge $q_s$ within subsystem $s=A_1,A_2$, as defined in the previous subsection. By definition, the symmetric separable states also form a convex set, and because of the above constraint, this set is a subset of generic separable states. An immediate implication of this property is that a symmetric state can have a zero log negativity although it may not be realized by symmetry preserving LOCCs. For example, consider a $\mathbb{Z}_2$ symmetric system of two qubits (or equivalently, two fermions) described by the state
 \begin{align}
     \hat \rho_A = \frac{1}{4} (\mathbb{1} + p\,\hat\sigma^x_1 \otimes \hat\sigma^x_2).
 \end{align}
The local $\mathbb{Z}_2$ symmetry operators are $\hat\sigma^z_1$ and $\hat\sigma^z_2$. Clearly, this is a symmetric state as it commutes with symmetry operator on $A$, i.e., $\hat\sigma^z_1\otimes \hat\sigma^z_2$, while it cannot be written as a symmetric separable state. However, $\rTa = \rho_A$ and ${\cal E}(\hat\rho_A) =0$. As mentioned in the introduction, a possible resolution is to use the symmetry averaged logarithmic negativity (\ref{eq:logneg-symmetry}), where we obtain ${\cal E}(\hat\rho_A^{(\pm)})=
\log(1+p)$ leading to $\overline{\cal E}_{A_1:A_2}=\log (1+p)$. To understand this physically, we see that it vanishes at $p=0$ as expected since it is a fully mixed state with zero entanglement, whereas $\overline{\cal E}_{A_1:A_2}=\log 2$ at $p=1$ where the state is an equal superposition of two Bell-pair type states with different $\mathbb{Z}_2$ charges which are farthest away from states prepared by symmetry preserving LOCCs.

An alternative way to see why $\rTa$ misidentifies the entanglement in symmetric mixed states is the fact that the action of partial transpose does not commute with symmetry charge projection operators (\ref{eq:additivity}). In other words, matrix elements of different diagonal blocks in a symmetric $\hat\rho_A$ are exchanged as a result of the partial transpose. 
Therefore, we propose to apply partial transpose to every block $\hat\rho_A^{(q_A)}$ separately, compute the associated logarithmic negativity and then take the average according to the Born probabilities $p_q$ as in (\ref{eq:logneg-symmetry}). 

It is easy to see that the commutation relation~\eqref{eq:symmetry_condition_symm_LOCC} implies that we can always write a symmetric separable state such that each term $\hat \rho_{s,j}$ in the expansion~\eqref{eq:separable}
has a definite symmetry charge of the corresponding subsystem, in other words, we may write
\begin{align}
   \hat \rho_{A}^{(q_A)} = \sum_{\substack{i,j \\ q_1+q_2=q_A}} p^{(q_A)}_{ij}\,\hat\rho^{(q_1)}_{A_1,i}\otimes \hat\rho^{(q_2)}_{A_2,j},
\end{align}
for every block in $\hat \rho_{A}^{(q_A)}$ in Eq.~(\ref{eq:rho-block}).
This means that all blocks  in a state generated by symmetric LOCCs are individually separable. Thus, if the symmetry averaged negativity in Eq.~\eqref{eq:logneg-symmetry} is nonzero for a symmetric state, then it is not separable under symmetric LOCCs; this makes the symmetry averaged negativity a suitable measure for this matter.

In the rest of this paper, we focus on the entanglement negativity spectrum of each block obtained from a symmetric random pure state.

%%%%%%%%%%%%%%%%%%%%%%%%%%%%%%%%%%%%%%%%%%%%%%%%%%%%%
%%%%%%%%%%%%%%%%%%%%%%%%%%%%%%%%%%%%%%%%%%%%%%%%%%%%%
%%%%%%%%%%%%%%%%%%%%%%%%%%%%%%%%%%%%%%%%%%%%%%%%%%%%%

\subsection{Setup of the problem}
We consider a tripartite system of $A_1, A_2$ and $B$, and they comprise $N_{A_1},N_{A_2},N_{B}$ qudits which are $R$ dimensional. We assume that the system harbors an Abelian symmetry with the symmetry group $\mathcal{G}$. We start from a random pure state $\ket{\Psi^{(Q)}}$ that has a definite quantum number $Q$ under the symmetry. One can write the components in a tensor product basis, as follows
\begin{align}
\label{eq:componenets_psi_Gamma}
\ket{\Psi^{(Q)}  } = \sum_{\substack{i,\alpha \\
                  q_A + q_B = Q}} \, X_{i_{q_A}, \alpha_{q_B}}^{(q_A, q_B)} \ket{e_{i_{q_A}}^{(q_A)}}\otimes \ket{e_{\alpha_{q_B}}^{(q_B)}},
\end{align}
where $i_{q_A}, \alpha_{q_B}$ are indices enumerating states in the sectors given by quantum numbers $q_A, q_B$ in $A$ and $B$ subsystems and  $X_{i_{q_A}, \alpha_{q_B}}^{(q_A, q_B)}$ is a Gaussian random variable. Here we make use of the fact that the total quantum number is equal to the sum of the quantum numbers of the constituents, i.e.~Eq.~(\ref{eq:additivity}).

As mentioned, the fact that $\ket{ \Psi^{(Q)}}$ has a definite quantum number leads to a block diagonal structure density matrix as in Eq.~(\ref{eq:rho-block}). 
As discussed above, we study the entanglement negativity of different sectors denoted by $\hat \rho_{A}^{(q_A)}$ separately.
At this point, we require that each one of $\hat \rho_{A}^{(q_A)}$ has a unit trace on average: 
\begin{equation}\label{eq:X_variance_correlation}
    \left\langle X_{i_{q_A}, \alpha_{q_B}}^{(q_A, q_B)\ast} \
    X_{i'_{q_A'}, \alpha'_{q_B'}}^{(q_A', q_B')} \right\rangle =  \frac{\delta_{q_A q_A'} \, \delta_{q_B q_B'} \, \delta_{i_{q_A} i_{q_A}'} \, \delta_{\alpha_{q_B} \alpha_{q_B}'}}{L_{q_A} L_{q_B} },
\end{equation}
where $L_{q_s}$ denotes the dimension of the subsystem $s= A,B$ of charge $q_s$. Note that the relation $q_A+q_B=Q$ is also assumed in the above equation.
In principle, for every disorder realization, $\hat \rho_{A}^{(q_{A})}$ needs to be normalized to have a unit trace; however, the fluctuations in the denominator lead to $\frac{1}{L_{q_A} L_{q_B} }$ corrections which will be neglected throughout this paper (see Refs.~\cite{ChenLudwig,Shapourian2021} for more discussion on this).
Note that in this notation $\left\langle \Psi^{(Q)}  \big| \Psi^{(Q)} \right\rangle$ is equal to the number of different symmetry sectors in subsystem $A$ (or $B$) on average.

With the above normalization for the different blocks of the density matrix, the Born weights in the symmetry averaged log negativity (\ref{eq:logneg-symmetry}) 
are given by $\frac{L_{q_A} L_{q_B}}{\sum_{\tilde{q}_{A}} L_{\tilde{q}_A } L_{\tilde{q}_B}  }$.

\begin{table}[]
    \centering
{\footnotesize 
\renewcommand{\arraystretch}{1.2}
\begin{tabular}{cc}
    \hline
    Variable & Description \\
    \hline
    \hline
    $R$ &  Hilbert space dim. of a qudit
     \\
    $L_s$ & Hilbert space dim. of subsys.~$s$
     \\
    $N_s$ & Number of qudits ($\log_R L_s$)
     \\
    $q_s$ & Symmetry charge of subsys.~$s$
     \\
    $\nu_s$ & Filling factor of $s$ ($q_s/N_s$, only for $U(1)$)
     \\
    \hline
$\hat\rho_A^{(q_A)}$  & Projected reduced density matrix of $A$       \\
$P(\xi)$  & Spectral density of $(\hat\rho_A^{(q_A)})^{T_2}$       \\
$\braket{{\cal N}(\hat\rho_A^{(q_A)})}$  & Ensemble-averaged negativity of $\hat\rho_A^{(q_A)}$       \\
$\braket{{\cal E}(\hat\rho_A^{(q_A)})}$  & Ensemble-averaged logarithmic negativity of $\hat\rho_A^{(q_A)}$       \\
${\cal N}_i(\hat\rho_A^{(q_A)})$ & (off-)diagonal contribution to $\braket{{\cal N}(\hat\rho_A^{(q_A)})}$ with $i=2(1)$ \\
$\mathfrak{n}_1(q_1,q_2)$ &  Charge-off-diagonal contribution to ${\cal N}_1(\hat\rho_A^{(q_A)})$  \\
$\mathfrak{n}_2(q_1)$ &  Charge-diagonal contribution to ${\cal N}_2(\hat\rho_A^{(q_A)})$ \\
    \hline
$\alpha_{q_1q_2}$ & $\frac{L_{A_1,q_1} L_{A_2,q_2}}{L_{q_A} L_{q_B}}$
\\
    % \hline
$\beta_{q_1}$  & $\frac{ L_{A_1,q_1} }{ L_{q_A} L_{q_B} }$      \\
     \hline
\end{tabular}
}
\renewcommand{\arraystretch}{1}
    \caption{Summary of notations in the paper.}
    \label{tab:notation}
\end{table}

\section{Renyi entropy and Renyi negativity}
\label{sec:moments}

We first explain the $1/L$ diagrammatic perturbation theory by calculating the Renyi entropies and Renyi negativities in this section. 
We start by deriving Renyi entropies.
The reduced density matrix in the sector given by $q_A$, i.e.~$\hat \rho_{A}^{(q_A)}$ can be written as 
\begin{equation}\label{eq:rho_A_expansion}
\begin{aligned}
    \hat \rho_{A}^{(q_A)}  &= \sum_{i_{q_A}, j_{q_A}} \sum_{\alpha_{q_B}} 
    X_{i_{q_A}, \alpha_{q_B}}^{(q_A, q_B)} \; X_{j_{q_A}, \alpha_{q_B}}^{(q_A, q_B)\ast} \ \left| e_{i_{q_A}}^{(q_A)} \right\rangle \left\langle  e_{j_{q_A}}^{(q_A)} \right|,
\end{aligned}
\end{equation}
with $q_B = Q - q_A$.
We represent the matrix elements of the density matrix $\hat \rho_{A}^{(q_A)}$ in the following way:
\begin{equation}
    \left[ \hat \rho_{A}^{(q_A)}\right]_{i_{q_A}, j_{q_A}} =
    \,
    \tikz[scale=0.8,baseline=-0.5ex]{
    \draw[dashed] (0.,0.2)-- (0,0);
    \draw[dashed] (0,0) -- (2,0);
    \draw[dashed]  (2,0.2) -- (2,0);
    \draw (-0.3,0.2)-- (-0.3,-0.15);
    \draw (2.3,-0.15)-- (2.3,0.2);
    \draw (-0.5,0.5) circle (0.) node[anchor=center] {\footnotesize$i_{q_A}$};
    \draw (2.5,0.5) circle (0.) node[anchor=center] {\footnotesize$j_{q_A}$};
    \draw (0.2,0.5) circle (0.) node[anchor=center] {\footnotesize$ \alpha_{q_B} $};
    \draw (1.8,0.5) circle (0.) node[anchor=center] {\footnotesize$ \alpha_{q_B} $};
    }\ .
\end{equation}
We will be interested in calculating averaged quantities over random realizations in the remainder of this work; since $X$ is a Gaussian random variable, all its moments can be decomposed in terms of its second moment. Noting this, in our diagrammatics, we denote an averaging over different random realizations of products of two $X$'s with curves such as:
\begin{equation}
\left\langle X X  \right\rangle \to
\ 
    \tikz[baseline=0ex]{
    \draw[dashed] (1.0,0) arc (0:180:0.5);
    \draw (1.2,0.) arc (0:180:0.7);
    } \ ,
\end{equation}
the actual value of the above second moment when appropriate indices for $X$'s are included can be read from Eq.~\eqref{eq:X_variance_correlation}. Let us start with the first moment, 
\begin{equation}
     \left\langle\text{Tr}(\hat\rho_{A}^{(q_A)}) \right\rangle = \ 
    \tikz[baseline=0ex]{
    \draw[dashed] (1.0,0) arc (0:180:0.5);
    \draw[dashed] (0,0) -- (1,0);
    \draw (1.2,0.) arc (0:180:0.7);
    \draw (-0.2,0.0)-- (-0.2,-0.15)--(1.2,-0.15)-- (1.2,0.0);
    } \  = L_{q_A}\, L_{q_B} \ \frac1{L_{q_A} \, L_{q_B}} = 1.
\end{equation}
where it is understood that all possible indices should be summed over when there is a closed loop; the factor $L_{q_A}\ L_{q_B}$ comes from this rule. On the other hand, the factor $\frac1{L_{q_A} \, L_{q_B}}$ comes from the variance of the Gaussian variables.

We next calculate the ensemble-average of Renyi entropies, as follows,
\begin{align}
    \left\langle \Tr \left[ \left( \hat\rho_{A}^{(q_A)} \right)^2 \right] \right\rangle 
    &=
    \,
    \tikz[scale=0.7,baseline=0.5ex]{
    \draw[dashed] (0,0) -- (1,0);
    \draw (-0.2,0.)-- (-0.2,-0.25);
    \draw (1.2,-0.1)-- (1.2,0.);
    %%%%%
    \draw[dashed] (2,0) -- (3,0);
    \draw (1.8,0.)-- (1.8,-0.1);
    \draw (3.2,-0.25)-- (3.2,0.);
    %%%% contractions
    \draw (1.2,-0.1)-- (1.8,-0.1);
    \draw (-0.2,-0.25) -- (3.2,-0.25);
    %%%% average
    \draw[dashed] (1.0,0) arc (0:180:0.5);
    \draw[dashed] (3.0,0) arc (0:180:0.5);
    \draw (1.2,0) arc (0:180:0.7);
    \draw (3.2,0) arc (0:180:0.7);
    }
\ \,
+
\ \,
    \tikz[scale=0.6,baseline=0.5ex]{
    \draw[dashed] (0,0) -- (1,0);
    \draw (-0.2,0.)-- (-0.2,-0.25);
    \draw (1.2,-0.1)-- (1.2,0.);
    %%%%%
    \draw[dashed] (2,0) -- (3,0);
    \draw (1.8,0.)-- (1.8,-0.1);
    \draw (3.2,-0.25)-- (3.2,0.);
    %%%%
    \draw (1.2,-0.1)-- (1.8,-0.1);
    \draw (-0.2,-0.25) -- (3.2,-0.25);
    %%%%%%% average
    \draw[dashed] (2.0,0) arc (0:180:0.5);
    \draw[dashed] (3.0,0) arc (0:180:1.5);
    \draw (1.8,0) arc (0:180:0.3);
    \draw (3.2,0) arc (0:180:1.7);
    }
\ \,
\nonumber
\\
&=\frac{1}{L_{q_A}} + \frac{1}{L_{q_B}},
\end{align}

\begin{align}
\label{eq:tr_r3}
     \left\langle \Tr \left[ \left( \hat\rho_{A}^{(q_A)} \right)^3 \right] \right\rangle 
    =&
    \
    \tikz[scale=0.4,baseline=0.5ex]{
    \draw[dashed] (0,0) -- (1,0);
    \draw (-0.2,0.)-- (-0.2,-0.35);
    \draw (1.2,-0.15)-- (1.2,0.);
    %%%%%
    \draw[dashed] (2,0) -- (3,0);
    \draw (1.8,0.)-- (1.8,-0.15);
    \draw (3.2,-0.15)-- (3.2,0.);
    %%%%%
    \draw[dashed] (4,0) -- (5,0);
    \draw (3.8,0.)-- (3.8,-0.15);
    \draw (5.2,-0.35)-- (5.2,0.);
    %%%% contractions
    \draw (1.2,-0.15)--(1.8,-0.15);
    \draw (3.2,-0.15)--(3.8,-0.15);
    \draw (-0.2,-0.35)-- (5.2,-0.35);
    %%%% average
    \draw[dashed] (4.0,0) arc (0:180:0.5);
    \draw[dashed] (2.0,0) arc (0:180:0.5);
    \draw[dashed] (5.0,0) arc (0:180:2.5);
    \draw (3.8,0) arc (0:180:0.3);
    \draw (1.8,0) arc (0:180:0.3);
    \draw (5.2,0) arc (0:180:2.7);
    }  \,
    +
    3\times\,
\tikz[scale=0.4,baseline=0.5ex]{
    \draw[dashed] (0,0) -- (1,0);
    \draw (-0.2,0.)-- (-0.2,-0.35);
    \draw (1.2,-0.15)-- (1.2,0.);
    %%%%%
    \draw[dashed] (2,0) -- (3,0);
    \draw (1.8,0.)-- (1.8,-0.15);
    \draw (3.2,-0.15)-- (3.2,0.);
    %%%%%
    \draw[dashed] (4,0) -- (5,0);
    \draw (3.8,0.)-- (3.8,-0.15);
    \draw (5.2,-0.35)-- (5.2,0.);
    %%%% contractions
    \draw (1.2,-0.15)--(1.8,-0.15);
    \draw (3.2,-0.15)--(3.8,-0.15);
    \draw (-0.2,-0.35)-- (5.2,-0.35);
    %%%% average
    \draw[dashed] (2.0,0) arc (0:180:0.5);
    \draw[dashed] (3.0,0) arc (0:180:1.5);
    \draw[dashed] (5.0,0) arc (0:180:0.5);
    \draw (1.8,0) arc (0:180:0.3);
    \draw (3.2,0) arc (0:180:1.7);
    \draw (5.2,0) arc (0:180:0.7);
    }  
    \nonumber \\ \nonumber \\
   & +
    \,
    \tikz[scale=0.5,baseline=-0.5ex]{
    \draw[dashed] (0,0) -- (1,0);
    \draw (-0.2,0.)-- (-0.2,-0.35);
    \draw (1.2,-0.15)-- (1.2,0.);
    %%%%%
    \draw[dashed] (2,0) -- (3,0);
    \draw (1.8,0.)-- (1.8,-0.15);
    \draw (3.2,-0.15)-- (3.2,0.);
    %%%%%
    \draw[dashed] (4,0) -- (5,0);
    \draw (3.8,0.)-- (3.8,-0.15);
    \draw (5.2,-0.35)-- (5.2,0.);
    %%%% contractions
    \draw (1.2,-0.15)--(1.8,-0.15);
    \draw (3.2,-0.15)--(3.8,-0.15);
    \draw (-0.2,-0.35)-- (5.2,-0.35);
    %%%% average
    \draw[dashed] (1.0,0) arc (0:180:0.5);
    \draw[dashed] (3.0,0) arc (0:180:0.5);
    \draw[dashed] (5.0,0) arc (0:180:0.5);
    \draw (1.2,0) arc (0:180:0.7);
    \draw (3.2,0) arc (0:180:0.7);
    \draw (5.2,0) arc (0:180:0.7);
    } \,
     \nonumber \\ 
     %\nonumber \\
    =& 
    \frac{1}{ L_{q_B}^2} 
    + \frac{3}{L_{q_A} L_{q_B}}
    + \frac{1}{L_{q_A}^2}.
\end{align}

One can also obtain dominant contributions in certain limits using the diagrammatic approach, first we consider the limit $L_{q_A} \gg L_{q_B}$:
\begin{align}
    \left\langle \Tr \left[ \left( \hat\rho_{A}^{(q_A)} \right)^n \right] \right\rangle 
    \approx
    \tikz[scale=0.45,baseline=0.5ex]{
    \draw[dashed] (0,0) -- (1,0);
    \draw (-0.2,0.)-- (-0.2,-0.35);
    \draw (1.2,-0.15)-- (1.2,0.);
    %%%%%
    \draw[dashed] (2,0) -- (3,0);
    \draw (1.8,0.)-- (1.8,-0.15);
    \draw (3.2,-0.15)-- (3.2,0.);
    %%%%%
    \draw[dashed] (4,0) -- (4.35,0);
    \draw (3.8,0.)-- (3.8,-0.15);
    % \draw (5.2,-0.15)-- (5.2,0.);
    %%%%%
    \draw[dashed] (7,0) -- (8,0);
    \draw[dashed] (5.7,0) -- (6,0);
    \draw (6.8,0.)-- (6.8,-0.15);
    \draw (6.2,0.)-- (6.2,-0.15);
    \draw (8.2,-0.35)-- (8.2,0.);
    %%%% contractions
    \draw (1.2,-0.15)--(1.8,-0.15);
    \draw (3.2,-0.15)--(3.8,-0.15);
    \draw (6.2,-0.15)--(6.8,-0.15);
    \draw (-0.2,-0.35)-- (8.2,-0.35);
    %%%% average
    \draw[dashed] (4.0,0) arc (0:180:0.5);
    \draw[dashed] (2.0,0) arc (0:180:0.5);
    \draw[dashed] (7,0) arc (0:180:0.5);
    \draw[dashed] (8,0) arc (0:180:4);
    \node[] at (5,0.2) {$\cdots$};
    \draw (3.8,0) arc (0:180:0.3);
    \draw (1.8,0) arc (0:180:0.3);
    \draw (6.8,0) arc (0:180:0.3);
    \draw (8.2,0) arc (0:180:4.2);
    }  
    = L_{q_B}^{1-n},
    \label{eq:LAgg_Renyi}
\end{align}
and in the opposite limit of $L_{q_A} \ll L_{q_B}$:
while in the opposite regime  $L_A \ll L_B$, we obtain
\begin{align}
    \left\langle \Tr \left[ \left( \hat\rho_{A}^{(q_A)} \right)^n \right] \right\rangle 
    \approx
    \tikz[scale=0.5,baseline=-0.5ex]{
    \draw[dashed] (0,0) -- (1,0);
    \draw (-0.2,0.)-- (-0.2,-0.35);
    \draw (1.2,-0.15)-- (1.2,0.);
    %%%%%
    \draw[dashed] (2,0) -- (3,0);
    \draw (1.8,0.)-- (1.8,-0.15);
    \draw (3.2,-0.15)-- (3.2,0.);
    %%%%%
    \draw[dashed] (5,0) -- (6,0);
    \draw (4.8,0.)-- (4.8,-0.15);
    \draw (6.2,-0.35)-- (6.2,0.);
    %%%% contractions
    \draw (1.2,-0.15)--(1.8,-0.15);
    \draw (3.2,-0.15)--(3.45,-0.15);
    \draw (-0.2,-0.35)-- (6.2,-0.35);
    \draw (4.5,-0.15)--(4.8,-0.15);
    \node[] at (4.,0.2) {$\cdots$};
    %%%% average
    \draw[dashed] (1.0,0) arc (0:180:0.5);
    \draw[dashed] (3.0,0) arc (0:180:0.5);
    \draw[dashed] (6,0) arc (0:180:0.5);
    \draw (1.2,0) arc (0:180:0.7);
    \draw (3.2,0) arc (0:180:0.7);
    \draw (6.2,0) arc (0:180:0.7);
    } 
    = L_{q_A}^{1-n}.
 \label{eq:LAll_Renyi}
\end{align}

The same technique can be used to calculate Renyi negativities as well, which are defined as the moments of $\left[ \hat \rho_{A}^{(q_A)} \right]^{T_2}$; at this point we need to decompose the lines corresponding to the $A$ subsystem into $A_1$ and $A_2$ constituents; 
in other words, we write the index  $ i_{q_A}$ (and the like) in \eqref{eq:rho_A_expansion} as a collection of the two indices $\left( i_{1,q_1} \ i_{2, \bar{q}_1} \right)$ where $q_1$ and $\bar{q}_1$ are quantum numbers for $A_1$ and $A_2$ such that $q_1 + q_2 = q_A$. As a result $\hat \rho_{A}^{(q_A)}$
has the following form in the diagrammatic notation:
\begin{equation}
\begin{aligned}
     &\left[\hat \rho_{A}^{(q_A)}\right]_{ \left( i_{1,q_1} i_{2,\bar{q}_1} \right) , \left( j_{1,r_1} j_{2,\bar{r}_1} \right) } 
    \\
    & \qquad = \sum_{\alpha_{q_B}} 
     X_{i_{1,q_1} i_{2,\bar{q}_1} , \alpha_{q_B}}^{(q_A, q_B)}
     \;
      X_{j_{1,r_1} j_{2,\bar{r}_1} , \alpha_{q_B}}^{(q_A, q_B)\ast}
      \\
    &\qquad =
    %\frac{1}{{\cal Z}} \ 
    \,
    \tikz[scale=0.8,baseline=-0.5ex]{
    % \node [int, pin={[init]above:$v_0$}]
    % \filldraw[black] (0,0) circle (2pt) node[anchor=west] {Intersection point};
    \draw[dashed] (0.,0.2)-- (0,0);
    \draw[dashed] (0,0) -- (2,0);
    \draw[dashed]  (2,0.2) -- (2,0);
    \draw (-0.2,0.2)-- (-0.2,-0.15);
    \draw (2.2,-0.15)-- (2.2,0.2);
    \draw[densely dotted,thick] (-0.5,0.2)-- (-0.5,-0.15);
    \draw[densely dotted,thick] (2.5,-0.15)-- (2.5,0.2);
    \draw (-0.6,0.5) circle (0.) node[anchor=center] {\footnotesize$r_1$};
    \draw (-0.2,0.5) circle (0.) node[anchor=center] {\footnotesize$\bar{r}_1$};
    \draw (2.2,0.5) circle (0.) node[anchor=center] {\footnotesize$\bar{q}_1$};
    \draw (2.6,0.5) circle (0.) node[anchor=center] {\footnotesize$q_1$};
    \draw (0.3,0.5) circle (0.) node[anchor=center] {\footnotesize$ q_B $};
    }\ ,
\end{aligned}
\end{equation}
where this time instead of the actual indices for each line, only quantum numbers for each subsystem are denoted. Note that separate quantum numbers for $A_1$ and $A_2$ subsystems are introduced on the diagram; dotted and solid lines correspond to subsystems $A_1$ and $A_2$, respectively. Furthermore quantum numbers are shown on corresponding lines, and it is required by the symmetry charge relations that $q_1 +\bar{q}_1 = q_A$, $r_1+ \bar{r}_1 = q_A$, and $q_A + q_B = Q$.

The above notation makes it clear that for a given $\hat \rho_{A}^{(q_A)}$, although the values of $q_A$ and $q_B$ are fixed, $A_1$ and $A_2$ subsystems can have different quantum numbers and in fact as we will see below, one needs to sum over them when evaluating diagrams. As a first check with this notation we calculate the trace of $\hat\rho_{A}^{(q_A)}$ using the diagrammatic notation; it follows as:
\begin{equation}
\begin{aligned}
    \braket{\Tr\hat\rho_{A}^{(q_A)}} &= \sum_{q_1} \ \
    \,
    \tikz[baseline=0ex]{
    \draw[dashed] (1.0,0) arc (0:180:0.5);
    \draw[dashed] (0,0) -- (1,0);
    \draw (1.2,0.) arc (0:180:0.7);
    \draw (-0.2,0.0)-- (-0.2,-0.15)--(1.2,-0.15)-- (1.2,0.0);
    \draw[densely dotted,thick] (1.6,0) arc (0:180:1.1);
    \draw[densely dotted,thick] (-0.6,0.0)-- (-0.6,-0.3)--(1.6,-0.3)-- (1.6,0.0);
    \draw (1.8,0.4) circle (0.) node[anchor=center] {\footnotesize$q_1$};
    \draw (-0.3,0.17) circle (0.) node[anchor=center] {\footnotesize$\bar{q}_1$};
    }\ \\
    &= \sum_{q_1} \frac{1}{L_{q_A} L_{q_B}} L_{A_1,q_1} L_{A_2,\bar{q}_1} \ L_{q_B}  =  1,
\end{aligned}
\end{equation}
with the quantum numbers of each subsystem shown explicitly on the lines. Since the qunatum number of the $B$ subsystem only takes one value we do not include it in the diagrams. For every closed loop, one should multiply by the size of the sector given by the quantum number in the given subsystem. Moreover, each time ensemble averaging is performed, as before, a factor of $\frac{1}{L_{q_A} L_{q_B}}$ should be multiplied according to Eq.~\eqref{eq:X_variance_correlation}. Note that all the contractions in the above diagram dictates that the quantum number of the $A_2$ subsystem should be equal to $\bar q_1 = q_A - q_1$. On the second row, we have used $\sum_{q_1} L_{A_1,q_1} L_{A_2,\bar{q}_1} = L_{q_A}$. 

In order to take partial transpose with respect to the $A_2$ subsystem in the diagrammatic approach,
we need one further step that is swapping the $A_1$ and $A_2$ legs for each density matrix insertion which will be depicted as
\begin{equation}
    \left[ \hat\rho_{A}^{(q_A)} \right]^{T_2} \to 
    \tikz[scale=0.8,baseline=-0.5ex]{
    \draw[dashed] (0.,0.2)-- (0,0);
    \draw[dashed] (0,0) -- (2,0);
    \draw[dashed]  (2,0.2) -- (2,0);
    \draw (-0.2,0.2)-- (-0.2,-0.15) -- (2.2,-0.5);
    \draw (2.2,0.2) -- (2.2,-0.15) -- (-0.2 , -0.5);
    \draw[densely dotted,thick] (-0.5,0.2)-- (-0.5,-0.15);
    \draw[densely dotted,thick] (2.5,-0.15)-- (2.5,0.2);
    \draw (-0.6,0.5) circle (0.) node[anchor=center] {\footnotesize$r_1$};
    \draw (-0.2,0.5) circle (0.) node[anchor=center] {\footnotesize$\bar{r}_1$};
    \draw (2.2,0.5) circle (0.) node[anchor=center] {\footnotesize$\bar{q}_1$};
    \draw (2.6,0.5) circle (0.) node[anchor=center] {\footnotesize$q_1$};
    \draw (0.3,0.5) circle (0.) node[anchor=center] {\footnotesize$ q_B $};
    }\ .
\end{equation}
Note that similar to above, it is still required that $q_1+\bar{q}_1 = q_A$ and $r_1+\bar{r}_1 = q_A$; this constraint should be imposed in every diagram containing leg crossings such as the above. One can now calculate ensemble averaged moments of the partially transposed density matrix to obtain Renyi negativities of different orders. The  first nontrivial one is the third moment, which without ensemble averaging takes the form:
\begin{equation}
\Tr  \left( \left[ \hat\rho_{A}^{(q_A)} \right]^{T_2} \right)^3
=
    \tikz[scale=0.8,baseline=-1.5ex]{
    \draw[dashed] (0,0.2)-- (0,0);
    \draw[dashed] (0,0) -- (1,0);
    \draw[dashed]  (1,0.2) -- (1,0);
    \draw (-0.2,0.2)-- (-0.2,-0.15);
    \draw (-0.2,-0.15)-- (1.,-0.7);
    \draw (1.2,-0.15)-- (1.2,0.2);
    \draw (1.2,-0.15)-- (0,-0.8);
    \draw[densely dotted,thick] (-0.3,0.2)-- (-0.3,-0.95);
    \draw[densely dotted,thick] (1.3,-0.1)-- (1.3,0.2);
    %%%%%
    \draw[dashed] (2,0.2)-- (2,0);
    \draw[dashed] (2,0) -- (3,0);
    \draw[dashed]  (3,0.2) -- (3,0);
    \draw (1.8,0.2)-- (1.8,-0.15);
    \draw (1.8,-0.15)-- (3.,-0.7);
    \draw (3.2,-0.15)-- (3.2,0.2);
    \draw (3.2,-0.15)-- (2,-0.7);
    \draw[densely dotted,thick] (1.7,0.2)-- (1.7,-0.1);
    \draw[densely dotted,thick] (3.3,-0.1)-- (3.3,0.2);
    %%%%%
    \draw[dashed] (4,0.2)-- (4,0);
    \draw[dashed] (4,0) -- (5,0);
    \draw[dashed]  (5,0.2) -- (5,0);
    \draw (3.8,0.2)-- (3.8,-0.15);
    \draw (3.8,-0.15)-- (5.,-0.8);
    \draw (5.2,-0.15)-- (5.2,0.2);
    \draw (5.2,-0.15)-- (4,-0.7);
    \draw[densely dotted,thick] (3.7,0.2)-- (3.7,-0.1);
    \draw[densely dotted,thick] (5.3,-0.95)-- (5.3,0.2);
    %%%% contractions
    \draw[densely dotted,thick] (1.3,-0.1)--(1.7,-0.1);
    \draw[densely dotted,thick](3.3,-0.1)--(3.7,-0.1);
    \draw[densely dotted,thick](-0.3,-0.95)--(5.3,-0.95);
    \draw (3,-0.7)--(4.,-0.7);
    \draw (0,-0.8)--(5.,-0.8);
    \draw (1.0,-0.7)-- (2,-0.7);
    }
\end{equation}
For illustration purposes, let us only focus on one of the terms appearing in the ensemble average of the above quantity, i.e.~the term given by
\begin{equation}
    \tikz[scale=0.55,baseline=0.5ex]{
    \draw[dashed] (0,0) -- (1,0);
    \draw (-0.2,0.)-- (-0.2,-0.15);
    \draw (-0.2,-0.15)-- (1.,-0.7);
    \draw (1.2,-0.15)-- (1.2,0.);
    \draw (1.2,-0.15)-- (0,-0.8);
    \draw[densely dotted,thick] (-0.3,0)-- (-0.3,-1.45);
    \draw[densely dotted,thick] (1.3,-0.1)-- (1.3,0.);
    %%%%%
    \draw[dashed] (2,0) -- (3,0);
    \draw (1.8,0.)-- (1.8,-0.15);
    \draw (1.8,-0.15)-- (3.,-0.7);
    \draw (3.2,-0.15)-- (3.2,0.);
    \draw (3.2,-0.15)-- (2,-0.7);
    \draw[densely dotted,thick] (1.7,0.)-- (1.7,-0.1);
    \draw[densely dotted,thick] (3.3,-0.1)-- (3.3,0.);
    %%%%%
    \draw[dashed] (4,0) -- (5,0);
    \draw (3.8,0.)-- (3.8,-0.15);
    \draw (3.8,-0.15)-- (5.,-0.8);
    \draw (5.2,-0.15)-- (5.2,0.);
    \draw (5.2,-0.15)-- (4,-0.7);
    \draw[densely dotted,thick] (3.7,0.)-- (3.7,-0.1);
    \draw[densely dotted,thick] (5.3,-1.45)-- (5.3,0.);
    %%%% contractions
    \draw[densely dotted,thick] (1.3,-0.1)--(1.7,-0.1);
    \draw[densely dotted,thick](3.3,-0.1)--(3.7,-0.1);
    \draw[densely dotted,thick](-0.3,-1.5)--(5.3,-1.5);
    \draw (3,-0.7)--(4.,-0.7);
    \draw (0,-0.8)--(5.,-0.8);
    \draw (1.0,-0.7)-- (2,-0.7);
    %%%% average
    \draw[densely dotted,thick] (5.3,0) arc (0:180:2.8);
    \draw[densely dotted,thick] (1.7,0) arc (0:180:0.2);
    \draw[densely dotted,thick] (3.7,0) arc (0:180:0.2);
    \draw[dashed] (4.0,0) arc (0:180:0.5);
    \draw[dashed] (2.0,0) arc (0:180:0.5);
    \draw[dashed] (5.0,0) arc (0:180:2.5);
    \draw (3.8,0) arc (0:180:0.3);
    \draw (1.8,0) arc (0:180:0.3);
    \draw (5.2,0) arc (0:180:2.7);
    \draw (-0.7,-0.2) circle (0.) node[anchor=center] {\scriptsize$q_1$};
    \draw (1.55,-0.35) circle (0.) node[anchor=center] {\scriptsize$q_1$};
    \draw (3.55,-0.35) circle (0.) node[anchor=center] {\scriptsize$q_1$};
    \draw (1.,-1.1) circle (0.) node[anchor=center] {\scriptsize$\bar{q}_1$};
    } \ .
\end{equation}
Quantum numbers of subsystems $A_1$ and $A_2$ are denoted on each of the loops corresponding to these two subsystems. Quantum number relations, as discussed above, have dictated all quantum numbers on the $A_1$ subsystem loops to be equal, while the quantum number on the single $A_2$ loop is its complementary $\bar{q}_1 = q-q_1$. To compute the total contribution of this diagram, we note that every closed loop with a given quantum number brings about one factor of the Hilbert space size corresponding to that quantum number, i.e.~the total contribution is given by $\sum_{q_1} L_{A_1,q_1}^3 L_{A_2,\bar{q}_1} L_{B,q_B} \frac{1}{\left( L_{A,q_A} L_{B,q_B} \right)^3}$; the summations takes the effect of all possible quantum numbers into account here.

A similar diagram for the fourth moment can also be drawn, which is only one of the terms among many: 
\begin{equation}\label{eq:fourth_moment_dominant_large_A1}
    \tikz[scale=0.55,baseline=1.5ex]{
    \draw[dashed] (0,0) -- (1,0);
    \draw (-0.2,0.)-- (-0.2,-0.15);
    \draw (-0.2,-0.15)-- (1.,-0.7);
    \draw (1.2,-0.15)-- (1.2,0.0);
    \draw (1.2,-0.15)-- (0,-1.1);
    \draw[densely dotted,thick] (-0.3,0.)-- (-0.3,-1.5);
    \draw[densely dotted,thick] (1.3,-0.1)-- (1.3,0.);
    %%%%%
    \draw[dashed] (2,0) -- (3,0);
    \draw (1.8,0.)-- (1.8,-0.15);
    \draw (1.8,-0.15)-- (3.,-0.7);
    \draw (3.2,-0.15)-- (3.2,0.);
    \draw (3.2,-0.15)-- (2,-0.7);
    \draw[densely dotted,thick] (1.7,0.)-- (1.7,-0.1);
    \draw[densely dotted,thick] (3.3,-0.1)-- (3.3,0.);
    %%%
    \draw[dashed] (4,0) -- (5,0);
    \draw (3.8,0.)-- (3.8,-0.15);
    \draw (3.8,-0.15)-- (5.,-0.7);
    \draw (5.2,-0.15)-- (5.2,0.);
    \draw (5.2,-0.15)-- (4,-0.7);
    \draw[densely dotted,thick] (3.7,0.)-- (3.7,-0.1);
    \draw[densely dotted,thick] (5.3,-0.1)-- (5.3,0.);
    %%%
    \draw[dashed] (6,0) -- (7,0);
    \draw (5.8,0.)-- (5.8,-0.15);
    \draw (5.8,-0.15)-- (7.,-1.1);
    \draw (7.2,-0.15)-- (7.2,0.);
    \draw (7.2,-0.15)-- (6,-0.7);
    \draw[densely dotted,thick] (5.7,0.)-- (5.7,-0.1);
    \draw[densely dotted,thick] (7.3,-1.5)-- (7.3,0.);
    %%%% contractions
    \draw[densely dotted,thick] (1.3,-0.1)--(1.7,-0.1);
    \draw[densely dotted,thick] (3.3,-0.1)--(3.7,-0.1);
    \draw[densely dotted,thick] (5.3,-0.1)--(5.7,-0.1);
    \draw[densely dotted,thick] (-0.3,-1.5)--(7.3,-1.5);
    \draw (1.0,-0.7)-- (2,-0.7);
    \draw (3.0,-0.7)-- (4,-0.7);
    \draw (5.0,-0.7)-- (6,-0.7);
    \draw (0.0,-1.1)-- (7,-1.1);
    %%%% average
    \draw[densely dotted,thick] (7.3,0) arc (0:180:3.8);
    \draw[densely dotted,thick] (1.7,0) arc (0:180:0.2);
    \draw[densely dotted,thick] (3.7,0) arc (0:180:0.2);
    \draw[densely dotted,thick] (5.7,0) arc (0:180:0.2);
    \draw[dashed] (6.0,0) arc (0:180:0.5);
    \draw[dashed] (4.0,0) arc (0:180:0.5);
    \draw[dashed] (2.0,0) arc (0:180:0.5);
    \draw[dashed] (7.0,0) arc (0:180:3.5);
    \draw (5.8,0) arc (0:180:0.3);
    \draw (3.8,0) arc (0:180:0.3);
    \draw (1.8,0) arc (0:180:0.3);
    \draw (7.2,0) arc (0:180:3.7);
    \draw (-0.7,-0.2) circle (0.) node[anchor=center] {\scriptsize$q_1$};
    \draw (1.55,-0.4) circle (0.) node[anchor=center] {\scriptsize$\bar{q}_2$};
    \draw (3.55,-0.35) circle (0.) node[anchor=center] {\scriptsize$q_1$};
    \draw (5.55,-0.4) circle (0.) node[anchor=center] {\scriptsize$\bar{q}_2$};
    \draw (.1,-.5) circle (0.) node[anchor=center] {\scriptsize$\bar{q}_1$};
    \draw (1.,-1.25) circle (0.) node[anchor=center] {\scriptsize$q_2$};
 }\, ,
\end{equation}
the specific structure of the diagram dictates two sets of independent quantum numbers, i.e.~$q_1$ and $q_2$ corresponding to $A_1$ and $A_2$ loops. Also, $\bar{q}_1 = q_A-q_1$ and $\bar{q}_2 = q_A-q_2$ label the loops in $A_2$ and $A_1$ subsystems respectively. As can be seen in the above two cases, each diagram can be labeled by its independent loop quantum numbers; for some diagrams there is only one independent quantum number and for some there are two. As a result, we can label each term contributing to the Renyi negativities by its independent quantum numbers; in the following we label the diagrams by the quantum numbers on the lines used for evaluating the overall trace, or in other words the line or lines at the bottom of each diagram. The contribution from this diagram take the form 
$ \frac{(L_{A_1,q_1} L_{A_1,\bar{q}_2})^2}{ (L_{A,q_A}L_{B,q_B})^4 } L_{B,q_B} L_{A_2,q_2} L_{A_2,\bar{q}_1} . $ 
Note that there are equal numbers of $A_1$ subsystem loops with indices $q_1$ and $\bar{q}_2$.

We now consider different regimes and discuss the dominant diagrams that appear in the expansion of each of the moments (see Table~\ref{tab:dictionary} for naming conventions in this paper). We can then use the moments in each regime to calculate entanglement negativity using the replica limit:
\begin{equation}
    \braket{{\cal E} (\hat\rho)}= \lim_{k\to \frac{1}{2}} \log \braket{\Tr(\hat\rho^{T_2})^{2k}}.
    \label{eq:replicalimit}
\end{equation}
As a result of this, we only consider order $n=2k$ moments in the following.

We start by considering the case where the $B$ subsystem is larger than $A$; we have seen above that factors of Hilbert space dimensions of different quantum number sectors appear when loops in the diagrams appear; as discussed earlier, one can think about diagrams with certain quantum numbers separately and depending on the ratio between Hilbert space dimensions for the specified quantum numbers decide to which regime they correspond. In this language, large $B$ subsystem is understood as having $L_{B,q_B} \gg L_{A_1,q_1} L_{A_2,q_2}$ for given $q_1$ and $q_2$ values. The dominant contribution comes from a diagram of the form:
\begin{equation}\label{eq:renyi_negativity_large_LB}
\begin{aligned}
     &
    \tikz[scale=0.5,baseline=-1ex]{
    \draw[dashed] (0,0) -- (1,0);
    \draw (-0.2,0.)-- (-0.2,-0.15);
    \draw (-0.2,-0.15)-- (1.,-0.7);
    \draw (1.2,-0.15)-- (1.2,0.);
    \draw (1.2,-0.15)-- (0,-0.8);
    \draw[densely dotted,thick] (-0.3,0)-- (-0.3,-1.45);
    \draw[densely dotted,thick] (1.3,-0.1)-- (1.3,0.);
    %%%%%
    \draw[dashed] (2,0) -- (3,0);
    \draw (1.8,0.)-- (1.8,-0.15);
    \draw (1.8,-0.15)-- (3.,-0.7);
    \draw (3.2,-0.15)-- (3.2,0.);
    \draw (3.2,-0.15)-- (2,-0.7);
    \draw[densely dotted,thick] (1.7,0.)-- (1.7,-0.1);
    \draw[densely dotted,thick] (3.3,-0.1)-- (3.3,0.);
    %%%%%
    \draw[dashed] (6,0) -- (7,0);
    \draw (5.8,0.)-- (5.8,-0.15);
    \draw (5.8,-0.15)-- (7.,-0.8);
    \draw (7.2,-0.15)-- (7.2,0.);
    \draw (7.2,-0.15)-- (6,-0.7);
    \draw (5.4,-0.7)-- (6,-0.7);
    \draw[densely dotted,thick] (5.7,0.)-- (5.7,-0.1);
    \draw[densely dotted,thick] (5.4,-0.1)-- (5.7,-0.1);
    \draw[densely dotted,thick] (7.3,-1.45)-- (7.3,0.);
    %%%% contractions
    \draw[densely dotted,thick] (1.3,-0.1)--(1.7,-0.1);
    \draw[densely dotted,thick](3.3,-0.1)--(3.7,-0.1);
    \draw[densely dotted,thick](-0.3,-1.45)--(7.3,-1.45);
    \draw (3,-0.7)--(3.7,-0.7);
    \draw (0,-0.8)--(7.,-0.8);
    \draw (1.0,-0.7)-- (2,-0.7);
    %%%% average
    \draw[densely dotted,thick] (1.3,0) arc (0:180:0.8);
    \draw[densely dotted,thick] (3.3,0) arc (0:180:0.8);
    \draw[densely dotted,thick] (7.3,0) arc (0:180:0.8);
    \draw[dashed] (1.0,0) arc (0:180:0.5);
    \draw[dashed] (3.0,0) arc (0:180:0.5);
    \draw[dashed] (7.0,0) arc (0:180:0.5);
    \draw (1.2,0) arc (0:180:0.7);
    \draw (3.2,0) arc (0:180:0.7);
    \draw (7.2,0) arc (0:180:0.7);
    \node[] at (4.6,0.3) {$\cdots$};
    \draw (1.55,-1.75) circle (0.) node[anchor=center] {\scriptsize$q_1$};
    \draw (1.,-1.1) circle (0.) node[anchor=center] {\scriptsize$\bar{q}_1$};
    } \\
    & \quad = %\sum_{q_1}
    \frac{L_{A_1,q_1} L_{A_2,\bar{q}_1}}{L_{A,q_A}^n}.  %\\
    % &=L_{A,q_A}^{1-n} \, ,
\end{aligned}
\end{equation}
If for all values of $q_1$, the relation $L_{B,q_B} \gg L_{A_1,q_1} L_{A_2,q_2}$ holds, one can sum the above result over all possible $q_1$ values to get $\sum_{q_1}
    \frac{L_{A_1,q_1} L_{A_2,\bar{q}_1}}{L_{A,q_A}^n} =L_{A,q_A}^{1-n}$.

The opposite regime of $L_{B,q_B} \ll L_{A_1,q_1} L_{A_2,q_2}$ consists of two different subregimes on its own; we first consider the subregime of $L_{A_1,q_1} \gg L_{A_2,q_2} L_{B,q_B}$: instead of the general case, we focus on the fourth moment from which the general result for even moments of the partially transposed density matrix can be deduced. One can check that the dominant diagram in this limit is given by the diagram in Eq.~\eqref{eq:fourth_moment_dominant_large_A1}. In general, for an even moment the contribution from a similar diagram will be given by:
\begin{equation}
    \frac{(L_{A_1,q_1} L_{A_1,\bar{q}_2})^{n/2} (L_{A_2,q_2} L_{A_2,\bar{q}_1} )}{ L_{A,q_A}^n L_{B,q_B}^{n-1}}
\end{equation}

Finally, there is another regime of interest where the dominant diagrams can be considered and it is given by requiring  $L_{B,q_B} \ll L_{A_1,q_1} L_{A_2,q_2}$ and $L_{A_s,q_s} \ll L_{A_{\bar{s}},q_{\bar{s}}} L_{B,q_B}$, simultaneously. This regime essentially means that none of the subsystems, within the specific symmetry charge sectors, is larger than the half of the whole sector. One of candidate diagrams that contributes most dominantly in this regime has the following form:
\begin{equation}
    \tikz[scale=0.45,baseline=1.5ex]{
    \draw[dashed] (0,0) -- (1,0);
    \draw (-0.2,0.)-- (-0.2,-0.15);
    \draw (-0.2,-0.15)-- (1.,-0.7);
    \draw (1.2,-0.15)-- (1.2,0.0);
    \draw (1.2,-0.15)-- (0,-1.45);
    \draw[densely dotted,thick] (-0.3,0.)-- (-0.3,-2);
    \draw[densely dotted,thick] (1.3,-0.1)-- (1.3,0.);
    %%%%%
    \draw[dashed] (2,0) -- (3,0);
    \draw (1.8,0.)-- (1.8,-0.15);
    \draw (1.8,-0.15)-- (3.,-0.7);
    \draw (3.2,-0.15)-- (3.2,0.);
    \draw (3.2,-0.15)-- (2,-0.7);
    \draw[densely dotted,thick] (1.7,0.)-- (1.7,-0.1);
    \draw[densely dotted,thick] (3.3,-0.1)-- (3.3,0.);
    %%%
    \draw[dashed] (4,0) -- (5,0);
    \draw (3.8,0.)-- (3.8,-0.15);
    \draw (3.8,-0.15)-- (5.,-0.7);
    \draw (5.2,-0.15)-- (5.2,0.);
    \draw (5.2,-0.15)-- (4,-0.7);
    \draw[densely dotted,thick] (3.7,0.)-- (3.7,-0.1);
    \draw[densely dotted,thick] (5.3,-0.1)-- (5.3,0.);
    %%%
    \draw[dashed] (6,0) -- (7,0);
    \draw (5.8,0.)-- (5.8,-0.15);
    \draw (5.8,-0.15)-- (7.,-0.8);
    \draw (7.3,-0.8)-- (7.,-0.8);
    \draw (7.2,-0.15)-- (7.2,0.);
    \draw (7.2,-0.15)-- (6,-0.7);
    \draw[densely dotted,thick] (5.7,0.)-- (5.7,-0.1);
    \draw[densely dotted,thick] (7.3,0)-- (7.3,-0.1);
    \draw[densely dotted,thick] (7.3,-0.1)-- (7.7,-0.1);
     %%%% contractions
    \draw[densely dotted,thick] (1.3,-0.1)--(1.7,-0.1);
    \draw[densely dotted,thick] (3.3,-0.1)--(3.7,-0.1);
    \draw[densely dotted,thick] (5.3,-0.1)--(5.7,-0.1);
    \draw (1.0,-0.7)-- (2,-0.7);
    \draw (3.0,-0.7)-- (4,-0.7);
    \draw (5.0,-0.7)-- (6,-0.7);
    %%%% average
    \draw[densely dotted,thick] (3.3,0) arc (0:180:1.8);
    \draw[densely dotted,thick] (1.7,0) arc (0:180:0.2);
    \draw[dashed] (2.0,0) arc (0:180:0.5);
    \draw[dashed] (3.0,0) arc (0:180:1.5);
    \draw (1.8,0) arc (0:180:0.3);
    \draw (3.2,0) arc (0:180:1.7);
    \draw[densely dotted,thick] (7.3,0) arc (0:180:1.8);
    \draw[densely dotted,thick] (5.7,0) arc (0:180:0.2);
    \draw[dashed] (6.0,0) arc (0:180:0.5);
    \draw[dashed] (7.0,0) arc (0:180:1.5);
    \draw (5.8,0) arc (0:180:0.3);
    \draw (7.2,0) arc (0:180:1.7);
    %%%%%%%%%% third block %%%%%%%
    \draw[dashed] (10,0) -- (11,0);
    \draw (9.8,0.)-- (9.8,-0.15);
    \draw (9.8,-0.15)-- (11,-0.7);
    \draw (11.2,-0.15)-- (11.2,0.0);
    \draw (11.2,-0.15)-- (10,-0.8);
    \draw (9.7,-0.8)-- (10,-0.8);
    \draw[densely dotted,thick] (9.7,0.)-- (9.7,-0.1);
    \draw[densely dotted,thick] (9.3,-0.1)-- (9.7,-0.1);
    \draw[densely dotted,thick] (11.3,-0.1)-- (11.3,0.);
    %%%%%
    \draw[dashed] (12,0) -- (13,0);
    \draw (11.8,0.)-- (11.8,-0.15);
    \draw (11.8,-0.15)-- (13.,-1.45);
    \draw (13.2,-0.15)-- (13.2,0.);
    \draw (13.2,-0.15)-- (12,-0.7);
    \draw[densely dotted,thick] (11.7,0.)-- (11.7,-0.1);
    \draw[densely dotted,thick] (13.3,-2.)-- (13.3,0.);
     %%%% contractions
    \draw[densely dotted,thick] (11.3,-0.1)--(11.7,-0.1);
    \draw (11.0,-0.7)-- (12,-0.7);
    %%%% average
    \draw[densely dotted,thick] (13.3,0) arc (0:180:1.8);
    \draw[densely dotted,thick] (11.7,0) arc (0:180:0.2);
    \draw[dashed] (12.0,0) arc (0:180:0.5);
    \draw[dashed] (13.0,0) arc (0:180:1.5);
    \draw (11.8,0) arc (0:180:0.3);
    \draw (13.2,0) arc (0:180:1.7);
    %%%%%%%%%%%
    \node[] at (8.6,0.5) {$\cdots$};
    %%% trace
    \draw[densely dotted,thick] (-0.3,-2)-- (13.3,-2);
    \draw (0,-1.45) -- (13,-1.45);
    \draw (-0.7,-0.4) circle (0.) node[anchor=center] {\scriptsize$q_1$};
    \draw (1.55,-0.4) circle (0.) node[anchor=center] {\scriptsize$\bar{q}_2$};
    \draw (5.55,-0.4) circle (0.) node[anchor=center] {\scriptsize$\bar{q}_2$};
    \draw (11.55,-0.4) circle (0.) node[anchor=center] {\scriptsize$\bar{q}_2$};
    \draw (2.,-1.) circle (0.) node[anchor=center] {\scriptsize$\bar{q}_1$};
    \draw (6.,-1.) circle (0.) node[anchor=center] {\scriptsize$\bar{q}_1$};
    \draw (12.,-1.) circle (0.) node[anchor=center] {\scriptsize$\bar{q}_1$};
    \draw (1.,-1.65) circle (0.) node[anchor=center] {\scriptsize$q_2$};
 }\, 
 ,
     \label{eq:LAgg}
\end{equation}
resulting in the following contribution
\begin{equation}\label{eq:a_contribution_ren_neg_semicircle}
    \frac{L_{A_1,q_1} L_{A_2,q_2} \ \left( L_{A_1,\bar{q}_2} L_{A_2,\bar{q}_1} \right)^{n/2} }{ L_{A,q_A}^n L_{B,q_B}^{n/2} }
\end{equation}
to the corresponding moment.
However, different diagrams with different structures can also be considered whose contributions are not subdominant compared to the one shown above. This fact along with the different combinations of Hilbert space dimensions corresponding to different charge sectors makes calculating the dominant contribution to Renyi negativities a not so straightforward matter in this regime. 

The change in dominant diagram as a function of replica index makes the analytic continuation to the replica limit (\ref{eq:replicalimit}) ambiguous. A standard way to avoid this ambiguity is to study the {\it spectrum} of the partially transposed  density matrix in a more systematic way and derive the resolvent function; this will be done in the next section.
We consider two concrete cases of symmetry groups, that is $\mathbb{Z}_R$ and $U(1)$ and apply our general results to these two cases.

\begin{table}[]
    \centering
{\footnotesize 
\renewcommand{\arraystretch}{1.2}
\begin{tabular}{cc}
    \hline
    Entanglement phase & Description \\
    \hline
    \hline
Positive partial transpose & $B$ is larger than $A$
\\
Replica symmetry breaking & No party is larger than half total system.
\\
Maximal entanglement  & $A_1$ or $A_2$ is larger than half total system.     \\
     \hline
\end{tabular}
}
\renewcommand{\arraystretch}{1}
    \caption{Dictionary of various entanglement phases.}
    \label{tab:dictionary}
\end{table}

\section{Entanglement negativity spectrum}
\label{sec:spectraldensity}

We now turn to calculating the negativity spectrum of different charge sectors of the density matrix, i.e.~components shown as $\hat \rho_{A}^{(q_A)}$. To this end, we make use of a Green function (or resolvent function) $G(z)$ defined as:
\begin{equation}
\label{eq:resolvent}
\begin{aligned}
  &  G(z) = \left\langle\frac{1}{ z - H } \right\rangle \\
    &= \left\langle \frac{1}{z} + \frac{1}{z} H \frac{1}{z}  + \frac{1}{z} H  \frac{1}{z} H  \frac{1}{z} + \cdots \right\rangle,
\end{aligned}
\end{equation}
where $H$ is taken to be $(\hat\rho_{A}^{(q_A)})^{T_2}$ the partial transpose of $\hat \rho_{A}^{(q_A)}$ with respect to $A_2$ for our purposes. Note that $G(z)$ is a matrix with the same set of indices as $\hat \rho_{A}^{(q_A)}$. The spectral density of $(\hat\rho_{A}^{(q_A)})^{T_2}$ is then computed as:
\begin{equation}
    \label{eq:principal-value}
    P(\xi) = - \frac{1}{\pi} \text{Im} \lim_{\epsilon \to 0} \Tr\left( G(z) \right) \big|_{z=\xi+i\epsilon},
\end{equation}
where the identity $\lim_{\epsilon \to 0} \frac{1}{\lambda+i\epsilon} = \text{PV} (\frac1\lambda) -i\pi \delta(\lambda)$ has been used. 

In order to calcualte $G(z)$ in different situations, we use a diagrammatic approach; the set of Feynman rules read:
\begin{itemize}
    \item Every closed loop for each subsystem brings in one factor of the size of that subsystem with the specified quantum number.
    \item On the diagrams, we will not specify the quantum number for the loops corresponding to the $B$ subsystem, as it is always equal to $q_B$ when we are calculating properties of $\hat \rho_{A}^{(q_A)}$, where $q_A + q_B = Q$.
    \item When a dotted line (corresponding to $A_1$) and a solid line (corresponding to $A_2$) originate from the same point in a diagram, a situation such as
    \begin{equation*}
        \tikz[scale=0.8,baseline=-0.5ex]{
        \draw[dashed] (0,0.5) -- (0,0) -- (-0.5,0);
        \draw (0.3,0.5)-- (0.3,-0.3) -- (-0.5,-0.3);
        \draw[densely dotted,thick] (0.6,0.5)-- (0.6,-0.6);
        \draw (0.7,0.7) circle (0.) node[anchor=center] {\footnotesize$q_1$};
        \draw (0.3,0.7) circle (0.) node[anchor=center] {\footnotesize$q_2$};
        \draw (-0.25,0.4) circle (0.) node[anchor=center] {\footnotesize$q_B$};
        } \quad ,
    \end{equation*}
    their quantum numbers are not independent, i.e.~$q_2=\bar{q}_1$, where $q_1 + \bar{q}_1 = q_A$.
\end{itemize}

As mentioned earlier the resolvent function $G(z)$ has the same indices as those of $\hat \rho_{A}^{(q_A)}$, i.e.~one needs to specify indices $q_i$ to address a component of $G(z)$. Interestingly, the structure of $G(z)$ is not very complicated. By examining diagrams in the expansion (\ref{eq:resolvent}), it is easy to see that $G$ has a block diagonal form as 
in:
\begin{align}
    G = G_1 \oplus G_2,
\end{align}
where
\begin{equation}
\label{eq:structure_of_G}
\begin{aligned}
    G_1 &=  \bigoplus_{q_1,q_2\neq \bar{q}_1}  G_{1,q_1q_2}  \mathbb{1}_{A_1,q_1} \otimes \mathbb{1}_{A_2,q_2}, \\
    G_2 &= \bigoplus_{q_1}  G_{2,q_1} \mathbb{1}_{A_1,q_1} \otimes \mathbb{1}_{A_2,\bar{q}_1}.
\end{aligned}
\end{equation}
In fact, every block of $G$ is proportional to the identity as reflected above (as a result of ensemble averaging). Using the principal value relation, we also define two components for the spectral density
\begin{align}
    \label{eq:spec-decomposition}        
    P(\xi) &= \sum_{q_1,q_2 \neq \bar{q}_1} P_{1,q_1 q_2}(\xi) + \sum_{q_1} P_{2,q_1}(\xi).
\end{align}
Accordingly, the total negativity of $\hat\rho_A^{(q_A)}$ can be written as a sum,
\begin{align}
    \label{eq:neg-qA-contributions}
    \braket{{\cal N}(\hat\rho_A^{(q_A)})} =
    {\cal N}_1(\hat\rho_A^{(q_A)})+
        {\cal N}_2(\hat\rho_A^{(q_A)}), 
\end{align}
where 
\begin{align}
    {\cal N}_1(\hat\rho_A^{(q_A)}) &=
    \sum_{q_1,q_2\neq \bar q_1} \mathfrak{n}_{1}(q_1,q_2), \\
    {\cal N}_2(\hat\rho_A^{(q_A)}) &=
        \sum_{q_1} \mathfrak{n}_{2}(q_1).
\end{align}
Here, $\mathfrak{n}_{1}(q_1,q_2)$ and $\mathfrak{n}_{2}(q_1)$ are simply introduced to denote the contributions of $ P_{1,q_1 q_2}(\xi)$ and $P_{2,q_1}(\xi)$ to the total negativity through Eq.~(\ref{eq:neg_dist}).

\subsection{Replica symmetry breaking regime }
We first consider the regime in which the conditions $\frac{ L_{A_1,q_1} }{ L_{A_{2,q_2 } } L_{q_B}  } \ll 1 $ and $\frac{ L_{A_2,q_2} }{ L_{A_{1,q_{1} } } L_{q_B}  } \ll 1 $ (for all $q_1$ and $q_2$) hold. We call this regime the semicircle (or replica symmetry breaking) regime. It roughly corresponds to requiring $N_{A_1},N_{A_2}<\frac{N}{2}$ as will be discussed further in the following.

In this regime, the dominant terms contributing to the resolvent function consist of the following diagrams:
\begin{widetext}
\begin{equation}
\begin{aligned}
\tikz[baseline=-0.5ex]{
    \draw[densely dotted,thick] (0.35,0.05)--(0.65,0.05);
    \draw[densely dotted,thick] (-0.35,0.05)--(-0.65,0.05);
    \draw (0.35,-0.05)--(0.65,-0.05);
    \draw (-0.35,-0.05)--(-0.65,-0.05);
    \draw[thick] (0,0) circle (10pt) node[anchor=center] {$G$};
    }
\ \, &= \ \,   \tikz[baseline=-0.5ex]{
    \draw[densely dotted,thick] (-0.5,0.05)--(0.5,0.05);
    \draw (-0.5,-0.05)--(0.5,-0.05);
    \draw (-0.5,0.2) circle (0.) node[anchor=center] {\footnotesize$q_1$};
    \draw (-0.5,-0.2) circle (0.) node[anchor=center] {\footnotesize$q_2$};
    }
 + \ \,
\tikz[baseline=-1.5ex]{
    \draw[dashed] (0,0) -- (1,0);
    \draw (-0.2,0.0)-- (-0.2,-0.15);
    \draw (-0.2,-0.15)-- (1.,-0.7);
    \draw (1.2,-0.15)-- (1.2,0.0);
    \draw (1.2,-0.15)-- (0,-0.7);
    \draw[densely dotted,thick] (-0.3,0.0)-- (-0.3,-0.4);
    \draw[densely dotted,thick] (1.3,-0.4)-- (1.3,0.0);
    %%%% average
    \draw[dashed] (1,0) arc (0:180:0.5);
    \draw (1.2,0) arc (0:180:0.7);
    \draw[densely dotted,thick] (1.3,0) arc (0:180:0.8);
    \draw (-0.,-0.9) circle (0.) node[anchor=center] {\footnotesize$\bar{q}_1$};
    \draw (-0.3,-0.55) circle (0.) node[anchor=center] {\footnotesize$q_1$};
    }
\ \,
+
\ \,
\tikz[scale=0.8,baseline=-0.5ex]{
    \draw[dashed] (0,0) -- (1,0);
    \draw (-0.2,0.)-- (-0.2,-0.15);
    \draw (-0.2,-0.15)-- (1.,-0.7);
    \draw (1.2,-0.15)-- (1.2,0.0);
    \draw (1.2,-0.15)-- (0,-0.8);
    \draw[densely dotted,thick] (-0.3,0.)-- (-0.3,-0.4);
    \draw[densely dotted,thick] (1.3,-0.1)-- (1.3,0.);
    %%%%%
    \draw[dashed] (2,0) -- (3,0);
    \draw (1.8,0.)-- (1.8,-0.15);
    \draw (1.8,-0.15)-- (3.,-0.8);
    \draw (3.2,-0.15)-- (3.2,0.);
    \draw (3.2,-0.15)-- (2,-0.7);
    \draw[densely dotted,thick] (1.7,0.)-- (1.7,-0.1);
    \draw[densely dotted,thick] (3.3,-0.4)-- (3.3,0.);
     %%%% contractions
    \draw[densely dotted,thick] (1.3,-0.1)--(1.7,-0.1);
    \draw (1.0,-0.7)-- (2,-0.7);
    %%%% average
    \draw[dashed] (1,0) arc (0:180:0.5);
    \draw (1.2,0) arc (0:180:0.7);
    \draw[densely dotted,thick] (1.3,0) arc (0:180:0.8);
    \draw[dashed] (3,0) arc (0:180:0.5);
    \draw (3.2,0) arc (0:180:0.7);
    \draw[densely dotted,thick] (3.3,0) arc (0:180:0.8);
    \draw (-0.,-1) circle (0.) node[anchor=center] {\footnotesize$\bar{q}_1$};
    \draw (-0.3,-0.55) circle (0.) node[anchor=center] {\footnotesize$q_1$};
 }
 \ \,
+
\cdots
 \\
&+ 
\ \,
\tikz[scale=0.7,baseline=-0.5ex]{
    \draw[dashed] (0,0) -- (1,0);
    \draw (-0.2,0.)-- (-0.2,-0.15);
    \draw (-0.2,-0.15)-- (1.,-0.7);
    \draw (1.2,-0.15)-- (1.2,0.0);
    \draw (1.2,-0.15)-- (0,-0.8);
    \draw[densely dotted,thick] (-0.3,0.)-- (-0.3,-0.4);
    \draw[densely dotted,thick] (1.3,-0.1)-- (1.3,0.);
    %%%%%
    \draw[dashed] (2,0) -- (3,0);
    \draw (1.8,0.)-- (1.8,-0.15);
    \draw (1.8,-0.15)-- (3.,-0.8);
    \draw (3.2,-0.15)-- (3.2,0.);
    \draw (3.2,-0.15)-- (2,-0.7);
    \draw[densely dotted,thick] (1.7,0.)-- (1.7,-0.1);
    \draw[densely dotted,thick] (3.3,-0.4)-- (3.3,0.);
     %%%% contractions
    \draw[densely dotted,thick] (1.3,-0.1)--(1.7,-0.1);
    \draw (1.0,-0.7)-- (2,-0.7);
    %%%% average
    \draw[densely dotted,thick] (3.3,0) arc (0:180:1.8);
    \draw[densely dotted,thick] (1.7,0) arc (0:180:0.2);
    \draw[dashed] (2.0,0) arc (0:180:0.5);
    \draw[dashed] (3.0,0) arc (0:180:1.5);
    \draw (1.8,0) arc (0:180:0.3);
    \draw (3.2,0) arc (0:180:1.7);
    \draw (-0.,-1.1) circle (0.) node[anchor=center] {\footnotesize$q_2$};
    \draw (-0.3,-0.55) circle (0.) node[anchor=center] {\footnotesize$q_1$};
    \draw (1.5,-1.1) circle (0.) node[anchor=center] {\footnotesize$\bar{q}_1$};
    \draw (1.5,-0.35) circle (0.) node[anchor=center] {\scriptsize$\bar{q}_2$};
 }
\ \,
+ 
 \ \,
 \tikz[scale=0.6,baseline=1.5ex]{
    \draw[dashed] (0,0) -- (1,0);
    \draw (-0.2,0.)-- (-0.2,-0.15);
    \draw (-0.2,-0.15)-- (1.,-0.7);
    \draw (1.2,-0.15)-- (1.2,0.0);
    \draw (1.2,-0.15)-- (0,-0.8);
    \draw[densely dotted,thick] (-0.3,0.)-- (-0.3,-0.4);
    \draw[densely dotted,thick] (1.3,-0.1)-- (1.3,0.);
    %%%%%
    \draw[dashed] (2,0) -- (3,0);
    \draw (1.8,0.)-- (1.8,-0.15);
    \draw (1.8,-0.15)-- (3.,-0.7);
    \draw (3.2,-0.15)-- (3.2,0.);
    \draw (3.2,-0.15)-- (2,-0.7);
    \draw[densely dotted,thick] (1.7,0.)-- (1.7,-0.1);
    \draw[densely dotted,thick] (3.3,-0.1)-- (3.3,0.);
    %%%
    \draw[dashed] (4,0) -- (5,0);
    \draw (3.8,0.)-- (3.8,-0.15);
    \draw (3.8,-0.15)-- (5.,-0.7);
    \draw (5.2,-0.15)-- (5.2,0.);
    \draw (5.2,-0.15)-- (4,-0.7);
    \draw[densely dotted,thick] (3.7,0.)-- (3.7,-0.1);
    \draw[densely dotted,thick] (5.3,-0.1)-- (5.3,0.);
    %%%
    \draw[dashed] (6,0) -- (7,0);
    \draw (5.8,0.)-- (5.8,-0.15);
    \draw (5.8,-0.15)-- (7.,-0.8);
    \draw (7.2,-0.15)-- (7.2,0.);
    \draw (7.2,-0.15)-- (6,-0.7);
    \draw[densely dotted,thick] (5.7,0.)-- (5.7,-0.1);
    \draw[densely dotted,thick] (7.3,-0.4)-- (7.3,0.);
     %%%% contractions
    \draw[densely dotted,thick] (1.3,-0.1)--(1.7,-0.1);
    \draw[densely dotted,thick] (3.3,-0.1)--(3.7,-0.1);
    \draw[densely dotted,thick] (5.3,-0.1)--(5.7,-0.1);
    \draw (1.0,-0.7)-- (2,-0.7);
    \draw (3.0,-0.7)-- (4,-0.7);
    \draw (5.0,-0.7)-- (6,-0.7);
    %%%% average
    \draw[densely dotted,thick] (3.3,0) arc (0:180:1.8);
    \draw[densely dotted,thick] (1.7,0) arc (0:180:0.2);
    \draw[dashed] (2.0,0) arc (0:180:0.5);
    \draw[dashed] (3.0,0) arc (0:180:1.5);
    \draw (1.8,0) arc (0:180:0.3);
    \draw (3.2,0) arc (0:180:1.7);
    \draw[densely dotted,thick] (7.3,0) arc (0:180:1.8);
    \draw[densely dotted,thick] (5.7,0) arc (0:180:0.2);
    \draw[dashed] (6.0,0) arc (0:180:0.5);
    \draw[dashed] (7.0,0) arc (0:180:1.5);
    \draw (5.8,0) arc (0:180:0.3);
    \draw (7.2,0) arc (0:180:1.7);
    \draw (-0.,-1.1) circle (0.) node[anchor=center] {\footnotesize$q_2$};
    \draw (-0.3,-0.55) circle (0.) node[anchor=center] {\footnotesize$q_1$};
    \draw (1.5,-1.1) circle (0.) node[anchor=center] {\footnotesize$\bar{q}_1$};
    \draw (1.5,-0.35) circle (0.) node[anchor=center] {\scriptsize$\bar{q}_2$};
    \draw (5.5,-1.1) circle (0.) node[anchor=center] {\footnotesize$\bar{q}_1$};
    \draw (5.5,-0.35) circle (0.) node[anchor=center] {\scriptsize$\bar{q}_2$};
 }
\ \,
+ \cdots \\
&+ 
 \ \,
\tikz[scale=0.5,baseline=2ex]{
    \draw[dashed] (0,0) -- (1,0);
    \draw (-0.2,0.)-- (-0.2,-0.15);
    \draw (-0.2,-0.15)-- (1.,-0.7);
    \draw (1.2,-0.15)-- (1.2,0.0);
    \draw (1.2,-0.15)-- (0,-0.8);
    \draw[densely dotted,thick] (-0.3,0.)-- (-0.3,-0.4);
    \draw[densely dotted,thick] (1.3,-0.1)-- (1.3,0.);
    %%%%%
    \draw[dashed] (2,0) -- (3,0);
    \draw (1.8,0.)-- (1.8,-0.15);
    \draw (1.8,-0.15)-- (3.,-0.7);
    \draw (3.2,-0.15)-- (3.2,0.);
    \draw (3.2,-0.15)-- (2,-0.7);
    \draw[densely dotted,thick] (1.7,0.)-- (1.7,-0.1);
    \draw[densely dotted,thick] (3.3,-0.1)-- (3.3,0.);
    %%%
    \draw[dashed] (4,0) -- (5,0);
    \draw (3.8,0.)-- (3.8,-0.15);
    \draw (3.8,-0.15)-- (5.,-0.7);
    \draw (5.2,-0.15)-- (5.2,0.);
    \draw (5.2,-0.15)-- (4,-0.7);
    \draw[densely dotted,thick] (3.7,0.)-- (3.7,-0.1);
    \draw[densely dotted,thick] (5.3,-0.1)-- (5.3,0.);
    \draw[densely dotted,thick] (5.3,-0.4)-- (5.3,0.);
     %%%% contractions
    \draw[densely dotted,thick] (1.3,-0.1)--(1.7,-0.1);
    \draw[densely dotted,thick] (3.3,-0.1)--(3.7,-0.1);
    \draw (1.0,-0.7)-- (2,-0.7);
    \draw (3.0,-0.7)-- (4,-0.7);
    %%%% average
    \draw[densely dotted,thick] (3.7,0) arc (0:180:1.2);
    \draw[densely dotted,thick] (3.3,0) arc (0:180:0.8);
    \draw[densely dotted,thick] (5.3,0) arc (0:180:2.8);
    \draw[dashed] (3.0,0) arc (0:180:.5);
    \draw[dashed] (5.,0) arc (0:180:2.5);
    \draw[dashed] (4.,0) arc (0:180:1.5);
    \draw (3.2,0) arc (0:180:0.7);
    \draw (3.8,0) arc (0:180:1.3);
    \draw (5.2,0) arc (0:180:2.7);
    %%%%
    \draw (-0.,-1.1) circle (0.) node[anchor=center] {\footnotesize$\bar{q}_1$};
    \draw (-0.3,-0.55) circle (0.) node[anchor=center] {\footnotesize$q_1$};
    \draw (1.5,-1.1) circle (0.) node[anchor=center] {\footnotesize$\bar{q}_1$};
    \draw (1.5,-0.38) circle (0.) node[anchor=center] {\scriptsize$q_1$};
 } \ \,
 + \cdots \\
&+ 
 \ \,
\tikz[scale=0.5,baseline=2ex]{
    \draw[dashed] (0,0) -- (1,0);
    \draw (-0.2,0.)-- (-0.2,-0.15);
    \draw (-0.2,-0.15)-- (1.,-0.7);
    \draw (1.2,-0.15)-- (1.2,0.0);
    \draw (1.2,-0.15)-- (0,-0.8);
    \draw[densely dotted,thick] (-0.3,0.)-- (-0.3,-0.4);
    \draw[densely dotted,thick] (1.3,-0.1)-- (1.3,0.);
    %%%%%
    \draw[dashed] (2,0) -- (3,0);
    \draw (1.8,0.)-- (1.8,-0.15);
    \draw (1.8,-0.15)-- (3.,-0.7);
    \draw (3.2,-0.15)-- (3.2,0.);
    \draw (3.2,-0.15)-- (2,-0.7);
    \draw[densely dotted,thick] (1.7,0.)-- (1.7,-0.1);
    \draw[densely dotted,thick] (3.3,-0.1)-- (3.3,0.);
    %%%
    \draw[dashed] (4,0) -- (5,0);
    \draw (3.8,0.)-- (3.8,-0.15);
    \draw (3.8,-0.15)-- (5.,-0.7);
    \draw (5.2,-0.15)-- (5.2,0.);
    \draw (5.2,-0.15)-- (4,-0.7);
    \draw[densely dotted,thick] (3.7,0.)-- (3.7,-0.1);
    \draw[densely dotted,thick] (5.3,-0.1)-- (5.3,0.);
    %%%
    \draw[dashed] (6,0) -- (7,0);
    \draw (5.8,0.)-- (5.8,-0.15);
    \draw (5.8,-0.15)-- (7.,-0.8);
    \draw (7.2,-0.15)-- (7.2,0.);
    \draw (7.2,-0.15)-- (6,-0.7);
    \draw[densely dotted,thick] (5.7,0.)-- (5.7,-0.1);
    \draw[densely dotted,thick] (7.3,-0.4)-- (7.3,0.);
     %%%% contractions
    \draw[densely dotted,thick] (1.3,-0.1)--(1.7,-0.1);
    \draw[densely dotted,thick] (3.3,-0.1)--(3.7,-0.1);
    \draw[densely dotted,thick] (5.3,-0.1)--(5.7,-0.1);
    \draw (1.0,-0.7)-- (2,-0.7);
    \draw (3.0,-0.7)-- (4,-0.7);
    \draw (5.0,-0.7)-- (6,-0.7);
    %%%% average
    \draw[densely dotted,thick] (3.7,0) arc (0:180:0.2);
    \draw[densely dotted,thick] (7.3,0) arc (0:180:3.8);
    \draw[densely dotted,thick] (5.7,0) arc (0:180:2.2);
    \draw[densely dotted,thick] (5.3,0) arc (0:180:1.8);
    \draw[dashed] (6.0,0) arc (0:180:2.5);
    \draw[dashed] (5.,0) arc (0:180:1.5);
    \draw[dashed] (4.,0) arc (0:180:0.5);
    \draw[dashed] (7.0,0) arc (0:180:3.5);
    \draw (5.8,0) arc (0:180:2.3);
    \draw (5.2,0) arc (0:180:1.7);
    \draw (3.8,0) arc (0:180:0.3);
    \draw (7.2,0) arc (0:180:3.7);
    \draw (-0.,-1.1) circle (0.) node[anchor=center] {\footnotesize$q_2$};
    \draw (-0.3,-0.55) circle (0.) node[anchor=center] {\footnotesize$q_1$};
    \draw (1.5,-1.1) circle (0.) node[anchor=center] {\footnotesize$\bar{q}_1$};
    \draw (1.5,-0.38) circle (0.) node[anchor=center] {\scriptsize$\bar{q}_2$};
    \draw (3.5,-1.1) circle (0.) node[anchor=center] {\footnotesize$q_2$};
    \draw (3.5,-0.38) circle (0.) node[anchor=center] {\scriptsize$q_1$};
 } \ \,
 + \cdots \ \ .
\label{eq:Gz_expansion}
\end{aligned}
\end{equation}
\end{widetext}
In the above, whenever the external lines are labeled with $q_1$ and $\bar{q}_1$ instead of $q_1$ and $q_2$, it is implied that such a term only contributes to components of $G$, with external lines labels satisfying $q_1 + q_2 = q_A$, which we call $G_2$ component; this requirement is dictated by the structure of theses terms and the connectivity of the vertices. However, other terms with two independent external labels $q_1$ and $q_2$ contribute to both $G_1$ and $G_2$.

\begin{figure*}
    \centering
    \includegraphics[scale=1.2]{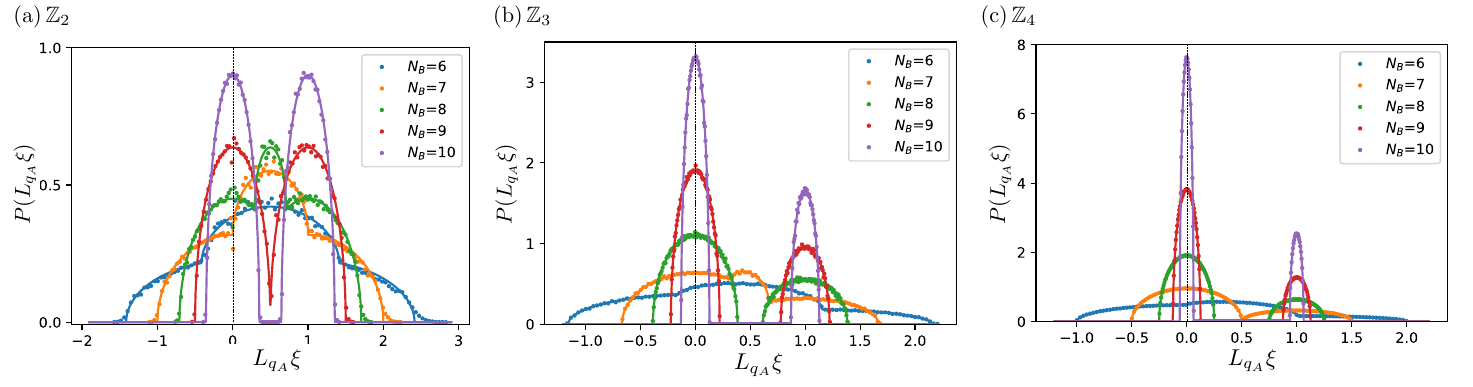}
    \caption{ Entanglement negativity spectrum of projected density matrices $\hat \rho_A^{(q_A)}$ for (a) $\Z_2$ symmetric qubit systems $R=2$, (b) $\Z_3$ symmetric qutrit systems $R=3$, and (c) $\Z_4$ symmetric ququart systems $R=4$, in the semicircle regime. 
    Solid lines are the random matrix theory result given in Eq.~(\ref{eq:spectral_ZR_projected}).
    Numerical data are represented by colored circles. The agreement between theory and numerics is evident. Here, $N_{A_1}=N_{A_2}=3$ and ensemble average is performed over $10^4$ samples. }
    \label{fig:Zr-sc}
\end{figure*}

Similar to the non-symmetric states, we shall call this regime  replica symmetry breaking~\cite{Shapourian2021} as explained below. Let us recall that the replica symmetry in our setup is a matrix identity corresponding to the invariance of the R\'enyi negativity $\Tr(\rTa)^n$ under a $\Z_n$ cyclic permutation of density matrices (or replicas). The replica symmetry action can diagrammatically be represented as a discrete translation symmetry. Given this diagrammatic definition, we can check whether a diagram is symmetric or not. As we see in Eq.(\ref{eq:Gz_expansion}), the dominant diagrams in the regime of interest in this subsection are not replica symmetric, and hence we call this regime replica symmetry breaking. We should however note that the overall summation is replica symmetric since the action of replica symmetry on diagrams only shuffles them within the sum.

One can write the function $G(z)$ in the following way as a geometric series:
\begin{equation}\label{eq:G_in_terms_of_sigma}
\begin{aligned}
    \tikz[baseline=-0.5ex]{
    \draw[densely dotted,thick] (0.35,0.05)--(0.65,0.05);
    \draw[densely dotted,thick] (-0.35,0.05)--(-0.65,0.05);
    \draw (0.35,-0.05)--(0.65,-0.05);
    \draw (-0.35,-0.05)--(-0.65,-0.05);
    \draw[thick] (0,0) circle (10pt) node[anchor=center] {$G$};
    }
    &=
   \tikz[baseline=-0.5ex]{
    \draw[densely dotted,thick] (-0.4,0.05)--(0.35,0.05);
    \draw (-0.4,-0.05)--(0.35,-0.05);
    }
    +
    \tikz[baseline=-0.5ex]{
    \draw[densely dotted,thick] (0.35,0.05)--(0.65,0.05);
    \draw[densely dotted,thick] (-0.35,0.05)--(-0.65,0.05);
    \draw (0.35,-0.05)--(0.65,-0.05);
    \draw (-0.35,-0.05)--(-0.65,-0.05);
    \draw[thick] (0,0) circle (10pt) node[anchor=center] {$\Sigma$};
    }
    +
    \tikz[baseline=-0.5ex]{
    \draw[densely dotted,thick] (0.35,0.05)--(0.65,0.05);
    \draw[densely dotted,thick] (-0.35,0.05)--(-0.65,0.05);
    \draw (0.35,-0.05)--(0.65,-0.05);
    \draw (-0.35,-0.05)--(-0.65,-0.05);
    \draw[thick] (0,0) circle (10pt) node[anchor=center] {$\Sigma$};
    \draw[thick] (1.05,0) circle (10pt) node[anchor=center] {$\Sigma$};
    \draw[densely dotted,thick] (1.4,0.05)--(1.7,0.05);
    \draw (1.4,-0.05)--(1.7,-0.05);
    }
    + \cdots  \\
      &= \frac{1}{z-\Sigma(z)},
\end{aligned}
\end{equation}
in terms of the self-energy function $\Sigma(z)$. Similar to $G$, $\Sigma$ has a block diagonal form as well:
\begin{equation}
\begin{aligned}
    \Sigma =  \bigoplus_{q_1,q_2\neq \bar{q}_1} & \Sigma_{1,q_1q_2} \ \ \mathbb{1}_{A_1,q_1} \otimes \mathbb{1}_{A_2,q_2} \\
     \oplus  \bigoplus_{q_1}\ \ \, & \Sigma_{2,q_1} \ \ \mathbb{1}_{A_1,q_1} \otimes \mathbb{1}_{A_2,\bar{q}_1}.
\end{aligned}
\end{equation}
$\Sigma(z)$ has an expansion in terms of $G(z)$ on its own:
\begin{equation}
\begin{aligned}
\tikz[baseline=-0.5ex]{
    \draw[densely dotted,thick] (0.35,0.05)--(0.65,0.05);
    \draw[densely dotted,thick] (-0.35,0.05)--(-0.65,0.05);
    \draw (0.35,-0.05)--(0.65,-0.05);
    \draw (-0.35,-0.05)--(-0.65,-0.05);
    \draw[thick] (0,0) circle (10pt) node[anchor=center] {$\Sigma$};
    }
\ \, &= \ \,
\tikz[baseline=-1.5ex]{
    \draw[dashed] (0,0) -- (1,0);
    \draw (-0.2,0.0)-- (-0.2,-0.15);
    \draw (-0.2,-0.15)-- (1.,-0.7);
    \draw (1.2,-0.15)-- (1.2,0.0);
    \draw (1.2,-0.15)-- (0,-0.7);
    \draw[densely dotted,thick] (-0.3,0.0)-- (-0.3,-0.4);
    \draw[densely dotted,thick] (1.3,-0.4)-- (1.3,0.0);
    %%%% average
    \draw[dashed] (1,0) arc (0:180:0.5);
    \draw (1.2,0) arc (0:180:0.7);
    \draw[densely dotted,thick] (1.3,0) arc (0:180:0.8);
    \draw (-0.,-0.9) circle (0.) node[anchor=center] {\footnotesize$\bar{q}_1$};
    \draw (-0.3,-0.55) circle (0.) node[anchor=center] {\footnotesize$q_1$};
    }
\ \
+
 \ \,
\tikz[scale=0.5,baseline=2ex]{
    \draw[dashed] (0,0) -- (1,0);
    \draw (-0.2,0.)-- (-0.2,-0.15);
    \draw (-0.2,-0.15)-- (1.,-0.5);
    \draw (1.2,-0.15)-- (1.2,0.0);
    \draw (1.2,-0.15)-- (0,-0.8);
    \draw[densely dotted,thick] (-0.3,0.)-- (-0.3,-0.4);
    \draw[densely dotted,thick] (1.3,-0.1)-- (1.3,0.);
    %%%%
    \draw[dashed] (4,0) -- (5,0);
    \draw (3.8,0.)-- (3.8,-0.15);
    \draw (3.8,-0.15)-- (5.,-0.7);
    \draw (5.2,-0.15)-- (5.2,0.);
    \draw (5.2,-0.15)-- (4,-0.5);
    \draw[densely dotted,thick] (3.7,0.)-- (3.7,-0.1);
    \draw[densely dotted,thick] (5.3,-0.1)-- (5.3,0.);
    \draw[densely dotted,thick] (5.3,-0.4)-- (5.3,0.);
     %%%% contractions
    \draw[densely dotted,thick] (1.3,-0.1)--(2,-0.1);
    \draw[densely dotted,thick] (3,-0.1)--(3.7,-0.1);
    \draw (1.0,-0.5)-- (2,-0.5);
    \draw (3.0,-0.5)-- (4,-0.5);
    %%%% average
    \draw[densely dotted,thick] (3.7,0) arc (0:180:1.2);
    % \draw[densely dotted,thick] (3.3,0) arc (0:180:0.8);
    \draw[densely dotted,thick] (5.3,0) arc (0:180:2.8);
    % \draw[dashed] (3.0,0) arc (0:180:.5);
    \draw[dashed] (5.,0) arc (0:180:2.5);
    \draw[dashed] (4.,0) arc (0:180:1.5);
    % \draw (3.2,0) arc (0:180:0.7);
    \draw (3.8,0) arc (0:180:1.3);
    \draw (5.2,0) arc (0:180:2.7);
    %%%%%
    \draw (2.5,-0.3) circle (.52) node[anchor=center] {\footnotesize$G$};
    %%%%
    \draw (-0.,-1.1) circle (0.) node[anchor=center] {\footnotesize$q_2$};
    \draw (-0.3,-0.55) circle (0.) node[anchor=center] {\footnotesize$q_1$};
    \draw (1.65,-0.85) circle (0.) node[anchor=center] {\footnotesize$\bar{q}_1$};
    \draw (1.7,0.2) circle (0.) node[anchor=center] {\scriptsize$\bar{q}_2$};
 } \ \, ,
\end{aligned}
\end{equation}
forming a Schwinger-Dyson (self-consistent) equation. 

\begin{widetext}
Next, we plug in the resolvent function \eqref{eq:structure_of_G} and get
\begin{equation}
\begin{aligned}
    \Sigma = \bigoplus_{q_1}\ \ \, &  \ \ \mathbb{1}_{A_1,q_1} \otimes \mathbb{1}_{A_2,\bar{q}_1}  \ \left[ \frac{L_{q_B}}{L_{q_A} L_{q_B}} 
     +\frac{L_{q_B}}{(L_{q_A} L_{q_B})^2} \; L_{A_1,q_1} L_{A_2,\bar{q}_1} \, G_{2,q_1} \right]\\
     \oplus  \bigoplus_{q_1,q_2\neq \bar{q}_1} &  \ \ \mathbb{1}_{A_1,q_1} \otimes \mathbb{1}_{A_2,q_2} \frac{L_{q_B}}{(L_{q_A} L_{q_B})^2} L_{A_1,\bar{q}_2} L_{A_2,\bar{q}_1} \, G_{1,\bar{q}_2\bar{q}_1} .
\end{aligned}    
\end{equation}

On the other hand, using Eq.~\eqref{eq:G_in_terms_of_sigma}, and writing $G$ in terms of $\Sigma$ result in:
\begin{equation}
\begin{aligned}
    G_{1,q_1q_2} &= \frac{1}{z - \frac{L_{A_1,\bar{q}_2} L_{A_2,\bar{q}_1}}{L_{q_A}^2 L_{q_B}} \ G_{1,\bar{q}_2\bar{q}_1} },\\
    G_{2,q_1} &= \frac{1}{z - \left( \frac{1}{L_{q_A}} +\frac{L_{A_1,q_1}L_{A_2,\bar{q}_1}}{L_{q_A}^2 L_{q_B}} \ G_{2,q_1}  \right)}.
\end{aligned}    
\end{equation}
This yields a set of quadratic equations for $G_1$ and $G_2$:
\begin{equation}
\begin{aligned}
   & G_{1,q_1q_2}^2 \frac{L_{A_1,q_1} L_{A_2,q_2}}{L_{q_A}^2 L_{q_B}} \; z   - G_{1,q_1q_2} \left[z^2 + \frac{1}{L_{q_A}^2L_{q_B} } \left(  L_{A_1,q_1} L_{A_2,q_2} - L_{A_1,\bar{q}_2} L_{A_2,\bar{q}_1} \right) \right]  + z = 0, \\
   \phantom{\tiny -}\\
    & G_{2,q_1}^2 \frac{L_{A_1,q_1} L_{A_2,\bar{q}_1}}{L_{q_A}^2 L_{q_B}} + G_{2,q_1} \left( \frac1{L_{q_A}} - z \right)  + 1 =0.
\end{aligned}    
\end{equation}
The solutions to the above equations read:
\begin{equation}\label{eq:Gs_general_aolution_semicircle_regime}
\begin{aligned}
     G_{1,q_1q_2} &= \frac{L_{q_A} \ }{2 \, z \, \alpha_{q_1q_2} } \left\{ \left[ z^2 + \frac{1}{L_{q_A}} \left( \alpha_{q_1q_2} - \alpha_{\bar{q}_2 \bar{q}_1} \right) \right]   \pm \sqrt{z^4 -  \frac{2z^2}{L_{q_A}} \left( \alpha_{q_1q_2} + \alpha_{\bar{q}_2 \bar{q}_1} \right)  + \frac{1}{L_{q_A}^2} \left(  \alpha_{q_1q_2} - \alpha_{\bar{q}_2 \bar{q}_1} \right)^2} \,  \right\} \ ,  \\
     \phantom{-}\\
       G_{2,q_1}  &=  \frac{L_{q_A}   }{2 \, \alpha_{q_1 \bar{q}_1} } \left\{  \left( z - \frac1{L_{q_A}} \right) \pm \sqrt{ \left( z - \frac1{L_{q_A}}  \right)^2 - 4 \, \frac{\alpha_{q_1 \bar{q}_1} }{L_{q_A}}  } \,  \right\},
\end{aligned}
\end{equation}
\end{widetext}

where we have defined a family of parameters,
\begin{align}
    \alpha_{q_1q_2} = \frac{L_{A_1,q_1} L_{A_2,q_2}}{L_{q_A} L_{q_B}}.
\end{align}

\subsubsection{$\mathbb{Z}_R$ symmtery}
Instead of focusing on the general case first, we initially consider the simple case of taking the symmetry group to be $\mathbb{Z}_R$.
In this case, since symmetry sectors having different quantum numbers have identical dimensions, $G_1$ and $G_2$ become indepenedent of their quantum number indices. The spectral densities derived from imaginary parts of the solutions in Eq.~\eqref{eq:Gs_general_aolution_semicircle_regime} can be written as:
\begin{equation}
\label{eq:semicircle_equal}
\begin{aligned}
    P(\xi) &= \frac{1}{\pi}\sum_{q_1 , q_2\neq\bar{q_1}} \left(L_{A_1,q_1} L_{A_2,q_2}\right) \ \frac{L_{q_A} }{2\alpha} \  \sqrt{\frac{4\alpha}{L_{q_A}} - \xi^2 }'  \\
    &+ \frac{1}{\pi}  \ \sum_{q_1} \ \left(L_{A_1,q_1} L_{A_2,\bar{q}_1}\right)  \ \frac{L_{q_A}}{2\alpha} \ \sqrt{ \frac{4\alpha}{L_{q_A}} - \left(\xi-\frac{1}{L_{q_A}} \right)^2 }',
\end{aligned}    
\end{equation}
where the summations and the initial factors of the sector dimensions account for the trace. Furthermore, the primed square roots above and from here on are used to denote $\sqrt{\cdot}'=\theta(\cdot)\sqrt{\cdot}$ for brevity.

In the case of $\Z_R$ symmetric states, we may write the Hilbert space dimensions explicitly as $L_{A_s,q_s}=R^{N_{A_s}-1}$ (which is independent of $q_i$) and  simplify (\ref{eq:semicircle_equal}) further into 
\begin{equation}\label{eq:spectral_ZR_projected}
\begin{aligned}
    P(\xi) &= \frac{1}{2\pi} \  R^{2N_A+N_B-2}  \ \Bigg\{  \ (R-1) \ \sqrt{ \frac{4}{R^{N-1}} - \xi^2 }' 
    \\
    & \quad +  \quad   \ 
    \sqrt{ \frac{4}{R^{N-1}} - \left( \xi - \frac{1}{R^{N_A-1}} \right)^2 }' \  \Bigg\}.
\end{aligned}    
\end{equation}
It can be seen that the spectral density consists of two semicircles as shown Fig.~\ref{fig:Zr-sc}. One of the semicircles is always centered at $\xi=0$ irrespective of the size of different subsystems. This means that the partially transposed reduced density matrix always has negative eigenvalues and thus there is some residual entanglement between $A_1$ and $A_2$ irrespective of how large the rest of the system is. It is important to note that the spectral density in this regime is independent of the way subsystem $A$ is partitioned into $A_1$ and $A_2$. This phenomenon (which was dubbed entanglement saturation in Ref.~\cite{Shapourian2021}) is similar to the non-symmetric states as a result of which the entanglement negativity does not depend on the size of $A_1$ and $A_2$. As we see below, this similarity does not hold in general for symmetric states.

With the above form for the spectral density, one can calculate the negativity using Eq.~(\ref{eq:neg_dist}).
We note that the first term in \eqref{eq:spectral_ZR_projected}, always contributes to the negativity and thus we call it the residual contribution which has the following form:
\begin{equation}
    \label{eq:N1-Zr-sc}
    \mathcal{N}_{1}(\hat\rho_{A}^{(q_A)}) = \frac{2}{3\pi} (R-1) \; R^{\frac12 \left( N_A - N_B - 1 \right)}.
\end{equation}
The second term in \eqref{eq:spectral_ZR_projected}, on the other hand, contributes only if $R^{\left( N_A - N_B - 1 \right)} > \frac14$:
\begin{equation}
\begin{aligned}
    \label{eq:N2-Zr-sc}
    \mathcal{N}_{2}(\hat\rho_{A}^{(q_A)}) &= \frac{1}{2\pi} \left( \frac83 R^{\frac12 \left( N_A - N_B -1 \right)} + \frac13 R^{\frac12 \left( N_B - N_A + 1 \right)}  \right) \\
    & \times \sqrt{ 1 - \frac14 R^{ \left( N_B - N_A + 1 \right)} } \\
    & - \frac{1}{\pi} \cos^{-1} \left( \frac12 R^{\frac12 \left( N_B - N_A + 1 \right)}  \right).
\end{aligned}
\end{equation}
The dominant contribution of this term deep in the semicircle regime has the form $\frac{4}{3\pi} R^{\frac12 \left( N_A - N_B -1 \right)}$.

The logarithmic negativity as a result reads:
\begin{equation}
      \langle\mathcal{E}(\hat\rho_A^{(q_A)})\rangle = 
     \begin{cases} 
       \log_2\left(\frac{4}{3\pi} (R+1) \right) + \frac12 \left( N_A - N_B - 1 \right) \log_2R  \vspace{0.1cm}\\
       \qquad \qquad \text{ if} \qquad R^{\left( N_A - N_B - 1 \right)} \gg \frac14 \\
       \phantom{a}\\
      \frac{1}{\log2}\frac{4}{3\pi} (R-1) \; R^{\frac12 \left( N_A - N_B - 1 \right)} \vspace{0.1cm}\\
       \qquad \qquad \text{ if} \qquad R^{\left( N_A - N_B - 1 \right)} \ll \frac14 
   \end{cases}
\end{equation}

\begin{figure*}
    \centering
    \includegraphics[scale=1.3]{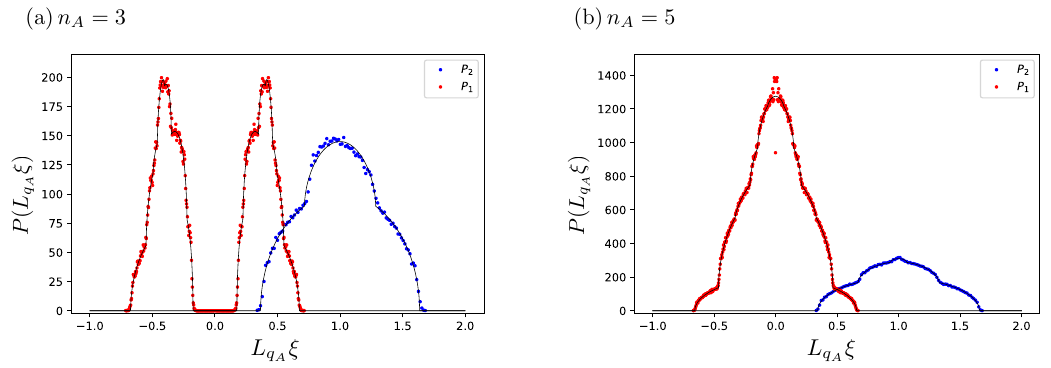}
    \caption{Entanglement negativity spectrum of $\hat \rho_A^{(q_A)}$ for a $U(1)$-symmetric system in the replica symmetry breaking regime. Here, $N_{A_1}=N_{A_2}=5$, and $N_B=12$. Total particle number is $q_A+q_B=11$ and projected sector is labeled by $q_A$. Solid lines are random matrix theory results given by Eq.~(\ref{eq:spectal_general_forms}) which are in good agreement with the exact numerical simulations (colored circles). }
    \label{fig:u1-semicircle}
\end{figure*}

\subsubsection{Generic symmetry group}
For a generic symmetry group,  the two components of the spectral density (\ref{eq:spec-decomposition}) can be similarly obtained from Eq.~\eqref{eq:Gs_general_aolution_semicircle_regime} as
\begin{equation}\label{eq:spectal_general_forms}
\begin{aligned}
    P_{1,q_1q_2}(\xi) & = \frac{1}{\pi} \frac{L_{q_A}^2 L_{q_B} }{2 \left|\xi \right|} \\
    &\sqrt{ \frac{4}{L_{q_A}^2}  \alpha_{q_1q_2}  \alpha_{\bar{q}_2 \bar{q}_1}  - \left[ \xi^2  - \frac{1}{L_{q_A}}  \left(  \alpha_{q_1q_2} + \alpha_{\bar{q}_2 \bar{q}_1} \right) \right]^2  }',\\
    P_{2,q_1}(\xi) & =  \frac{1}{\pi} \frac{L_{q_A}^2 L_{q_B} }{2} \sqrt{4 \, \frac{\alpha_{q_1 \bar{q}_1} }{L_{q_A}}  - \left( \xi - \frac1{L_{q_A}}  \right)^2 }' .
\end{aligned}    
\end{equation}
It can be seen above that $P_{2,q_1}$ always results in a semicircle law for the spectral density. These semicircles are centered at the same point (which only depends on total symmetry charge $A$), while their radii are different (determined by $q_1$). The general form of the $P_1$ contribution is more involved. In particular, its shape deviates from a semicircle and generally depends on symmetry charge sectors $q_1$ and $q_2$. Nevertheless, it is symmetric around $\xi = 0$. 
It is easy to check that $P_1$ is only non-zero over the following range 
\begin{equation}
    \left| \sqrt{\alpha_{q_1 q_2}} - \sqrt{\alpha_{\bar{q}_2 \bar{q}_1}} \right|  < \left| \xi \right| \sqrt{L_{q_A}}  <   \left| \sqrt{\alpha_{q_1 q_2}} + \sqrt{\alpha_{\bar{q}_2 \bar{q}_1}} \right|.
\end{equation} 
This means that if $\alpha_{q_1 q_2} \neq \alpha_{\bar{q}_2 \bar{q}_1}$ for all allowed values of $q_1$ and $q_2$, there will be a gap in the spectral density around $\xi=0$.
We examine this prediction in the case of $U(1)$ symmetry as plotted in Fig.~\ref{fig:u1-semicircle}. We first note that the Hilbert space dimension of subsystem $A_s$ is given by
$L_{A_s,q_s} = \binom{N_{A_s}}{q_{A_s}}$. 
Then, the condition $\alpha_{q_1 q_2} \neq \alpha_{\bar{q}_2 \bar{q}_1}$ is always met unless $q_A = N_{A_1}=N_{A_2}$. Therefore, for a generic charge, $P_1$ has a gap (as in Fig.~\ref{fig:u1-semicircle}(a)), and a non-zero continuous form of $P_1$ only appears at a fine tuned point (Fig.~\ref{fig:u1-semicircle}(b)).

\subsubsection{Thermodynamic limit for the U(1) case}\label{sec:thermo_limit_U(1)_plateau}

In this part, we describe a $U(1)$-symmetric system in terms of a system with conserved number of particles. To take the thermodynamic limit, we find it more convenient to 
characterize the symmetry charge sector in terms the filling factor $\nu_s = \frac{q_s}{N_s}$ where $q_s$ and $N_s$ denote the particle number and the number of sites in (or volume of) the subsystem $s$, respectively. Thermodynamic limit is then understood as taking $q_s,N_s\to \infty$ while $\nu_s$ is kept finite.

The relations for the spectral densities in \eqref{eq:spectal_general_forms} could be simplified in the thermodynamic limit, for example all functionalities become those of filling factors and the summations turn into integrals. Plugging in the spectral density~\eqref{eq:spectal_general_forms} (after the continuum approximation) to Eq.~(\ref{eq:neg_dist}) for the entanglement negativity, we obtain
\begin{widetext}
\begin{align}
    \label{eq:neg-u1-decomp}
    \braket{\mathcal{N}(\hat\rho_A^{(q_A)})} 
    &= N_{A_1} N_{A_2} \int d\nu_1 d\nu_2 \; \mathfrak{n}_{1}(\nu_1,\nu_2) + N_{A_1} \int d\nu_1 \; \mathfrak{n}_{2}(\nu_1), \\
    \label{eq:negativity_saddle}
    \mathfrak{n}_{1}(\nu_1,\nu_2) & =  \frac{L_{\nu_A}^2 L_{\nu_B} }{2\pi} \; \int_{\xi<0} d\xi \;  \sqrt{ \frac{4}{L_{\nu_A}^2}  \alpha_{\nu_1 \nu_2}  \alpha_{\bar{\nu}_2 \bar{\nu}_1}  - \left[ \xi^2  - \frac{1}{L_{\nu_A}}  \left(  \alpha_{\nu_1 \nu_2} + \alpha_{\bar{\nu}_2 \bar{\nu}_1} \right) \right]^2  }', \nonumber \\
    \mathfrak{n}_{2}(\nu_1) & =   \frac{L_{\nu_A}^2 L_{\nu_B} }{2\pi} \int_{\xi<0} d\xi \; (-\xi) \; \sqrt{4 \, \frac{\alpha_{\nu_1 \bar{\nu}_1} }{L_{\nu_A}}  - \left( \xi - \frac1{L_{\nu_A}}  \right)^2 }' .
\end{align}
\end{widetext}
In the above expressions, the first line is a continuum version of Eq.~(\ref{eq:neg-qA-contributions}), and it is assumed that the symmetry sector of $A$ that we are considering is characterized by a particle number equal to $N_A \nu_A$ and that the total particle number in the system is given by $N_A \nu_A + N_B \nu_B$. Given a value for $\nu_1$, the complimentary $\bar{\nu}_1$ is chosen so that $N_{A_1}\nu_1 + N_{A_2}\bar{\nu}_1 = N_{A} \nu_A$. Furthermore, $\alpha_{\nu_1 \nu_2} = \frac{L_{A_1,\nu_1} \; L_{A_2,\nu_2}}{ L_{\nu_A} L_{\nu_B} }$, and in the thermodynamic limit, 
\begin{equation}
    L_s = \frac{1}{\sqrt{2\pi N_s \nu_s \left( 1 - \nu_s \right)  }} \; e^{N_s \, f(\nu_s)},
\end{equation}
with 
\begin{equation} \label{eq:definition_f}
f(\nu_s) = -\nu_s \log\nu_s - (1-\nu_s) \log\left( 1 - \nu_s \right).
\end{equation}

In the following, we show how a saddle point approximation can be applied to calculate the negativity in the forms presented above. We first focus on the component given by $\mathfrak{n}_1$:
\begin{equation}
\begin{aligned}
    \mathcal{N}_1(\hat\rho_A^{(\nu_A)}) &=  N_{A_1} N_{A_2} \frac{L_{\nu_A}^2 L_{\nu_B} }{2\pi} \int d\nu_1 d\nu_2 \;  \left( \frac{  \alpha_{\nu_1 \nu_2} + \alpha_{\bar{\nu}_2 \bar{\nu}_1} }{L_{\nu_A}}   \right)^{3/2} \\
    & \qquad \qquad \int_{-\sqrt{1+a}}^{-\sqrt{1-a}} d\tilde{\xi} \; \sqrt{a^2 - \left( 1 -  \tilde{\xi}^2 \right)^2},
\end{aligned}    
\end{equation}
where $a^2 = \frac{4 \; \alpha_{\nu_1\nu_2} \alpha_{\bar{\nu}_2 \bar{\nu}_1}}{ \left( \alpha_{\nu_1\nu_2} + \alpha_{\bar{\nu}_2 \bar{\nu}_1} \right)^2 } \leq 1$ and a change of variable to $\tilde{\xi}$ is performed. Using another change of variable $1 -  \tilde{\xi}^2  = a \cos\theta$, the integral on the second row takes the form $a^2 \frac{1}{4} \int_0^{2\pi}d\theta \; \frac{\sin^2\theta}{\sqrt{1-a \cos\theta}}$: 
\begin{equation}\label{eq:int_to_be_saddle_point_approxed}
\begin{aligned}
    \mathcal{N}_1(\hat\rho_A^{(\nu_A)})
    &= \frac{  L_{\nu_A}^{1/2} L_{\nu_B} }{2\pi} I_a \int d\nu_1 d\nu_2 \;  \frac{ \alpha_{\nu_1\nu_2} \alpha_{\bar{\nu}_2 \bar{\nu}_1}}{ \left( \alpha_{\nu_1\nu_2} + \alpha_{\bar{\nu}_2 \bar{\nu}_1} \right)^{1/2} }. 
\end{aligned}    
\end{equation}
The integral $I_a=\int_0^{2\pi} d\theta \; \frac{\sin^2\theta}{\sqrt{1-a \cos\theta}}$ varies monotonically over $[\pi,\frac{8 \sqrt{2}}{3})$ for $0<a<1$.
We now make a saddle point approximation in the integral over $\nu_1, \nu_2$ for a given $\nu_A$. The important part of the integrand turns out to be the following due to its exponential dependence on $N_{A,i}$:
\begin{equation}
     \frac{ \alpha_{\nu_1\nu_2} \alpha_{\bar{\nu}_2 \bar{\nu}_1}}{ \left( \alpha_{\nu_1\nu_2} + \alpha_{\bar{\nu}_2 \bar{\nu}_1} \right)^{1/2} } 
     = \frac{1}{\left( L_{\nu_A} L_{\nu_B} \right)^{3/2}} \ e^{\mathcal{F}(\nu_1,\nu_2) }.
\end{equation}
The single exponential on the right hand side has the exponent:
\begin{equation}
\begin{aligned}
    \mathcal{F}(\nu_1,\nu_2) = N_{A_1} \left[  f(\nu_1) + f(\bar{\nu}_2) \right] + N_{A_2} \left[  f(\nu_2) + f(\bar{\nu}_1) \right] \\
     - \frac12 \log \left[  
     e^{N_{A_1} f(\nu_1) + N_{A_2} f(\nu_2)} + e^{N_{A_1} f(\bar{\nu}_2) + N_{A_2} f(\bar{\nu}_1)} 
     \right],
\end{aligned}
\end{equation}
where terms slower than $N_{A,i}$ in the thermodynamic limit are neglected for the process of evaluating the integral in the saddle point approximation but will be restored eventually.
One then needs to find the $\nu_1$ and $\nu_2$ values that result in the maximum value for the exponent; the integral in \eqref{eq:int_to_be_saddle_point_approxed} can thus be approximated by a Gaussian integral around these values of $\nu_1$ and $\nu_2$. Given the form introduced above for the function $f$ and the $\bar{\nu}_i$ values, it can be shown that $\mathcal{F}$ acquires its maximum value when $\nu_1 = \nu_2 = \nu_A$. This maximal value reads:
\begin{equation}
    \mathcal{F}(\nu_A, \nu_A) = \frac{1}{2} \left[ -\log 2 +  3 \; (N_{A_1}+N_{A_2}) \; f(\nu_A)    \right].
\end{equation}
One then needs to also calculate the second derivatives of $\mathcal{F}$ at this point. These derivatives for a general value of $\nu_A$ take a very involved form which we do not show here (but will use below), but for the special case of $\nu_A = \frac12$, they take a simpler form; $\nu_A = \frac12$, the matrix $\partial_{\nu_i} \partial_{\nu_j} \mathcal{F}(\nu_1, \nu_2)$ assumes a diagonal form with the elements $-3\frac{N_2}{N_1} (N_1+N_2)$ and $-3\frac{N_1}{N_2} (N_1+N_2)$ on the diagonal. If we now take all the exponential factors together, we obtain the following relation:
\begin{widetext}
\begin{equation}
    % \int d\nu_1 d\nu_2 \; \mathfrak{n}_1(\nu_1,\nu_2)
    \mathcal{N}_1(\hat\rho_A^{(\nu_A)})
    \sim \frac{1}{L_{\nu_A}  L_{\nu_B}^{1/2} } \ e^{\frac32  N_A f(\nu_A) } = e^{\frac12 \left[ N_A f(\nu_A) - N_B f(\nu_B) \right]}.
\end{equation}
The above expression applies to all values of $\nu_A$.

Taking all the prefactors into account for the case of $\nu_A = \frac12$, we get the contribution to negativity due to $\mathfrak{n}_1$ as:
\begin{equation}
\begin{aligned}
      \mathcal{N}_1(\hat\rho_A^{(\nu_A=\frac12)})
    = \frac{16}9 \; \left(\frac2\pi \right)^{3/4} \frac{\left( N_{A_1} N_{A_2} N_B \right)^{1/4} }{ N_A^{1/2} }
    \;
     \left[ \nu_B (1-\nu_B)  \right]^{1/4}
  \; e^{\frac12 \left[  N_A f(\frac12) - N_B f(\nu_B) \right]}.
\end{aligned}    
\end{equation}
For general $\nu_A$, this conribution to negativity takes the following form:
\begin{equation}
\begin{aligned}
     \mathcal{N}_1(\hat\rho_A^{(\nu_A)})
    =\frac{\frac{16}9 \; \left(\frac2\pi \right)^{3/4} \frac{\left( N_{A_1} N_{A_2} N_B \right)^{1/4} }{ N_A^{1/2} }
    \;
     \left[ \nu_B (1-\nu_B)  \right]^{1/4}}{\sqrt{ 1+ \frac43 \frac{N_{A_1} N_{A_2}}{N_A}  \, (1-\nu_A) \, \nu_A \, \log^2\left(\frac{1}{\nu_A}-1\right)    }}
  \; e^{\frac12 \left[  N_A f(\nu_A) - N_B f(\nu_B) \right]}.
\end{aligned}    
\end{equation}

A similar saddle point approximation could be employed for the other term in the negativity \eqref{eq:negativity_saddle} that is given by $\mathfrak{n}_2$; we first note that $\mathfrak{n}_2(\nu_1)$ vanishes if     $4 \alpha_{\nu_1 \bar{\nu}_1 } L_{\nu_A} =  4 \frac{L_{A_1,\nu_1} L_{A_2,\bar{\nu}_1} }{L_{\nu_B}} < 1$. On the other hand, if we assume that we are away from this limit, and actually are deep within the replica symmetry breaking regime (where $\frac{L_{A_1,\nu_1} L_{A_2,\bar{\nu}_1} }{L_{\nu_B}} \gg 1$ for the most dominant terms contributing to the $\nu_1$ integral over $\mathfrak{n}_2$) one can write this contribution to the negativity as:
\begin{equation}
    \mathcal{N}_2(\hat\rho_A^{(\nu_A)})
    = N_{A_1} \int d\nu_1 \; \mathfrak{n}_{2}(\nu_1) = N_{A_1} \int d\nu_1 \; \frac{4}{3\pi} \frac{1}{\sqrt{L_{\nu_B}} L_{\nu_A}} \left( L_{A_1,\nu_1} L_{A_2,\bar{\nu}_1} \right)^{3/2}
\end{equation}
The saddle point solution is given by $\nu_1 = \nu_A$. It takes the following form after straightforward manipulations:
\begin{equation}
    \mathcal{N}_2(\hat\rho_A^{(\nu_A)})
    = \frac{4}{3\sqrt{3} \pi} \; \left(\frac{2}{\pi}\right)^{1/4} \;
    \left(\frac{N_B }{N_{A_1} N_{A_2}}\right)^{1/4} \; \frac{ \left[\nu_B(1-\nu_B) \right]^{1/4} }{\left[\nu_A(1-\nu_A) \right]^{1/2}} \ 
    e^{\frac12 \left[ N_A f(\nu_A) - N_B f(\nu_B) \right]}.
\end{equation}
One should note that unlike $\mathcal{N}_{A_1:A_2; \; \nu_A}^{(1)}$ this result only holds if the exponent $\frac12 \left[ N_A f(\nu_A) - N_B f(\nu_B) \right]$ is positive, as discussed above.

\end{widetext}

\subsection{The general case (including maximal entanglement in $A$)}
In the general case, we take the condition on the subsystem sizes to be $N_{A_1} > N_{A_2}$; in this regime, more terms in the self energy should be taken into account as shown below in Eq.~\eqref{eq:self_energy_general_case}. It can be seen that if one takes only the first two terms in the self-energy, the previous result in the replica symmetry breaking regime is recovered.

Apart from the single term on the first row, the rest of the terms are grouped into two classes: one, the class of terms on the second row, which contains diagrams with odd numbers of resolvent function insertions which contributes to both $G_1$ and $G_2$; and two, the class on the third row whose terms contain even numbers of resolvent function insertions which only contributes to $G_2$.

\begin{widetext}

\begin{equation}\label{eq:self_energy_general_case}
\begin{aligned}
&\tikz[baseline=-0.5ex]{
    \draw[densely dotted,thick] (0.35,0.05)--(0.65,0.05);
    \draw[densely dotted,thick] (-0.35,0.05)--(-0.65,0.05);
    \draw (0.35,-0.05)--(0.65,-0.05);
    \draw (-0.35,-0.05)--(-0.65,-0.05);
    \draw[thick] (0,0) circle (10pt) node[anchor=center] {$\Sigma$};
    }
\ \, = \ \,
\tikz[baseline=-1.5ex]{
    \draw[dashed] (0,0) -- (1,0);
    \draw (-0.2,0.0)-- (-0.2,-0.15);
    \draw (-0.2,-0.15)-- (1.,-0.7);
    \draw (1.2,-0.15)-- (1.2,0.0);
    \draw (1.2,-0.15)-- (0,-0.7);
    \draw[densely dotted,thick] (-0.3,0.0)-- (-0.3,-0.4);
    \draw[densely dotted,thick] (1.3,-0.4)-- (1.3,0.0);
    %%%% average
    \draw[dashed] (1,0) arc (0:180:0.5);
    \draw (1.2,0) arc (0:180:0.7);
    \draw[densely dotted,thick] (1.3,0) arc (0:180:0.8);
    \draw (-0.,-0.9) circle (0.) node[anchor=center] {\footnotesize$\bar{q}_1$};
    \draw (-0.3,-0.55) circle (0.) node[anchor=center] {\footnotesize$q_1$};
    }
\\
&+
 \ \,
\tikz[scale=0.45,baseline=2ex]{
    \draw[dashed] (0,0) -- (1,0);
    \draw (-0.2,0.)-- (-0.2,-0.15);
    \draw (-0.2,-0.15)-- (1.,-0.5);
    \draw (1.2,-0.15)-- (1.2,0.0);
    \draw (1.2,-0.15)-- (0,-0.8);
    \draw[densely dotted,thick] (-0.3,0.)-- (-0.3,-0.4);
    \draw[densely dotted,thick] (1.3,-0.1)-- (1.3,0.);
    %%%%
    \draw[dashed] (4,0) -- (5,0);
    \draw (3.8,0.)-- (3.8,-0.15);
    \draw (3.8,-0.15)-- (5.,-0.7);
    \draw (5.2,-0.15)-- (5.2,0.);
    \draw (5.2,-0.15)-- (4,-0.5);
    \draw[densely dotted,thick] (3.7,0.)-- (3.7,-0.1);
    \draw[densely dotted,thick] (5.3,-0.1)-- (5.3,0.);
    \draw[densely dotted,thick] (5.3,-0.4)-- (5.3,0.);
     %%%% contractions
    \draw[densely dotted,thick] (1.3,-0.1)--(2,-0.1);
    \draw[densely dotted,thick] (3,-0.1)--(3.7,-0.1);
    \draw (1.0,-0.5)-- (2,-0.5);
    \draw (3.0,-0.5)-- (4,-0.5);
    %%%% average
    \draw[densely dotted,thick] (3.7,0) arc (0:180:1.2);
    \draw[densely dotted,thick] (5.3,0) arc (0:180:2.8);
    \draw[dashed] (5.,0) arc (0:180:2.5);
    \draw[dashed] (4.,0) arc (0:180:1.5);
    \draw (3.8,0) arc (0:180:1.3);
    \draw (5.2,0) arc (0:180:2.7);
    %%%%%
    \draw (2.5,-0.3) circle (.52) node[anchor=center] {\footnotesize$G$};
    %%%%
    \draw (-0.,-1.1) circle (0.) node[anchor=center] {\scriptsize $q_2$};
    \draw (-0.3,-0.55) circle (0.) node[anchor=center] {\scriptsize $q_1$};
    \draw (1.65,-0.85) circle (0.) node[anchor=center] {\scriptsize $\bar{q}_1$};
    \draw (1.7,0.2) circle (0.) node[anchor=center] {\tiny $\bar{q}_2$};
 } 
 \ \,  
 +  
 \ \, 
 \tikz[scale=0.36,baseline=2ex]{
    \draw[dashed] (0,0) -- (1,0);
    \draw (-0.2,0.)-- (-0.2,-0.15);
    \draw (-0.2,-0.15)-- (1.,-0.5);
    \draw (1.2,-0.15)-- (1.2,0.0);
    \draw (1.2,-0.15)-- (0,-0.8);
    \draw[densely dotted,thick] (-0.3,0.)-- (-0.3,-0.4);
    %%%%
    \draw[dashed] (4,0) -- (5,0);
    \draw (3.8,0.)-- (3.8,-0.15);
    \draw (3.8,-0.15)-- (5.,-0.5);
    \draw (5.2,-0.15)-- (5.2,0.);
    \draw (5.2,-0.15)-- (4,-0.5);
    %%%%
    \draw[dashed] (8,0) -- (9,0);
    \draw (7.8,0.)-- (7.8,-0.15);
    \draw (7.8,-0.15)-- (9.,-0.5);
    \draw (9.2,-0.15)-- (9.2,0.);
    \draw (9.2,-0.15)-- (8,-0.5);
    %%%%
    \draw[dashed] (12,0) -- (13,0);
    \draw (11.8,0.)-- (11.8,-0.15);
    \draw (11.8,-0.15)-- (13.,-0.7);
    \draw (13.2,-0.15)-- (13.2,0.);
    \draw (13.2,-0.15)-- (12,-0.5);
     %%%% contractions
    \draw[densely dotted,thick] (1.3,-0.1)--(2,-0.1);
    \draw[densely dotted,thick] (3,-0.1)--(3.7,-0.1);
    \draw[densely dotted,thick] (5.3,-0.1)--(6,-0.1);
    \draw[densely dotted,thick] (7,-0.1)--(7.7,-0.1);
    \draw[densely dotted,thick] (9.3,-0.1)--(10,-0.1);
    \draw[densely dotted,thick] (11,-0.1)--(11.7,-0.1);
    \draw (1.0,-0.5)-- (2,-0.5);
    \draw (3.0,-0.5)-- (4,-0.5);
    \draw (5.0,-0.5)-- (6,-0.5);
    \draw (7.0,-0.5)-- (8,-0.5);
    \draw (9.0,-0.5)-- (10,-0.5);
    \draw (11.0,-0.5)-- (12,-0.5);
     \draw[densely dotted,thick] (13.3,-0.4)-- (13.3,0.);
    %%%% average
    \draw[densely dotted,thick] (3.7,0) arc (0:180:1.2);
    \draw[densely dotted,thick] (7.7,0) arc (0:180:1.2);
    \draw[densely dotted,thick] (11.7,0) arc (0:180:1.2);
    \draw[densely dotted,thick] (13.3,0) arc (0:180:6.8);
    \draw[dashed] (4.,0) arc (0:180:1.5);
    \draw[dashed] (8.,0) arc (0:180:1.5);
    \draw[dashed] (12.,0) arc (0:180:1.5);
    \draw[dashed] (13.,0) arc (0:180:6.5);    
    % \draw (3.2,0) arc (0:180:0.7);
    \draw (3.8,0) arc (0:180:1.3);
    \draw (7.8,0) arc (0:180:1.3);
    \draw (11.8,0) arc (0:180:1.3);
    \draw (13.2,0) arc (0:180:6.7);
    %%%%%
    \draw (2.5,-0.3) circle (.52) node[anchor=center] {\footnotesize$G$};
    \draw (6.5,-0.3) circle (.52) node[anchor=center] {\footnotesize$G$};
    \draw (10.5,-0.3) circle (.52) node[anchor=center] {\footnotesize$G$};
    %%%%
    \draw (-0.,-1.2) circle (0.) node[anchor=center] {\scriptsize $q_2$};
    \draw (-0.7,-0.5) circle (0.) node[anchor=center] {\scriptsize $q_1$};
    \draw (1.65,-0.85) circle (0.) node[anchor=center] {\scriptsize $\bar{q}_1$};
    \draw (2,0.5) circle (0.) node[anchor=center] {\tiny $\bar{q}_2$};
    \draw (6.,0.5) circle (0.) node[anchor=center] {\tiny$q_1$};
    \draw (10.,0.5) circle (0.) node[anchor=center] {\tiny$\bar{q}_2$};
 } 
 \ \, + \cdots \\
  &+  \ \, 
  \tikz[scale=0.32,baseline=2ex]{
    \draw[dashed] (0,0) -- (1,0);
    \draw (-0.2,0.)-- (-0.2,-0.15);
    \draw (-0.2,-0.15)-- (1.,-0.5);
    \draw (1.2,-0.15)-- (1.2,0.0);
    \draw (1.2,-0.15)-- (0,-0.8);
    \draw[densely dotted,thick] (-0.3,0.)-- (-0.3,-0.4);
    %%%%
    \draw[dashed] (4,0) -- (5,0);
    \draw (3.8,0.)-- (3.8,-0.15);
    \draw (3.8,-0.15)-- (5.,-0.5);
    \draw (5.2,-0.15)-- (5.2,0.);
    \draw (5.2,-0.15)-- (4,-0.5);
    %%%%
    \draw[dashed] (8,0) -- (9,0);
    \draw (7.8,0.)-- (7.8,-0.15);
    \draw (7.8,-0.15)-- (9.,-0.7);
    \draw (9.2,-0.15)-- (9.2,0.);
    \draw (9.2,-0.15)-- (8,-0.5);
     %%%% contractions
    \draw[densely dotted,thick] (1.3,-0.1)--(2,-0.1);
    \draw[densely dotted,thick] (3,-0.1)--(3.7,-0.1);
    \draw[densely dotted,thick] (5.3,-0.1)--(6,-0.1);
    \draw[densely dotted,thick] (7,-0.1)--(7.7,-0.1);
    \draw (1.0,-0.5)-- (2,-0.5);
    \draw (3.0,-0.5)-- (4,-0.5);
    \draw (5.0,-0.5)-- (6,-0.5);
    \draw (7.0,-0.5)-- (8,-0.5);
     \draw[densely dotted,thick] (9.3,-0.4)-- (9.3,0.);
    %%%% average
    \draw[densely dotted,thick] (3.7,0) arc (0:180:1.2);
    \draw[densely dotted,thick] (7.7,0) arc (0:180:1.2);
    \draw[densely dotted,thick] (9.3,0) arc (0:180:4.8);
    \draw[dashed] (4.,0) arc (0:180:1.5);
    \draw[dashed] (8.,0) arc (0:180:1.5);
    \draw[dashed] (9.,0) arc (0:180:4.5);    
    \draw (3.8,0) arc (0:180:1.3);
    \draw (7.8,0) arc (0:180:1.3);
    \draw (9.2,0) arc (0:180:4.7);
    %%%%%
    \draw (2.5,-0.3) circle (.52) node[anchor=center] {\footnotesize$G$};
    \draw (6.5,-0.3) circle (.52) node[anchor=center] {\footnotesize$G$};
    %%%%
    \draw (-0.,-1.2) circle (0.) node[anchor=center] {\scriptsize $\bar{q}_1$};
    \draw (-0.7,-0.5) circle (0.) node[anchor=center] {\scriptsize $q_1$};
    \draw (2,0.5) circle (0.) node[anchor=center] {\tiny $q_1$};
    \draw (6.,0.5) circle (0.) node[anchor=center] {\tiny$q_1$};
}
 \ \, + \ \,
 \tikz[scale=0.28,baseline=2ex]{
    \draw[dashed] (0,0) -- (1,0);
    \draw (-0.2,0.)-- (-0.2,-0.15);
    \draw (-0.2,-0.15)-- (1.,-0.5);
    \draw (1.2,-0.15)-- (1.2,0.0);
    \draw (1.2,-0.15)-- (0,-0.8);
    \draw[densely dotted,thick] (-0.3,0.)-- (-0.3,-0.4);
    %%%%
    \draw[dashed] (4,0) -- (5,0);
    \draw (3.8,0.)-- (3.8,-0.15);
    \draw (3.8,-0.15)-- (5.,-0.5);
    \draw (5.2,-0.15)-- (5.2,0.);
    \draw (5.2,-0.15)-- (4,-0.5);
    %%%%
    \draw[dashed] (8,0) -- (9,0);
    \draw (7.8,0.)-- (7.8,-0.15);
    \draw (7.8,-0.15)-- (9.,-0.5);
    \draw (9.2,-0.15)-- (9.2,0.);
    \draw (9.2,-0.15)-- (8,-0.5);
    %%%%
    \draw[dashed] (12,0) -- (13,0);
    \draw (11.8,0.)-- (11.8,-0.15);
    \draw (11.8,-0.15)-- (13.,-0.5);
    \draw (13.2,-0.15)-- (13.2,0.);
    \draw (13.2,-0.15)-- (12,-0.5);
        %%%%
    \draw[dashed] (16,0) -- (17,0);
    \draw (15.8,0.)-- (15.8,-0.15);
    \draw (15.8,-0.15)-- (17.,-0.7);
    \draw (17.2,-0.15)-- (17.2,0.);
    \draw (17.2,-0.15)-- (16,-0.5);
     %%%% contractions
    \draw[densely dotted,thick] (1.3,-0.1)--(2,-0.1);
    \draw[densely dotted,thick] (3,-0.1)--(3.7,-0.1);
    \draw[densely dotted,thick] (5.3,-0.1)--(6,-0.1);
    \draw[densely dotted,thick] (7,-0.1)--(7.7,-0.1);
    \draw[densely dotted,thick] (9.3,-0.1)--(10,-0.1);
    \draw[densely dotted,thick] (11,-0.1)--(11.7,-0.1);
    \draw[densely dotted,thick] (13.3,-0.1)--(14,-0.1);
    \draw[densely dotted,thick] (15,-0.1)--(15.7,-0.1);
    \draw (1.0,-0.5)-- (2,-0.5);
    \draw (3.0,-0.5)-- (4,-0.5);
    \draw (5.0,-0.5)-- (6,-0.5);
    \draw (7.0,-0.5)-- (8,-0.5);
    \draw (9.0,-0.5)-- (10,-0.5);
    \draw (11.0,-0.5)-- (12,-0.5);
    \draw (13.0,-0.5)-- (14,-0.5);
    \draw (15.0,-0.5)-- (16,-0.5);
     \draw[densely dotted,thick] (17.3,-0.4)-- (17.3,0.);
    %%%% average
    \draw[densely dotted,thick] (3.7,0) arc (0:180:1.2);
    \draw[densely dotted,thick] (7.7,0) arc (0:180:1.2);
    \draw[densely dotted,thick] (11.7,0) arc (0:180:1.2);
    \draw[densely dotted,thick] (15.7,0) arc (0:180:1.2);
    \draw[densely dotted,thick] (17.3,0) arc (0:180:8.8);
    \draw[dashed] (4.,0) arc (0:180:1.5);
    \draw[dashed] (8.,0) arc (0:180:1.5);
    \draw[dashed] (12.,0) arc (0:180:1.5);
    \draw[dashed] (16.,0) arc (0:180:1.5);
    \draw[dashed] (17.,0) arc (0:180:8.5);    
    \draw (3.8,0) arc (0:180:1.3);
    \draw (7.8,0) arc (0:180:1.3);
    \draw (11.8,0) arc (0:180:1.3);
    \draw (15.8,0) arc (0:180:1.3);
    \draw (17.2,0) arc (0:180:8.7);
    %%%%%
    \draw (2.5,-0.3) circle (.52) node[anchor=center] {\scriptsize $G$};
    \draw (6.5,-0.3) circle (.52) node[anchor=center] {\scriptsize $G$};
    \draw (10.5,-0.3) circle (.52) node[anchor=center] {\scriptsize $G$};
    \draw (14.5,-0.3) circle (.52) node[anchor=center] {\scriptsize $G$};
    %%%%
    \draw (-0.,-1.4) circle (0.) node[anchor=center] {\scriptsize $\bar{q}_1$};
    \draw (-0.8,-0.5) circle (0.) node[anchor=center] {\scriptsize $q_1$};
    \draw (2,0.5) circle (0.) node[anchor=center] {\tiny $q_1$};
    \draw (6.,0.5) circle (0.) node[anchor=center] {\tiny$q_1$};
    \draw (10.,0.5) circle (0.) node[anchor=center] {\tiny$q_1$};
    \draw (14.,0.5) circle (0.) node[anchor=center] {\tiny$q_1$};
 } 
 \ \, + \cdots \ .
\end{aligned}
\end{equation}

The self energy, as a result, obeys the following equation:
\begin{equation}\label{eq:self_energy_general}
\begin{aligned}
    \Sigma = \bigoplus_{q_1}\ \ \, &  \ \ \mathbb{1}_{A_1,q_1} \otimes \mathbb{1}_{A_2,\bar{q}_1}  \,   \frac{1}{L_{q_A}} \left[  1 + \alpha_{q_1 \bar{q}_1 } G_{2,q_1} \right] \, 
    \frac{1}{1 - \beta_{q_1}^2 G_{2,q_1}^2 }
    \\
     \oplus  \bigoplus_{q_1,q_2\neq \bar{q}_1} &  \ \ \mathbb{1}_{A_1,q_1} \otimes \mathbb{1}_{A_2,q_2} 
     \frac{\alpha_{\bar{q}_2\bar{q}_1}}{L_{q_A}} \, 
     \frac{G_{1, \bar{q}_2 \bar{q}_1 } }{1 - \beta_{q_1} \beta_{\bar{q}_2} \, G_{1,q_1 q_2} G_{1, \bar{q}_2 \bar{q}_1 } } ,
\end{aligned}    
\end{equation}
where we have defined another family of parameters as $\beta_{q_1} = \frac{ L_{A_1,q_1} }{ L_{q_A} L_{q_B} }$. Putting these back into Eq.~\eqref{eq:G_in_terms_of_sigma}, one can derive the self-consistent equations for the resolvent function.

\subsubsection{$\mathbb{Z}_R$ symmetry}

Let us start with the case of the symmetry group $\mathbb{Z}_R$:
\begin{equation}\label{eq:cubic_eq_ZR}
\begin{aligned}
    & G_1^3 \, \frac{z}{R^{2\left( N_{A_2}+N_B-1 \right)}} 
    + G_1^2 \left[ \frac{1}{R^{N-1}} - \frac{1}{R^{2\left( N_{A_2}+N_B-1  \right)}} \right]
    - G_1  z 
    + 1 = 0, \\
    & G_2^3 \, \frac{z}{R^{2\left( N_{A_2}+N_B-1 \right)}} 
    + G_2^2 \left[ \frac{1}{R^{N-1}} - \frac{1}{R^{2\left( N_{A_2}+N_B-1  \right)}} \right]
    + G_2 \left[ \frac{1}{R^{N_A-1}}  - z \right]
    + 1 = 0.
\end{aligned}    
\end{equation}

\end{widetext}

\begin{figure}
    \centering
    \includegraphics[scale=1.05]{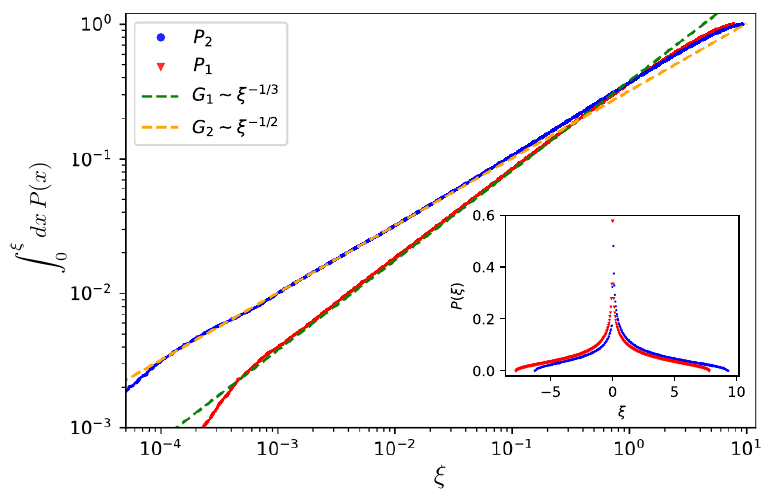}
    \caption{Critical exponents for the negativity spectrum at the transition point in $\Z_3$-symmetric states. For a better numerical accuracy, we calculate the cumulative distribution function instead of the spectral density. $N_{A_1}=2, N_{A_2}=6$, and $N_B = 5$. Inset shows the spectral density in linear scale.}
    \label{fig:Z3-exp}
\end{figure}

One can see from the above that both $G_1$ and $G_2$ diverge at $z=0$ if:
\begin{equation}
\begin{aligned}
    R^{N-1} = R^{2 \left( N_{A_2} + N_B - 1 \right)} \ \Rightarrow N_{A_1} = N_{A_2} + N_B - 1.
\end{aligned}    
\end{equation}
The leading divergence for $G_1$ and $G_2$ can be read from the cubic equations:
\begin{equation}
    G_1 \sim \frac{i}{z^{1/3}}, \qquad G_2 \sim \frac{i}{z^{1/2}}.
\end{equation}
This limit corresponds to the transition between the semicircle and the maximally entangled regimes similar to the non-symmetric states~\cite{Shapourian2021}, albeit the transition in the latter case occurs at a different point $N_{A_1}=N_{A_2}+N_B$. In addition, in contrast to the $\Z_R$ symmetry above, the critical exponent in the non-symmetric case is $1/2$. We confirm the critical exponents in the case of $\Z_3$ symmetric states in Fig.~\ref{fig:Z3-exp}.

Next, we discuss the behavior of the entanglement negativity on the two sides of this transition line. We have seen in the previous part (c.f.~Eqs.~(\ref{eq:N1-Zr-sc})-(\ref{eq:N2-Zr-sc})) that for $N_{A_s} < N_{A_{\bar{s}}} + N_B - 1$ the entanglement negativity shows a plateau as $N_{A_1}$ and $N_{A_2}$ are varied and $N_A$ is kept constant.
Using Eq.~(\ref{eq:cubic_eq_ZR}), one can work out the entanglement negativity in the limit where $N_{A_1} > N_{A_2} + N_B - 1$, i.e.~the regime of maximal entanglement in $A$ subsystem.

As shown in appendix \ref{app:negtivity_integration}, deep in the maximal entanglement regime, the imaginary parts of $G_1$ and $G_2$ for $z<0$ take the form:
\begin{equation}
    \mathrm{Im}(G_1) = \mathrm{Im}(G_2) = R^{N_{A_1}} \; \frac{\gamma^2}{2} \; \sqrt{-\left( R^{N_{A_1} } z + \frac{1}{\gamma} \right)^2 + \frac2\gamma}'
\end{equation}
where $\gamma = R^{-N_{A_1}+N_{A_2}+N_B - 1}$. This results in the following total value for the negativity in this regime:
\begin{equation}
    \langle \mathcal{N}(\hat\rho_A^{(q_A)}) \rangle = \frac{1}{2} R^{N_{A_2}}.
\end{equation}
This result is very similar to that in the non-symmetric case. In other words, when subsystem $A_1$ is much larger than its complement, the entanglement negativity between $A_1$ and $A_2$ is maximal and bounded by the volume of the smaller subsystem (in this case $A_2$).

\subsubsection{General symmetry group}
On the other hand, in general for an arbitrary symmetry group the self energy shown in \eqref{eq:self_energy_general}, results in the following equations for the $G_2$ component of the resolvent functions:
\begin{equation}\label{eq:general_SD_G2}
\begin{aligned}
     G_{2,q_1}^3 \ z \beta_{q_1}^2 +  G_{2,q_1}^2 \left( \frac{1}{L_{q_A}} \alpha_{q_1\bar{q}_1 } - \beta_{q_1}^2 \right) \\
     + G_{2,q_1} \left( \frac{1}{L_{q_A}} - z \right) + 1 = 0, 
\end{aligned}    
\end{equation}
while the equations for $G_1$ components read:
\begin{equation}\label{eq:general_SD_G1}
\begin{aligned}
     - z \, G_{1,q_1q_2} & \left( 1 - \beta_{q_1}\beta_{\bar{q}_2} \, G_{1,q_1 q_2} G_{1,\bar{q}_2 \bar{q}_1}  \right) \\
    & - G_{1,q_1 q_2} G_{1,\bar{q}_2 \bar{q}_1} \left( \beta_{q_1}\beta_{\bar{q}_2} - \frac{1}{L_{q_A}}\alpha_{\bar{q}_2\bar{q}_1} \right) + 1 = 0. 
\end{aligned}    
\end{equation}
One interesting property that can be derived from the above equations is the different behaviors and the criticality in the spectral density. We note that different components of $G_1$ and $G_2$ with different indices $q_i$ should be considered separately for this matter.

First, we study the $G_2$ component with a given quantum number index; it can be seen from \eqref{eq:general_SD_G2} that for $\alpha_{q_1\bar{q}_1} = L_{q_A} \beta_{q_1}^2$ or equivalently $L_{A_1,q_1} = L_{A_2,\bar{q}_1} L_{q_B}$, the spectral density diverges, and the leading divergence is given by $P_{2,q_1} \sim \frac{1}{\xi^{1/2}}$. By changing subsystem sizes while keeping the quantum numbers unchanged (when possible), one can check that the entanglement negativity shows two different behaviors on the two sides of this singular transition: for $L_{A_1,q_1} < L_{A_2,\bar{q}_1} L_{q_B}$ one is in the plateau regime and deep within that regime the entanglement negativity reads 
\begin{equation}
\mathfrak{n}_2(q_1) = \frac{4}{3\pi} \frac{1}{L_{q_A} \sqrt{L_{q_B}}}{ \left( L_{A_1,q_1 }  L_{A_2,\bar{q_1} }  \right)^{3/2} }
\end{equation}
(one should also have $\frac{L_{A_1,q_1 }  L_{A_2,\bar{q_1} }}{L_{q_B}}>1$ to avoid the PPT transition).
On the other hand, for $L_{A_1,q_1} > L_{A_2,\bar{q}_1} L_{q_B}$ one is in the maximal entanglement regime, where the negativity to the leading order is given by
\begin{equation}\label{eq:neg_sym_resolved_max_entanglement_G2}
    \mathfrak{n}_{2}(q_1)
    = \frac12 \frac{L_{A_1,q_1}  L_{A_2,\bar{q}_1} }{ L_{q_A}}  L_{A_2,\bar{q}_1}
\end{equation}
deep inside that regime (see the discussion in appendix \ref{app:negtivity_integration}).

One can check the above criterion for criticality in the case of the $U(1)$ symmetry. The relation $L_{A_1,q_1} = L_{A_2,\bar{q}_1} L_{q_B}$ takes the form \begin{align}\label{eq:criticiality_U(1)}
    N_{A_1} f(\nu_1) - N_{A_2} f(\bar{\nu}_1) = N_B f(\nu_B).
\end{align}
for the $U(1)$  symmetric case, up to subleading corrections.
Defining the ratios $r_1 = N_{A_1}/N_A$, $r_A = N_{A}/(N_A+N_B)$,
the equation takes the form 
\begin{equation}
    \begin{aligned}
        r_1 f(\nu_1) & -
        (1-r_1)  f\left(\frac{\nu_A-\nu_1 r_1}{1-r_1}\right)  \\
        &= \left(\frac{1}{r_A}-1\right) f\left(\frac{\nu-\nu_A r_A}{1-r_A}\right).
    \end{aligned}
\end{equation}
For a fixed partition given in term of the values of $r_1$ and $r_A$ and the filling fractions $\nu_A$ and $\nu$ (total filling fraction), the question of whether there is criticality in any of the sectors or not can be addressed by checking whether the above equation has a solution for $0\leq \nu_1\leq 1$.
Hence, the criticality is equivalent to the existence of a solution for $\nu_1$ subject to a linear constraint
\begin{align}
    \max \left(0, \frac{\nu_A-(1-r_1)}{r_1} \right) \leq \nu_1 \leq \min \left( \frac{1}{2}, \frac{\nu_A}{r_1} \right).
\end{align}
Using this, one can see that criticality is observed in a {\it critical region} as opposed to the non-symmetric states where there is a critical line; the extent of this critical region can be determined numerically. A few examples are shown in Fig.~\ref{fig:phasediag}.

Next, we turn to studying the $G_1$ component. Using the analogous equation of \eqref{eq:general_SD_G1} for $G_{1,\bar{q}_2\bar{q}_1}$, one can work out an equation solely containing $G_{1,q_1q_2}$:
\begin{equation}\label{eq:general_SD_G1_single_component}
\begin{aligned}
    & - G_{1,q_1q_2}^3 \; z^2 \alpha_{q_1 q_2}  \beta_{q_1}\beta_{\bar{q}_2}  \\
    & - G_{1,q_1q_2}^2 \; z \left( \alpha_{q_1q_2} \left[ \frac{\alpha_{\bar{q}_2\bar{q}_1}}{L_{q_A}}  - 2 \beta_{q_1} \beta_{\bar{q}_2} \right] + \alpha_{\bar{q}_2\bar{q}_1}   \beta_{q_1} \beta_{\bar{q}_2} \right) \\
    & + G_{1,q_1q_2} \left( z^2 \;  \alpha_{\bar{q}_2\bar{q}_1}  + \left[ \alpha_{q_1q_2}  - \alpha_{\bar{q}_2\bar{q}_1} \right] \left[  \frac{\alpha_{\bar{q}_2\bar{q}_1}}{L_{q_A}}  - \beta_{q_1} \beta_{\bar{q}_2} \right] \right) \\
    &- z \; \alpha_{\bar{q}_2\bar{q}_1}  = 0.
\end{aligned}    
\end{equation}
This equation shows that the only way for $G_{1,q_1q_2}$ (and as a result the spectral density) to have a nonzero value at $z=0$, is to have either 
$\alpha_{q_1q_2}  = \alpha_{\bar{q}_2\bar{q}_1}$ or
$\frac{\alpha_{\bar{q}_2\bar{q}_1}}{L_{q_A}}  = \beta_{q_1} \beta_{\bar{q}_2}$. 
One can simply check by imposing each of these two relations, that there is criticality in the system only when both of them are satisfied; under such conditions the spectral density diverges as $P_{1,q_1q_2} \sim \frac{1}{\xi^{1/3}}$. The conditions can also be written in terms of subspace dimensions as $L_{A_1,q_1} L_{A_2,q_2} = L_{A_1,\bar{q}_2} L_{A_2,\bar{q}_1}$ and $L_{A_1,q_1} = L_{A_2,q_2} L_{q_B}$. 

We note here that for the case of the $U(1)$ symmetry, the first of the above two criticality conditions can only be satisfied if $q_1 = \bar{q}_2$; since the indices on the $G_1$ contribution to the spectrum should strictly not satisfy this relation, therefore the latter type of criticality does not occur in $U(1)$ symmetric systems. As a result,  Eq.~\eqref{eq:criticiality_U(1)} solely determines the critical region for $U(1)$.
Note that in general, and with symmetries other than $U(1),$ this relation can be satisfied with $q_1 \neq \bar{q}_2$, such as in the $\mathbb{Z}_R$ case explained earlier.

\begin{figure*}
    \centering
    \includegraphics[scale=0.8]{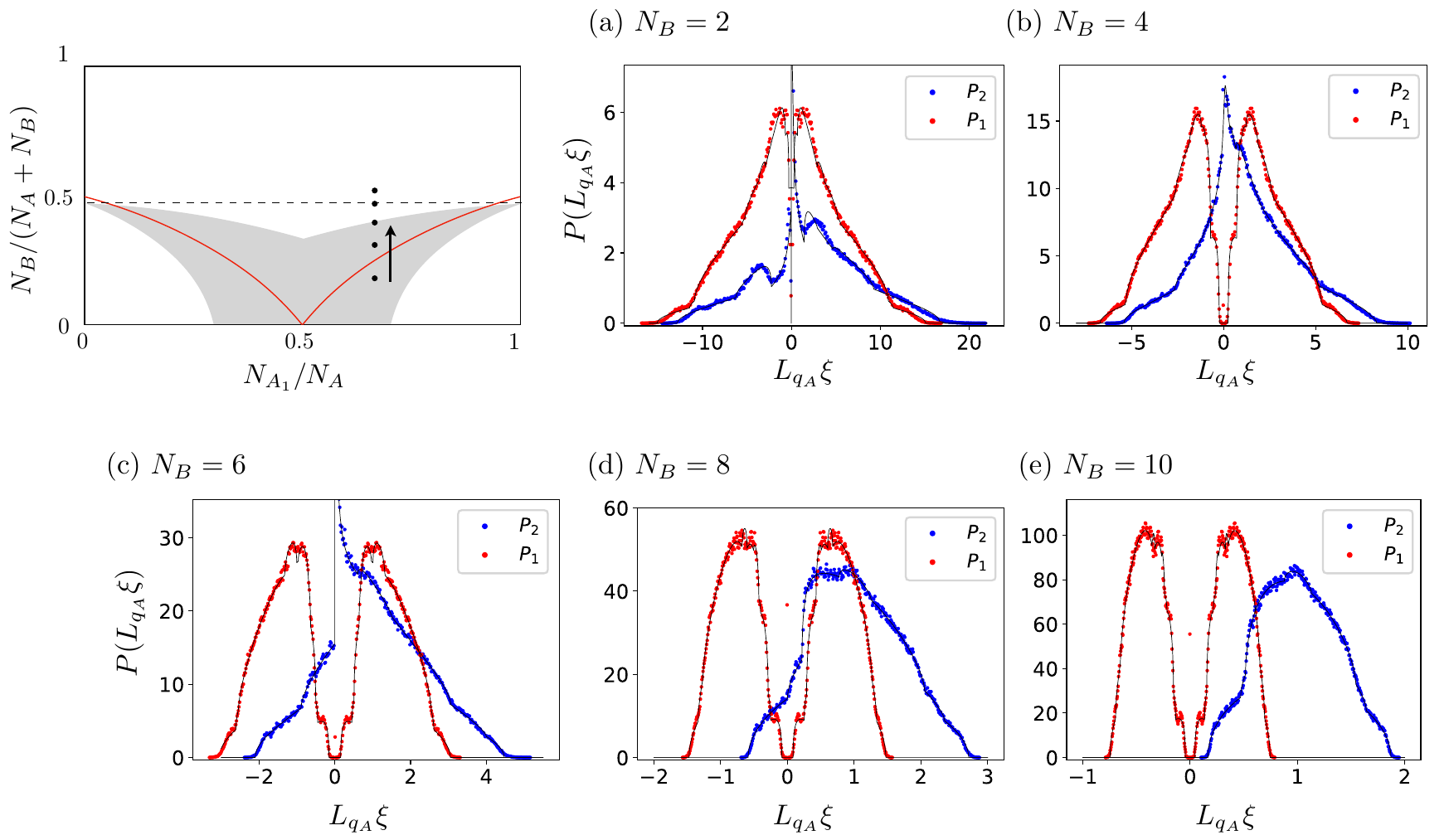}
    \caption{Evolution of the negativity spectrum as the size of subsystem $B$ is increased for $U(1)$ symmetric systems. This trend corresponds to sweeping a vertical path from bottom to top in the phase diagram as shown in the first panel. The colored circles in each panel are numerical simulations (averaged over $10^4$ samples) and solid lines correspond to the numerical solutions to Eqs.~(\ref{eq:general_SD_G1_single_component}) and ~(\ref{eq:general_SD_G2}) from random matrix theory. Here, we set $N_{A_2}=3$ and $N_{A_1}=6$ and the filling fractions are $\nu_A = \frac{1}{3}$ and $\nu_B = \frac{1}{2}$.
    Note that the points within the critical region (shaded area) of the phase diagram are characterized by a diverging spectral density at the origin (panels (a)-(c)).}
    \label{fig:NS_vs_Lb}
\end{figure*}

Figure~\ref{fig:NS_vs_Lb} shows some numerical simulations for various ratios of size of $A$ to that of $B$. The agreement between analytical results from the random matrix theory and the numerical results is evident. Importantly, we see in Fig.~\ref{fig:NS_vs_Lb}(a)-(c) that the $P_2$ contribution to the spectral density diverges at zero as explained earlier.

Apart from the criticality, one can also work out the contribtion to negativity from $G_1$ in the regime where the $A_1$ subsystem with the given quantum number constitutes more than half of the whole system given the specific quantum numbers in consideration. It can be shown using the solution of Eq.~\eqref{eq:general_SD_G1_single_component} that the symmetry resolved negativity is given by (see appendix \ref{app:negtivity_integration} for discussion):
\begin{equation}\label{eq:neg_sym_resolved_max_entanglement_G1}
    \mathfrak{n}_{1}(q_1,q_2)
    =
    \frac12 \frac{1}{L_{q_A}} \sqrt{L_{A_1,q_1} L_{A_1,\bar{q}_2}} \, L_{A_2,q_2} L_{A_2, \bar{q}_1}
\end{equation}

\subsubsection{Maximal entanglement for the $U(1)$ symmetry in thermodynamic limit}
We now turn to calculating the total negativity in the regime of maximal entanglement for a $U(1)$ symmetric system in the thermodynamic limit. This analysis is complementary to the one presented in Sec.~\ref{sec:thermo_limit_U(1)_plateau}, i.e.~together with the previous result we find the explicit form of the negativity to the leading order in the lower part of the phase diagram (c.f.~Figs.~\ref{fig:phasediag} and \ref{fig:NS_vs_Lb}).

We first note that in the thermodynamic limit and using a continuum approximation, the two contributions to the negativity (as in Eq.~(\ref{eq:neg-u1-decomp})) can be written as:
\begin{equation}\label{eq:negativity_saddle}
\begin{aligned}
    \mathfrak{n}_{1}(\nu_1,\nu_2) & =  \frac{ \left(L_{A_1,\nu_1} \, L_{A_1,\bar{\nu}_2}  \right)^{1/2} 
    L_{A_2,\nu_2} \, L_{A_2,\bar{\nu}_1}
    }{2 \, L_{\nu_A}} \\
    \mathfrak{n}_{2}(\nu_1) & =   \frac{ L_{A_1,\nu_1} \,
    L_{A_2,\bar{\nu}_1}^2
    }{2 \, L_{\nu_A}}  ,
\end{aligned}    
\end{equation}
where we make use of Eqs.~\eqref{eq:neg_sym_resolved_max_entanglement_G2} and \eqref{eq:neg_sym_resolved_max_entanglement_G1}. Note that by definition $N_{A_1} \nu_1 + N_{A_2}\bar\nu_1 = N_A \nu_A$. In order to find the saddle point solution, we write the leading exponential functionalities of $\mathfrak{n}_{1}$ and $\mathfrak{n}_{2}$ as follows:
\begin{equation}
\begin{aligned}
    \mathfrak{n}_{1} &\sim  
    \frac{
    e^{\left( \frac12 N_{A_1} f(\nu_1) + N_{A_2} f(\bar\nu_1) \right) +
    \left( \frac12 N_{A_1} f(\bar\nu_2) + N_{A_2} f(\nu_1) \right)}
    }{
    e^{N_A f(\nu_A)}
    }
    \\
    \mathfrak{n}_{2} &\sim 
    \frac{
    e^{N_{A_1} f(\nu_1) + 2 N_{A_2} f(\bar\nu_1)}
    }{
    e^{N_A f(\nu_A)}
    }
\end{aligned}
\end{equation}
where the function $f$ is defined in Eq.\eqref{eq:definition_f}. After a straightforward calculation, we find that both functions reach their maximum value at $\nu_1$ satisfying the relation:
\begin{equation}
    \log\left( \frac{1-\nu_1}{\nu_1}  \right)  = 2 \log \left( \frac{1-r_1 -  \nu_A +r_1 \nu_1}{ \nu_A - r_1 \nu_1}  \right).
\end{equation}
We find the solution by taking $\nu_1$ to have a form as $\nu_A + \delta \nu$ and assuming $\delta \nu $ to be small. This results in the following form for the saddle point value $\nu_1 = \nu_A - %\frac1{1+2 \zeta }
\frac{1-r_1}{1+r_1} \, \nu_A (1-\nu_A) \log(\frac{1-\nu_A}{\nu_A})$. 

Since at the saddle point $\nu_1 = \bar\nu_2$ and $\nu_2=\bar\nu_1$, the maximum values of $\mathfrak{n}_1$ and $\mathfrak{n}_2$ take the same form. This leads to the following form for the dominant contribution to the logarithmic negativity:
\begin{equation}
    \langle \mathcal{E}(\hat\rho_A^{(q_A)}) \rangle =
    N_{A_1} f(\nu_1) + 2 N_{A_2} f(\bar{\nu}_1) - N_A f(\nu_A)
\end{equation}
where $\nu_1$ and $\bar{\nu}_1$ should have their saddle point values. Exploiting the above form for the saddle point value of $\nu_1$, one arrives at the following dominant contribution to the logarithmic negativity:
\begin{equation}
\label{eq:logneg-u1-me}
\langle \mathcal{E}(\hat\rho_A^{(q_A)}) \rangle =
    N_{A_2} f(\nu_A) + N_{A_1} \frac{\nu_A (1-\nu_A)  
    \left[\log\left(\frac{1-\nu_A}{\nu_A} \right)\right]^2}
    {1+2N_{A_1}/N_{A_2}}.
\end{equation}
The second term is a correction to the first one which we get from the perturbative expansion for the saddle point value of $\nu_1$.

\section{Quantum circuit measuring the symmetry charge}
\label{sec:experiment}
In this section, we discuss a way in which the total charge in subsystem $B$ can be measured without totally collapsing the state of $B$ to a pure state. 
The output state can then be used to calculate the symmetry resolved R\'enyi entropy or R\'enyi negativity using the already established protocols in circuit-based quantum computer~\cite{Hastings2010,Johri2017,Yirka2021} or trapped ions~\cite{Elben2020}.
We present the case of the $\Z_2$ symmetry here. $\mathbb{Z}_R$ and $U(1)$ symmetry groups are similar and will be discussed in Appendix~\ref{app:measurement}.

\begin{figure}[!h]
    \centering
    \begin{quantikz}
    \lstick{$\ket{0}$} & \gate{H} & \ctrl{1} & \ctrl{2} & \ctrl{3} & \gate{H} & \meter{}\\
     \lstick[wires=3]{$B \ $} & \qw & \ctrl{-1} & \qw & \qw & \qw & \qw \\
     & \qw & \qw & \ctrl{-2} & \qw & \qw & \qw \\
     & \qw & \qw & \qw & \ctrl{-3} & \qw & \qw \\
     \lstick[wires=4]{$A \ $}& \qw & \qw & \qw & \qw & \qw & \qw \\
     & \qw & \qw & \qw & \qw & \qw & \qw \\
     & \qw & \qw & \qw & \qw & \qw & \qw \\
     & \qw & \qw & \qw & \qw & \qw & \qw 
    \end{quantikz}
    \caption{Quantum circuit to measure the $\Z_2$ charge of subsystem $B$.}
    \label{fig:circuit_measuring_party}
\end{figure}
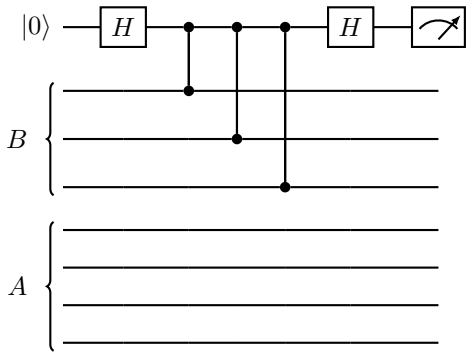

One needs to add one ancillary qubit to the system; it is initialized in the $\left|0\right\rangle$ state and then a Hadamard gate is applied to this qubit which turns its state to $\frac{\left|0\right\rangle + \left|1\right\rangle}{\sqrt{2}}$. Then, one acts with two-qubit control-$Z$ operators on the ancilla and all the qubits in the $B$ subsystem one by one. This entangles the ancilla with the $B$ subsystem; acting with another Hadamard gate on the ancilla and then measuring it in the computational basis, depending on the outcome of the measurement the $\Z_2$ charge of the $B$ subsystem is determined. This same construction is known to be capable of parity measurment in a system, note that the parity of $B$ is exactly its $\Z_2$ charge.

Let us now briefly discuss the cases of $\mathbb{Z}_R$ and $U(1)$ symmetry groups. For the case of $\mathbb{Z}_R$ symmetry, the same circuit when all the gates are generalized to qudit gates with $d=R$ could be used to measure the total charge of $B$. On the other hand, for the case of having a $U(1)$ symmetry within a qubit system, in order to measure the charge of $B$, one needs to perform a series of measurements similar to the one descibed above. Concretely, one first measures the parity of the $B$ subsystem charge using the same circuit as the one described above. Then, using similar circuits consisting of entangling gates and measurements one continues to determine the $B$ charge mod $4$, $8$, etc. A total number of $\lceil \log_2 (N_B + 1) \rceil$ measurements suffices to determine the charge in the $B$ subsystems, where $N_B$ is the number of qubits.
More details can be found in Appendix~\ref{app:measurement}.

\section{Discussion}
\label{sec:discussion}

In conclusion, we studied the entanglement negativity of symmetric random mixed states.
As we explained, the PPT criterion needs to be modified in the case of symmetric systems when only symmetric LOCCs are permitted. We introduced a more refined version of the entanglement negativity in terms of the average over the entanglement negativity associated with each symmetry charge sector (i.e., density matrix projected into a given symmetry quantum number). Our proposed quantity correctly captures a subset of separable states generated by symmetric LOCCs. Therefore, for the most of the paper, we focused on 
the entanglement negativity of a random mixed state with a fixed symmetry charge $\hat \rho_A^{(q_A)}$, which we call symmetry resolved (or projected) entanglement negativity.
To calculate this quantity, we generalized the diagrammatic approach for the partial transpose~\cite{Shapourian2021} to symmetric states and showed that charge conservation imposes several constraints on the diagrams. 
It turned out that the diagrams can be grouped into two types and calculations can be done in a systematic manner. 
As a result, we were able to fully characterize symmetry projected mixed states in terms of their entanglement negativity spectrum.
We illustrated our predictions via two examples of $\Z_R$ and $U(1)$ symmetry groups and explicitly derived the entanglement phase diagram in these cases.
The general structure of the phase diagram is similar to that of non-symmetric states, however, there are two notable differences.  
First, strictly speaking there is no PPT regime for $\hat \rho_A^{(q_A)}$. This property is manifest in the spectral density of $(\hat \rho_A^{(q_A)})^{T_2}$ as follows: We found that the spectral density is a sum of two distribution functions one of which is centered around zero; hence, $(\hat \rho_A^{(q_A)})^{T_2}$ has always some negative eigenvalues. Second, the critical line between the maximal entanglement and the replica symmetry breaking phases, where the spectral density diverges, may broaden into a critical phase. Furthermore, the divergence exponent is $1/2$ and $1/3$ for the  two contributions to the spectral density as opposed to the non-symmetric case where the critical exponent is $1/2$.
Finally, we designed a quantum circuit to perform the symmetry charge projection which can eventually be used to simulate the symmetry resolved entanglement negativity.

For the most part in this paper, we focused on characterizing the symmetry projected states $\hat\rho_A^{(q_A)}$. This result can then be used to compute the symmetry averaged entanglement negativity (\ref{eq:logneg-symmetry}) for randomly distributed  mixed states. Alternatively, one can use simple arguments to find the leading order contribution to this quantity.  We note that the Born weights $p_{q_A}$ in (\ref{eq:logneg-symmetry}) are proportional to the Hilbert space dimensions of symmetry charge sectors. This clearly needs to be determined case by case. For instance, in the case of $\Z_R$ symmetry both $p_{q_A}$ and $\braket{{\cal E}(\hat \rho_A^{(q_A)})}$ are independent of $q_A$ and we get $\overline{\cal E}_{A_1:A_2} = \braket{{\cal E}(\hat \rho_A^{(q_A)})}$. In contrast, the Hilbert space dimensions of various sectors of $U(1)$ symmetry charge depends on the filling factor, $p_{q_A}\sim e^{N_A f(\frac{q_A}{N_A})}$ where $f(.)$ is the Shannon entropy (\ref{eq:definition_f}); consequently, the sum (\ref{eq:logneg-symmetry}) is exponentially dominated by the sector having a homogeneous charge distribution among subsystems, i.e.~$\frac{q_A}{N_A} = \frac{Q}{N_A+N_B}$.

Throughout this paper, we discussed the entanglement negativity and its variants. It is worth comparing this quantity with other entanglement proxies such as the mutual information, albeit the latter is not technically an entanglement measure. As we have shown in Appendix~\ref{app:mutual information}, the mutual information to the leading order matches with that of the logarithmic  negativity through the usual relation ${\cal E}_{A_1:A_2} \sim \frac{1}{2} I_{A_1:A_2}$ in the case of $\Z_R$ symmetry group. In contrast, for $U(1)$ symmetric states, this relation does not hold (even to the leading order) in the maximal entanglement phase (compare Eqs.~(\ref{eq:logneg-u1-me}) and (\ref{eq:muinf-u1-me})). The physical significance of this difference remains to be understood. The lowest order corrections, however, are different between the mutual information and logarithmic negativity over the entire phase diagram. In particular, the mutual information is constant deep in the PPT regime, while the logarithmic negativity is exponentially small. This contrast is the result of the fact that classical correlations contribute to the mutual information, whereas they do not contribute to the logarithmic negativity.

Apart from the notion of mixed state entanglement measures in symmetric systems and incorporating global symmetry constraints in random matrix theory, our analysis is relevant to other physical systems as follows: Our $U(1)$ case study is applicable to a system of interacting 
complex Sachdev-Ye-Kitaev dots~\cite{Kitaev-talk,Gu2020}, where the hopping matrix is negligible and local filling fraction is fixed. Another example is a particle-number conserving system coupled to a bath in a canonical ensemble such that it can only exchange energy but not particle.

As we showed in this paper, a careful analysis of multipartite entanglement in symmetric systems requires introducing new entanglement measures. However, the block diagonal structure of the reduced density matrix is quite generic and not limited to symmetric systems. For instance, a similar constraint on local energy density appears in a tripartite state described by a microcanonical ensemble; in other words, we have $\sum_s N_s \epsilon_s = $const.~where $N_s$ and $\epsilon_s$ denote the volume and local energy density of subsystem $s$, respectively. In such scenarios, it would makes more sense to directly calculate the logarithmic negativity ${\cal E}(\hat\rho_A)$ rather than its symmetry averaged version. In principle, the formalism developed in this paper is applicable to this setup upon substituting the Hilbert space dimension $L_{q_A}$ by density of states $D_A(\epsilon_A)= e^{S_A(\epsilon_A)}$ using the Boltzmann's entropy formula, where $S_A(\epsilon_A)$ denotes the thermodynamic entropy at energy $\epsilon_A$~\cite{murthy2019structure, Grover2020}.
However, we realized that using our method to calculate the resolvent function in this case leads to a set of coupled non-linear Schwinger-Dyson equations where finding a closed-form solution is rather tedious (if not impossible), see Appendix~\ref{app:unprojected} for details. Nevertheless, one can numerically solve this set of equations. We postpone this analysis to a future publication. Alternatively, it may make sense in certain physical systems to approximate the three-body condition by reducing it to a two-body condition such as $N_{A_1}\epsilon_1 + N_{A_2}\epsilon_2= \text{const.}$ (where our method in this paper is directly applicable) or $N_{A_1}\epsilon_1 + N_{B}\epsilon_B= \text{const.}$ (see e.g., Ref.~\cite{Vardhan2021}). 

Let us wrap up our discussion with a few more directions for future research.
We considered Abelian global symmetry groups in this work. It would be interesting to generalize this formalism to non-Abelian symmetry groups and systems with local symmetry constraints such as in the gauge theories and anyon chains~\cite{PhysRevLett.124.050602} and possibly pinpoint the differences.
Recently, the entanglement negativity of random tensor networks~\cite{Dong-Qi-2021,Dong2021,KudlerFlam2021_neg} was investigated. Given the dramatic effects of global symmetries on entanglement properties of single tensors (which was studied in the current work), it may be worth exploring which universal properties in random tensor networks would change in the presence of global symmetries.

\acknowledgements

We acknowledge insightful discussions with Shang Liu, Ashvin Vishwanath, Pasquale Calabrese, Jonah Kudler-Flam, Hong Liu, Shreya Vardhan, Leon Balents, Chaitanya Murthy, Chetan Nayak, Bela Bauer, and Andreas Ludwig. K.H. in particular appreciates numerous fruitful discussions with Razieh Mohseninia.
K.H. was supported by the NSF CMMT program under Grants No. DMR-2116515.

%%%%%%%%%%%%%%%%%%%%%%%%%%%%%%%%%%%%%%%%%%%%%%%%%%%%%

\appendix

\section{Calculation of negativity deep in each phase}\label{app:negtivity_integration}

\renewcommand\theequation{B\arabic{equation}}

In this appendix, we discuss how the two components $\mathfrak{n}_1(q_1,q_2)$ and $\mathfrak{n}_2(q_1)$ are calculated deep in the maximal entanglement regime.

We start with $\mathbb{Z}_R$.
Using Eq.~(\ref{eq:cubic_eq_ZR}), one can work out the entanglement negativity in the limit where $N_{A_1} > N_{A_2} + N_B - 1$, i.e.~the regime of maximal entanglement in $A$ subsystem; 
In this limit, the set of equations in \eqref{eq:cubic_eq_ZR} can be rewritten as:
\begin{equation}
\begin{aligned}
   & \left( G_1 z - 1 \right)  \  \left(  G_1^2 \frac{1}{ R^{2(N_{A_2}+N_B-1)} } -1  \right) 
    + G_1^2 \frac{1}{R^{N-1}} = 0, \\
   & \left( G_2 z - 1 \right)  \  \left(  G_2^2 \frac{1}{ R^{2(N_{A_2}+N_B-1)} } -1  \right) \\
   & \qquad + G_2^2 \frac{1}{R^{N-1}}  + G_2 \frac{1}{R^{N_A-1}}= 0.
\end{aligned}   
\end{equation}
A redefinition of $G_i = R^{N_{A_1}} \tilde{G}_i$ along with $z=\tilde{z} \, R^{-N_{A_1}}$ results in:
\begin{equation}
\begin{aligned}
    &\left[\left( \tilde G_1 \tilde z - 1 \right)  \  \left(  \tilde G_1^2 \frac{1}{ \gamma^2 } -1  \right) \right]
    + \tilde G_1^2 \frac{1}{\gamma} = 0, \\
    &\left[ \left( \tilde G_2 \tilde z - 1 \right)  \  \left(  \tilde G_2^2 \frac{1}{ \gamma^2 } -1  \right) \right]
    + \tilde G_2^2 \frac{1}{ \gamma }  +  \frac{\tilde G_2}{R^{N_{A_2}-1}} = 0.
\end{aligned}   
\end{equation}
where $\gamma = \frac{\alpha}{\beta^2 L_A} = R^{-N_{A_1}+N_{A_2}+N_B - 1}$ is used in accordance with our previous definition of it. For the case of $\mathbb{Z}_R$, the charge indices of $\alpha,\beta, L_{A}$ are dropped. Note that $\gamma$ is a small number in the maximal entanglement regime, and as a result the terms outside the square brackets are subleading in both equations. The three solutions for $\tilde G_1$ and $\tilde G_2$ are simply found to the leading order: 
\begin{equation}
    \tilde G_i = 1/\tilde z, \ \pm \sqrt{\gamma} \,.
\end{equation}
We are interested in finding the imaginary parts of $\tilde G_i$ for negative values of $\tilde z$; as it turns out taking subleading terms into account when two, out of three, of the above roots are close to each other. As a result, with the substitutions $\tilde G_i = -\gamma + \delta G$ and $\tilde z = -\frac1\gamma + \delta z$, we expand the equations for the new variable $\delta G$ and $\delta z$ and to leading order, we get the following relation for the imaginary parts of $\delta G_i$ for $\tilde{z}<0$:
\begin{equation}
    \mathrm{Im}(\delta G_1) = \mathrm{Im}(\delta G_2) = \frac{\gamma^2}{2} \; \sqrt{-\delta z^2 + \frac2\gamma}'
\end{equation}

As a result, the imaginary parts of $G_1$ and $G_2$ for $z<0$ read:
\begin{equation}
    \mathrm{Im}(G_1) = \mathrm{Im}(G_2) = R^{N_{A_1}} \; \frac{\gamma^2}{2} \; \sqrt{-\left( R^{N_{A_1} } z + \frac{1}{\gamma} \right)^2 + \frac2\gamma}'
\end{equation}

This can be used to do the integral in Eq.~\eqref{eq:neg_dist} to obtain:
\begin{equation}
    \mathfrak{n}_1 =  \mathfrak{n}_2 = \frac12 R^{N_{A_2} - 2}.
\end{equation}
These correspond to single charge sectors, although the charge indices are dropped. Taking the effect of all charge sectors into account, we arrive at:
\begin{equation}
    \langle \mathcal{N}(\hat\rho_A^{(q_A)}) \rangle = \frac{1}{2} R^{N_{A_2}}.
\end{equation}

\bigskip

A general symmetry group can also be considered and very similar manipulations as above can be performed to calculate the negativity. The results for a general symmetry group are shown in the main text.

\section{Mutual information calculations}
\label{app:mutual information}

\renewcommand\theequation{C\arabic{equation}}

We will be calculating mutual information between $A_1$ and $A_2$, i.e.
\begin{equation}
    \left\langle I_{A_1,A_2}\right\rangle = \left\langle S_{A_1} \right\rangle + \left\langle S_{A_2} \right\rangle - \left\langle S_{A} \right\rangle
\end{equation}
for different settings in this appendix. For mutual information, similar to the cases in the main text, one needs {\it symmetry resolved} quantities and in particular in this case the von Neumann entanglement entropy. The latter quantity can be computed by taking the replica limit of the Renyi entropy as follows:
\begin{equation}
    S_{s} = \lim_{n \to 1} \frac1{1-n} \log S^{(n)}_s,
\end{equation}
where $s$ stands for either of $A_1,A_2,A$. With the same setting as the main text, where the symmetry charge of $A$ and $B$ are fixed to have values $q_A, q_B$, we can write the dominant contribution to the entanglement entropy of $A$ as:
\begin{equation}
    S_A = \begin{cases}
      \log L_{A,q_A} & \quad L_{A,q_A} \ll L_{B,q_B},\\
      \log L_{B,q_B} & \quad L_{A,q_A} \gg L_{B,q_B}.
    \end{cases}
\end{equation}

On the other hand, the cases of $A_1$ and $A_2$ entanglement entropies are less straightforward as there is a charge conservation constraint between them. For a general Renyi index $n$, one can work out the form of the Renyi entropy of $A_1$ subsystem as (similar relations hold for $A_2$):
\begin{equation}
    S_{A_1}^{(n)} = \sum_{q_1} S_{A_1,q_1}^{(n)},
\end{equation}
with the symmetry resolved Renyi enetropy defined as:
\begin{equation}
    S_{A_1,q_1}^{(n)} = \begin{cases}
      \frac{L_{A_1,q_1} L_{A_2,\bar{q}_1}^n}{L_{A,q_A}^n}  & \quad L_{A_1,q_1} \ll L_{A_2,\bar{q}_1} L_{B,q_B},\\
      \frac{L_{A_1,q_1}^n}{L_{A,q_A}^n} L_{A_2,\bar{q}_1} L_{B,q_B}^{1-n} & \quad L_{A_1,q_1} \gg L_{A_2,\bar{q}_1} L_{B,q_B}.
    \end{cases}
\end{equation}

Having the above relations at hand, one can now calculate the mutual information for the two symmetry cases of our interest in this work, i.e.~$\mathbb{Z}_R$ and $U(1)$; first, we consider $\mathbb{Z}_R$: this is a simple case, since in this case the size of symmetry resolved Hilbert spaces are equal regardless of the quantum number they correspond to. We consider three different cases here:
\begin{itemize}
    \item $N_A < N_B$:
    \begin{equation}
        \left\langle     I_{A_1,A_2}\right\rangle = \log  R.
    \end{equation}
    \item $N_A > N_B$ and $N_{A_{s}} < N_B + N_{A_{\bar{s}}} - 1$:
    \begin{equation}
        \left\langle     I_{A_1,A_2}\right\rangle = (N_A-N_B-1)\log  R.
    \end{equation}
    \item $N_A > N_B$ and $N_{A_{s}} > N_B + N_{A_{\bar{s}}} - 1$:
    \begin{equation}
        \left\langle     I_{A_1,A_2}\right\rangle = N_{A_2}\log  R.
    \end{equation}
\end{itemize}
  
At this point, we consider the $U(1)$ symmetry and calculate its mutual information. The subtlety here is that Hilbert space size depends on the quantum number. As a result of this, we consider the thermodynamic limit and perform a saddle point approximation. The first thing we need to calculate is the entanglement entropy of the $A_1$ subsystem. We consider first a case where the dominant contributions to the entanglement entropy are from those subspaces that satisfy the relation $L_{A_1,q_1} \ll L_{A_1,\bar{q}_1} L_{B,q_B}$:
\begin{widetext}
\begin{equation}
\begin{aligned}
    S_{A_1} &= \lim_{n \to 1} \frac{1}{1-n} \log \sum_{n_{A_1}} \frac{\binom{N_{A_1}}{n_{A_1}} \left[ \binom{N_{A_2}}{n_A-n_{A_1}} \right]^n}{ \left[ \binom{N_A}{n_A} \right]^n }\\
    & \approx \lim_{n \to 1} \frac{1}{1-n} \log \sum_{n_{A_1}}
    \frac{\binom{N_{A_1}}{n_{A_1}} \binom{N_{A_2}}{n_A-n_{A_1}} \left[ 1+ (n-1) \log\binom{N_{A_2}}{n_A-n_{A_1}} \right]
    }{     \left[ \binom{N_A}{n_A} \right]^n }\\
    & \approx \log\binom{N_A}{n_A} - \sum_{n_{A_1}} \frac{\binom{N_{A_1}}{n_{A_1}} \binom{N_{A_2}}{n_A-n_{A_1}}   \log\binom{N_{A_2}}{n_A-n_{A_1}} 
    }{      \binom{N_A}{n_A} } \\
    & \approx  -\frac12 \log N_A + N_{A} f(\nu_A) + \frac12 \log N_{A_2} - N_{A_2} f(\nu_A)  \\
    & = N_{A_1} f(\nu_A) - \frac12 \log \left(\frac{N_{A}}{N_{A_2}} \right)
\end{aligned}
\end{equation}
\end{widetext}
In going from the first row to the second we have taken $n$ to be close to one and Taylor expanded, form the second row to the third row we have used $\sum_{n_{A_1}} \binom{N_{A_1}}{n_{A_1}} \binom{N_{A_2}}{n_A-n_{A_1}} = \binom{N_{A}}{n_{A}}$ and Taylor expanded the outer logarithm. In going from the third row to the fourth row we have done a continuum approximation and replaced the sum by an integral and did a saddle point approximation, we are furthermore using the notation introduced in Sec.~\ref{sec:thermo_limit_U(1)_plateau}, where $\nu_s$ shows the filling factor of a subsystem and the function $f$ is defined in Eq.\eqref{eq:definition_f}. The saddle point solution is found to be $\nu_{A_1} = \nu_A$. Note that the calculation done above is similar in spirit to those in \cite{murthy2019structure}, where energy conservation considerations are taken into account.

Similarly, in the opposite limit where the  dominant contributions are from those subspaces with $L_{A_1,q_1} \gg L_{A_1,\bar{q}_1} L_{B,q_B}$, one can show that the entanglement entropy reads:
\begin{equation}
    S_{A_1} = N_{A_2} f(\nu_A) - \frac12 \log \left(  \frac{N_A}{ N_{A_1} } \right) + \log L_{B,q_B}.
\end{equation}
One can now calculate the mutual information in different regime given the above forms:
\begin{itemize}
    \item $N_A < N_B$
        \begin{equation}
        \begin{aligned}
            \left\langle     I_{A_1,A_2}\right\rangle &= - \frac12  \log\left( \frac{N_A}{N_{A_1} N_{A_2}}\right) \\
            &+ \frac12 \log\left( 2\pi \nu_A [1-\nu_A]   \right).
        \end{aligned}
        \end{equation}
    \item $NA>N_B$ and $N_{A_s} < N_B + N_{A_{\bar s}}$:
        \begin{equation}
        \begin{aligned}
            \left\langle     I_{A_1,A_2}\right\rangle &= 
            N_A f(\nu_A) - N_B f(\nu_B)\\
            &- \frac12  \log\left( \frac{N_A^2}{N_{A_1} N_{A_2} N_B}\right) \\
            &+ \frac12 \log\left( 2\pi \nu_A [1-\nu_A]   \right).
        \end{aligned}
        \end{equation}
    \item $NA>N_B$ and $N_{A_s} > N_B + N_{A_{\bar s}}$:
        \begin{equation}
        \begin{aligned}
        \label{eq:muinf-u1-me}
            \left\langle     I_{A_1,A_2}\right\rangle &= 
            2 N_{A_2} f(\nu_A) \\
            &- \frac12  \log\left( \frac{N_A}{N_{A_1} }\right) .
        \end{aligned}
        \end{equation}
\end{itemize}

\begin{figure*}[!t]
    \centering
    \begin{quantikz}
    \lstick{$\ket{0}$} & \gate{H} & \ctrl{1} & \ctrl{2} & \ctrl{3} & \gate{U_a} & \gate{H} & \meter{}\\
     \lstick[wires=3]{$B \ $} 
     & \qw & \gate{Z^{1/2^{n-1} }} & \qw & \qw & \qw & \qw & \qw \\
     & \qw & \qw & \gate{Z^{1/2^{n-1} }} & \qw & \qw & \qw & \qw \\
     & \qw & \qw & \qw & \gate{Z^{1/2^{n-1} }} & \qw & \qw & \qw \\
     \lstick[wires=4]{$A \ $}
     & \qw & \qw & \qw & \qw & \qw & \qw & \qw \\
     & \qw & \qw & \qw & \qw & \qw & \qw & \qw \\
     & \qw & \qw & \qw & \qw & \qw & \qw & \qw \\
     & \qw & \qw & \qw & \qw & \qw & \qw & \qw
    \end{quantikz}
    \caption{The circuit that is used to measure the charge of the $B$ subsystem modulo $2^n$.}
    \label{fig:circuit_U_1}
\end{figure*}
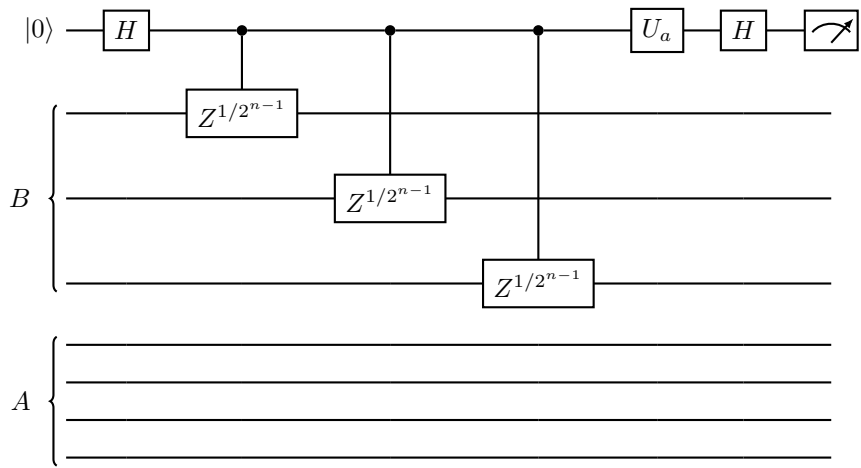
\section{Measuring $\mathbb{Z}_R$ and $U(1)$ charges}
\label{app:measurement}
\renewcommand\theequation{D\arabic{equation}}

In this section of the appendix we discuss how the $\mathbb{Z}_R$ charge of the $B$ subsystem in a qudit system ($d=R$) or the $U(1)$ charge of $B$ in a qubit system is measured.

We start by considering the $\mathbb{Z}_R$ symmetry first. The Hilbert space of each site is spanned by the basis $\left\{ |i\rangle \right\}$ with $i=0,1,\ldots,R-1$. In this case, one needs an ancillary qudit and the circuit discussed in the main text works provided that all the unitaries are generalized to qudit gates. In particular:
\begin{itemize}
    \item the control-$Z$ operator is generalized to have the form $CZ_{(R)} = \sum_{j} |j\rangle \langle j | \otimes Z^j  $, where now $Z = \mathrm{diag}(1, \omega, \omega^2, \ldots, \omega^{R-1} )$, and $\omega = e^{ \frac{2\pi i}{R} }$.
    \item The $\mathbb{Z}_R$ generalization of  Hadamard reads $H_R |j\rangle = \frac{1}{\sqrt{R}} \sum_{i=0}^{R-1} \omega^{ij} |i \rangle$. Noting that this general Hadamard gate is not Hermitian, one needs to modify the second Hadamard acting on the ancilla to $H^\dag$.
\end{itemize}
It is straightforward now to show that if the measurement outcome is $|k\rangle$, the $\mathbb{Z}_R$ charge of $B$ is also determined to be $k$. 

We now discuss how the charge of subsystem $B$ in the case of a $U(1)$ symmetry can be measured. This measurement is done in a series of steps where in consecutive steps the charge modulo $2, 4, 8, \ldots$ is measured determining the whole charge. A total of $\lceil \log_2 (N_B + 1) \rceil$ consecutive measurements is needed as explained below.
For each step an ancilla is utilized to implement the circuit shown in Fig.~\ref{fig:circuit_U_1}. The unitary $Z^{1/2^n}$ is defined as 
$\begin{pmatrix}
1 & 0 \\
0 & e^{\frac{\pi i}{2^n}}
\end{pmatrix}$.
We describe the steps below:
\begin{itemize}
    \item Using the same circuit shown in Fig.~\ref{fig:circuit_measuring_party}, the total charge of $B$ modulo $2$ is measured which we denote as $q_1$.
    \item In the second step, in order to measure the charge modulo $4$, the circuit in Fig.~\ref{fig:circuit_U_1} is implemented with control-$Z^{1/2}$ operators. Note that $Z^{1/2}$ is identical to the $S$ gate. After the controlled gates are applied, one acts with the unitary $U_a$ on the ancilla. This unitary depends on the outcome of the charge modulo $2$ measurement, i.e.~$U_a = (Z^{1/2})^{- q_1}$. This will result in the charge modulo $4$ which we denote as $q_2$.
    \item In general, for measuring the charge modulo $2^n$, which we denote as $q_n$, one utilizes control-$Z^{1/2^{n-1}}$ operators. Furthermore, $U_a$ should be chosen based on the outcome of all previous measurements as $U_a = (Z^{1/2})^{- q_{n-1} }$. This procedure is continued until the charge of subsystem $B$ is determined.
\end{itemize}

\section{Negativity spectrum without symmetry projection}
\label{app:unprojected}

\renewcommand\theequation{E\arabic{equation}}

In this appendix, we provide Schwinger-Dyson equations for the resolvent function associated with the partial transpose of the full block-diagonal density matrix. 
In other words, we consider the spectrum of $\hat\rho_A^{T_2}$ when the total charge of $A$ is not projected.
This calculation is mostly for completeness, since we believe that the partial transpose of the full density matrix is not a good indicator of quantum entanglement when classical correlations are only due to symmetric LOCCs. Furthermore, this result may be useful if one wants to study the entanglement negativity of random states at finite energy density or the microcanonical ensemble with the total energy constraint $N_{A} \epsilon_A + N_B \epsilon_B = \text{const}$. As usual, we provide two sets of self-consistent equations: One for the replica symmetry breaking (or semicircle) regime and Two, the more general case which not only contains the former regime but also the maximal entanglement regime.

\subsection{Replica symmetry breaking regime}
  Here, we need to take all blocks of $\hat\rho_{A}^{(q_A)}$ in $\hat\rho_{A}$ into account at the same time. Eq.~\eqref{eq:Gz_expansion} still holds, but since the charge in subsystem $A$ is not determined, one needs to sum over all possible charge values in subsystem $A$ (or subsystem $B$ since the total charge is fixed). This means that the equations for $\Sigma$ would be modified as follows:
\begin{widetext}
\begin{equation}
\begin{aligned}
    \Sigma &= \bigoplus_{q_1, q_2} \mathbb{1}_{A_1,q_1} \otimes \mathbb{1}_{A_2,q_2}  \bigg[ \frac{1}{\sum_{\tilde{q}_A} L_{\tilde{q}_A}  L_{\tilde{q}_B}  }    L_{B, Q - q_1 -q_2} \\
    &+ \left(\frac{1}{\sum_{\tilde{q}_A} L_{\tilde{q}_A}  L_{\tilde{q}_B} } \right)^2  \sum_{q_B} L_{B,q_B} \;
     L_{A_2, \; Q - q_1 - q_B } \; L_{A_1, \; Q -  q_2 - q_B } \; G_{Q - g_q - q_B , Q - q_1 -q_B }   \bigg].
\end{aligned}    
\end{equation}
\end{widetext}
Note that here unlike before, the equations for different components of the resolvent function depend on each other; in fact, the set of equations governing $G_{q_1,q_2}$ with $\Delta q = q_1 - q_2$ kept constant is closed. Hence, we can label the spectral density with the charge imbalance $\Delta q$ (See Refs.~\cite{Cornfeld2018, Murciano,Neven2021} for a similar observation). Unlike the case of $U(1)$ symmetry, it is easy solve the above equation for the $\mathbb{Z}_R$ symmetry group, as all symmetry sectors have the same Hilbert space size. Ultimately, the spectral density is found to be
\begin{equation}
\begin{aligned}
    P(\xi) &= \frac{1}{2\pi} \  R^{2N_A+N_B}  \  
    \sqrt{ \frac{4}{R^{N}} - \left( \xi - \frac{1}{R^{N_A}} \right)^2 }' .
\end{aligned}    
\end{equation}
This means that the negativity shows a plateau when $R^{\left( N_A - N_B \right)} > \frac14$.

\subsection{General case}
The self energy in the general case where there is no projection on $\hat\rho_A$ can be written as:
\begin{widetext}
\begin{equation}
\begin{aligned}
    \Sigma &= \bigoplus_{q_1, q_2} \mathbb{1}_{A_1,q_1} \otimes \mathbb{1}_{A_2,q_2} \\
    & \bigg[ \frac{1}{\sum_{\tilde{q}} L_{\tilde{q}_A}  L_{\tilde{q}_B  }}    L_{B, Q - q_1 - q_2}     \; \frac{1}{1-\left( \frac{L_{A_1,q_1}}{\sum_{\tilde{q}} L_{\tilde{q}_A}  L_{\tilde{q}_B  }} G_{q_1q_2} \right)^2}\\
    &+ \left(\frac{1}{\sum_{\tilde{q}} L_{\tilde{q}_A}  L_{\tilde{q}_B  } } \right)^2  \sum_{q_B} L_{B,q_B} 
    \frac{ L_{A_2,Q - q_1 - q_B } \; L_{A_1, Q - q_2 - q_B } \; G_{Q - q_2- q_B , Q -  q_1 - q_B } }
    {1 - \frac{L_{A_1,q_1} L_{A_1, Q - q_2 - q_B} }
    {\left( \sum_{\tilde{q}} L_{\tilde{q}_A}  L_{\tilde{q}_B  } \right)^2}
    G_{q_1q_2}  G_{Q -  q_2 -  q_B , Q -  q_1  - q_B }  }  \bigg].
\end{aligned}    
\end{equation}
\end{widetext}
Note that this includes both the replica symmetry breaking regime that was discussed above and also the regime of maximal entanglement between $A_1$ and $A_2$.

\bibliography{refs.bib}

\end{document}